\documentclass[12pt,english,longbibliography,nofootinbib,superscriptaddress,sort&compress,showkeys]{article}
\usepackage[T1]{fontenc}
\usepackage[utf8]{inputenc}
\usepackage{amsmath}
\usepackage{amssymb}
\usepackage{graphicx}
\usepackage[letterpaper]{geometry}
\usepackage[numbers]{natbib}

\makeatletter

\newcommand{\lyxmathsym}[1]{\ifmmode\begingroup\def\b@ld{bold}
  \text{\ifx\math@version\b@ld\bfseries\fi#1}\endgroup\else#1\fi}

\providecommand{\tabularnewline}{\\}

\usepackage{fancyhdr}
\usepackage{amsfonts}
\usepackage{tablefootnote}
\usepackage{caption}
\usepackage[english]{babel}
\usepackage{subcaption}

\makeatother

\usepackage{babel}
\begin{document}
\title{{\large\textbf{The effect of normal stress on stacking fault energy
in face-centered cubic metals}}}
\author{Yang Li and Yuri Mishin\\
{\normalsize Department of Physics and Astronomy, MSN 3F3},\\
 {\normalsize George Mason University, Fairfax, VA 22030, USA}}
\maketitle
\begin{abstract}
Plastic deformation and fracture of FCC metals involve the formation
of stable or unstable stacking faults (SFs) on (111) plane. Examples
include dislocation cross-slip and dislocation nucleation at interfaces
and near crack tips. The stress component normal to (111) plane can
strongly affect the SF energy when the stress magnitude reaches several
to tens of GPa. We conduct a series of DFT calculations of SF energies
in six FCC metals: Al, Ni, Cu, Ag, Au, and Pt. The results show that
normal compression significantly increases the stable and unstable
SF energies in all six metals, while normal tension decreases them.
The SF formation is accompanied by inelastic expansion in the normal
direction. The DFT calculations are compared with predictions of several
representative classical and machine-learning interatomic potentials.
Many potentials fail to capture the correct stress effect on the SF
energy, often predicting trends opposite to the DFT calculations.
Possible ways to improve the ability of potentials to represent the
stress effect on SF energy are discussed.
\end{abstract}
\date{}


\section{Introduction}\label{sec:Introduction}

A generalized stacking fault (GSF) is a planar defect created by sliding
one part of a single crystal past another along a given atomic plane
in a particular crystallographic direction \citep{anderson2017theory}.
The GSF energy (GSFE) describes the variation in potential energy
per unit area as a function of the translation vector \citep{zimmerman2000generalized,Lu00a,Yan04,Wang:2015aa}.
GSFE governs the dislocation behavior during plastic deformation and
fracture of crystalline materials \citep{anderson2017theory,bulatov1998connecting}.
It constitutes the critical input to the Peiers-Nabarro model \citep{anderson2017theory,Rice1992,Liu2017,Lu2000b,Schoeck01,Lu:2005aa}
and phase field dislocation dynamics simulations \citep{beyerlein2016understanding,Kim:2021aa,hunter2011influence}.

In face-centered cubic (FCC) metals, the most important GSF is obtained
by sliding (111) atomic planes past each other in a $\left\langle 211\right\rangle $
direction. The GSFE has a local minimum at the translation $\mathbf{u}=a[11\bar{2}]/6$,
where $a$ is the cubic lattice constant. This minimum is separated
from the perfect crystalline state ($\mathbf{u}=0$) by a local maximum.
The metastable state corresponding to the energy minimum represents
the intrinsic stacking fault (SF) whose energy (SFE) is one of the
most fundamental properties of FCC metals \citep{borovikov2016effects,anderson2017theory}.
The maximum corresponds to the unstable stacking fault (USF) whose
energy (USFE) also plays an important role in many mechanical processes.
For example, the SFE affects the dissociation width of full dislocations
into Shockley partials. This width largely controls the dislocation
cross-slip because the partials must recombine (thus eliminating the
SF) before the dislocation can change the glide plane. The USFE controls
the nucleation of dislocations at a crack tip \citep{Rice1992,AndersonR86}
and at surfaces and internal interfaces \citep{Xu:2003aa}. USFE is
especially important in dislocation nucleation-controlled plasticity,
which dominates the mechanical behavior of nanowires \citep{Cao2015,Li:2018ab,Mordehai:2018aa},
pristine nanoparticles \citep{amodeo2017mechanical,Sharma:2018aa,Liang:2024aa},
and similar defect-free nanoscale objects. 

GSFE can be altered by an applied stress \citep{yan2021effects,linda2022effect}.
The most accurate investigations of the stress effect on GSFE utilize
first-principles density functional theory (DFT) calculations. Brandl
et al.~\citep{brandl2007general} applied DFT calculations to study
the GSFE in Al, Ni, and Cu under isotropic and shear strains. They
found that under isotropic volume strain, the SFE of all three metals
decreases with increasing volume, with Cu and Ni being less sensitive
to the volume changes than Al. Under volume-conserving shear strain,
the USFE of Ni and Al increases while the SFE decreases. In contrast,
the GSFE curves for Cu exhibit only minor changes. Branicio et al.~\citep{branicio2013effect}
computed the SFE in Cu under volumetric, uniaxial, and shear strains.
It was found that under uniaxial strain applied along the {[}111{]}
direction, the SFE of Cu increases as the strain changes from tension
to compression. Andric et al.~\citep{andric2019stress} analyzed
the tensile stress dependence of the GSFE in Ni, Cu, Al, and Mg. Their
DFT calculations have shown that in all three metals, the SFE decreases
with increasing tensile stress normal to (111) planes.\footnote{Andric et al.~\citep{andric2019stress} considered two different
definitions of GSFE. The thermodynamically consistent definition of
GSFE in the presence of applied stresses (see section \ref{subsec:DFT-calculations}
below) corresponds to what they called the GSF enthalpy. } 

Interatomic potentials have also been employed to study GSFEs under
applied stresses. For example, calculations with an embedded-atom
potential for Cu have shown that compressive (tensile) stress normal
to the SF increases (decreases) the USFE and decreases (increases)
the SFE \citep{tschopp2008influence}. The latter trend contradicts
the DFT calculations \citep{branicio2013effect,andric2019stress}.
Similar results were obtained in a more recent study by Zhang et al.~\citep{zhang2015molecular},
who investigated the effect of stress on GSFE using embedded-atom
potentials for Cu, Al, and Ni. When uniaxial stress ranging from $-5$
GPa to $5$ GPa was applied along the {[}111{]} direction, the USFE
was found to increase as the stress changed from tension to compression.
As will be shown later in this paper, this response is qualitatively
consistent with DFT calculations. The SFE of Al also increased as
the stress changed from tension to compression, which is also consistent
with DFT calculations. However, the SFE in Cu and Ni decreased as
the stress changed from tension to compression instead of increasing.
Significant disagreements between potential predictions and DFT calculations
for the stress effect on SFE were also reported by other authors \citep{branicio2013effect,andric2019stress}. 

Such disagreements are highly alarming because the large-scale computer
simulations of deformation and fracture of materials rely on interatomic
potentials. The reliability of such simulations depends on the potential's
ability to faithfully represent the impact of applied stresses on
the GSFE and thus the dislocation behavior. This is especially important
for the simulations of nano-mechanical processes involving high stresses
on the level of tens of GPa or higher, including shock deformation
and the deformation and failure of nanowires and defect-free nanoparticles.
There is a pressing need to understand the extent and origins of the
mentioned disagreements and find ways to reduce them. This is especially
important for the classical interatomic potentials. Although they
are generally less accurate than the recently emerged machine-learning
potentials \citep{Mishin:2021uc}, they are computationally much more
efficient, giving access to molecular dynamics (MD) simulations of
large collections of atoms (e.g., up to $10^{8}$ or even $10^{9}$
atoms) or long MD simulations (e.g., up to microseconds). High stresses
often arise in MD simulations involving plastic deformation and fracture. 

In this work, we perform a systematic DFT investigation of the effect
of normal stress on GSFE in FCC metals, focusing on the energies of
the stable and unstable faults (SFE and USFE). We additionally investigate
the effect of normal stress on the shear modulus and Poisson's ratio
of the metal. The reason for including the elastic properties is that
the dislocation dissociation width is controlled by the balance between
the elastic repulsion between the Shockley partials and the SF tension.
Thus, predictions of the stress effect on the dissociation width require
the knowledge of both factors. 

The calculations are performed for six FCC metals representing simple
metals (Al), noble metals (Cu, Ag, Au, and Pt), and transition metals
(Ni). The DFT results are compared with a set of representative interatomic
potentials for these metals. We include several types of classical
(also known as traditional \citep{Mishin:2021uc}) potentials and
three types of machine-learning (ML) potentials \citep{Zhang:2025aa}.
Our goal is to evaluate the capabilities of interatomic potentials
to predict the correct SF energies in FCC metals, as well as the dislocation
dissociation width, under high-stress conditions. After a detailed
comparison of the potential predictions with DFT calculations, we
discuss possible approaches to improve the reliability of classical
potentials for modeling high-stress deformation of FCC metals.

\section{Computational Methodology}

\subsection{DFT calculations}\label{subsec:DFT-calculations}

Two types of models are commonly employed in the DFT calculations
of SFEs: the tilted-cell method~\citep{jhon2005computational,kibey2006generalized}
and the slab method~\citep{brandl2007general}. In the tilted-cell
method, a shear is applied to the lattice vectors in a direction parallel
to the prescribed slip plane to generate a single stacking fault while
preserving periodic boundary conditions. In the slab method, a crystal
slab with two open surfaces is constructed, and the stacking fault
is generated by rigidly shifting the upper half of the slab relative
to the lower half. The slab thickness must be large enough to minimize
the surface effects on the fault energy. To further remove surface
contributions to the energy, the faulted slab energy is compared with
that of a reference slab containing the same open surfaces but without
the stacking fault. Due to the presence of open surfaces, the stress
state of the cell cannot be controlled precisely. Therefore, in this
study, we adopted the tilted-cell method, which inherently avoids
the influence of free surfaces. 

\begin{figure}[!ht]
\vspace{0cm}
 \centering \includegraphics[height=8cm]{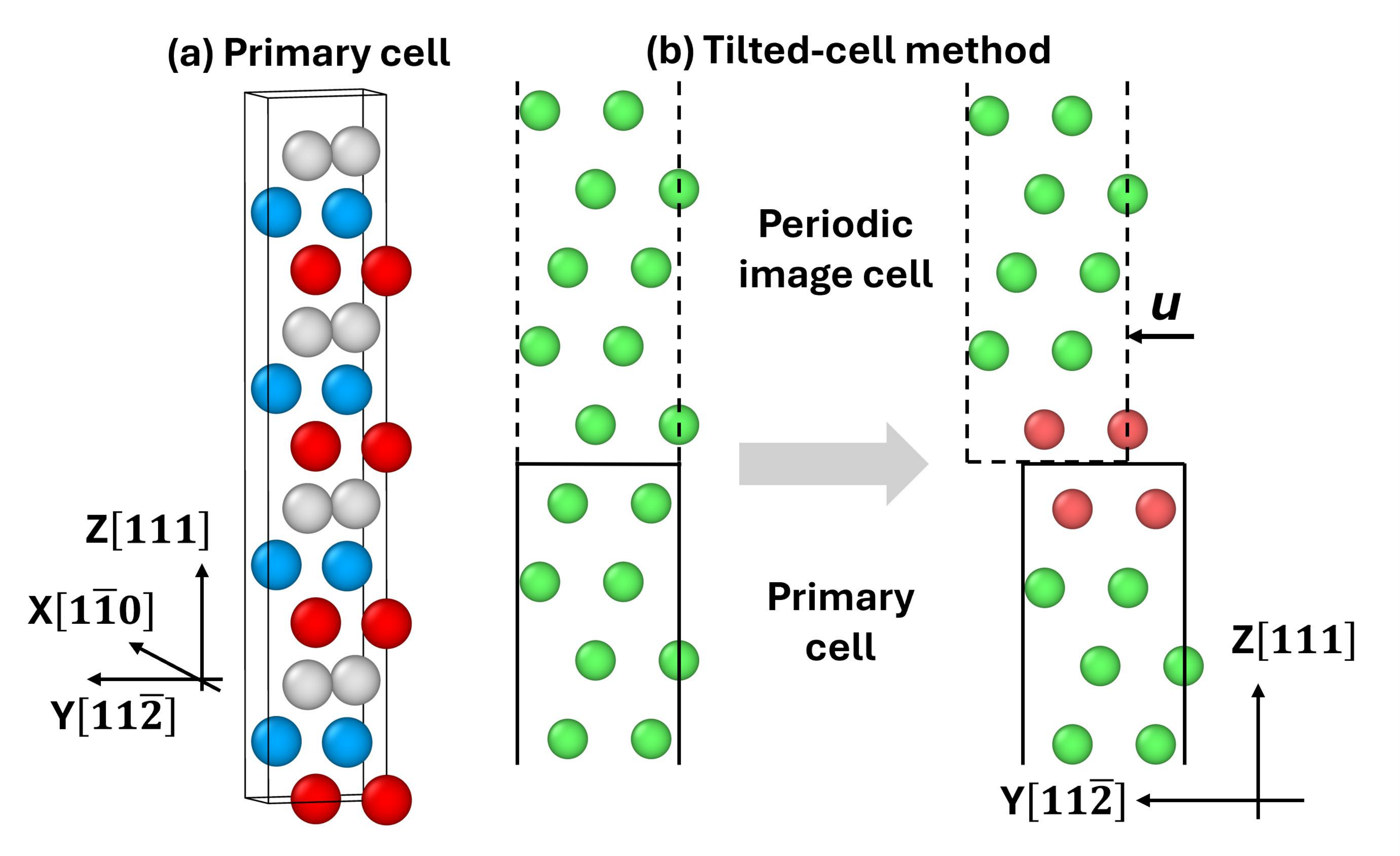} \caption[]{ (a) Atomic structure of the cell used in the GSFE calculations.
The cell contains 24 (111) layers, each with two atoms. The atoms
are colored by their $Z$ positions to provide better visualization
of the structure. (b) Illustration of the tilted-cell method. By tilting
the $Z$ axis towards $Y$ by an amount $u$, a stacking fault is
created at the boundary between the primary cell and its periodic
image. The solid rectangle marks the primary cell, and the dashed
rectangle marks its periodic image. The atoms in FCC and non-FCC environments
are shown in green and red, respectively.}
\label{fig:Supercell} 
\end{figure}

All DFT calculations were carried out with the Vienna \textit{Ab initio}
Simulation Package (VASP) \citep{kresse1996efficiency,kresse1996efficient}
using the projector-augmented-wave (PAW) method \citep{blochl1994projector}
and the Perdew--Burke--Ernzerhof (PBE) \citep{perdew1996generalized}
generalized gradient approximation for exchange--correlation. The
PAW datasets employed here were \texttt{PAW\_PBE} Al (04Jan2001),
Cu (22Jun2005), Au (04Oct2007), Pt (04Feb2005), Ni (02Aug2007), and
Ag (02Apr2005). (The release date is indicated in parentheses.) A
plane-wave kinetic-energy cutoff of 500~eV was used. To calculate
the GSFE, a periodic cell containing 24 atoms arranged in 12 (111)
atomic layers was constructed (Fig. \ref{fig:Supercell}(a)). The
cell was orthogonal with the following crystallographic orientations
of the Cartesian axes: $X\!\parallel\![1\bar{1}0]$, $Y\!\parallel\![11\bar{2}]$,
and $Z\!\parallel\![111]$. Brillouin-zone sampling employed Monkhorst--Pack
$48\times26\times4$ meshes. For the calculations of the shear modulus
and Poisson's ratio, a simulation cell containing 6 atoms arranged
in 3 (111) atomic layers was created, and the Brillouin-zone sampling
employed Monkhorst--Pack $64\times36\times26$ meshes. First-order
Methfessel--Paxton smearing with the width of 0.20 eV was applied
during the Brillouin-zone integration. The electronic self-consistency
criterion was set to $10^{-7}\,\mathrm{eV}$ or $10^{-8}\,\mathrm{eV}$,
depending on the material. The ionic relaxation was considered converged
when the maximum residual force fell below $10^{-3}\,\mathrm{eV}\,\text{\AA}^{-1}$
or $10^{-4}\,\mathrm{eV}\,\text{\AA}^{-1}$, depending on the material.
Calculations for Ni were spin-polarized with the initial magnetic
moments of 0.6~$\mu_{\mathrm{B}}$ per Ni atom. Other elements were
treated as non-magnetic.

The GSFE was computed by the following steps. 
\begin{enumerate}
\item The initial fault-free structure was optimized by a conjugate-gradient
scheme until forces and stresses met the convergence criteria, leaving
the principal stress components along $X$, $Y$, and $Z$ close to
zero. 
\item A stacking fault was created by deforming the cell shape to triclinic
(i.e., tilting the axis $Z$ toward $Y$) with the $YZ$ shear component
of a chosen magnitude $u$ while keeping the Cartesian coordinates
of all atoms fixed (no remapping) (Fig.~\ref{fig:Supercell}(b)).
Periodic boundary conditions were maintained in all directions. As
a result, a stacking fault formed at the boundary of the periodic
simulation cell, with the fault plane normal to $[111]$. 
\item With the cell shape and size held fixed, the total energy was minimized
by relaxing the atomic positions only along the $[111]$ direction
(with the lateral components fixed). The relaxation was performed
using a conjugate gradient algorithm and was terminated once the maximum
force on any atom fell below $10^{-3}\,\mathrm{eV}\,\text{\AA}^{-1}$
or $10^{-4}\,\mathrm{eV}\,\text{\AA}^{-1}$, depending on the material.
After the relaxation, the normal stress component $\sigma_{n}$ (along
the $[111]$ direction) was generally nonzero because introducing
a fault is usually accompanied by an inelastic change in the atomic
density in the faulted layer, whereas the cell length $L$ along $[111]$
was constrained to remain fixed.
\item To eliminate the residual stress, the cell length $L$ was varied
in increments of 0.02\%, repeating step 3 after each increment until
the normal stress $\sigma_{n}$ became negligible. 
\item The GSFE at the targeted normal stress $\sigma_{n}$ was calculated
using the relaxed perfect (from step 1) and faulted (from step 4)
cells as reference configurations. Both cells were subjected to incremental
uniaxial strains along $[111]$ in steps of $\pm0.02\%$, with step~3
repeated after each increment until $\sigma_{n}$ in \emph{both} cells
matched the target. The cross-sectional area $A$ perpendicular to
$[111]$ was held fixed (no lateral strain). 
\end{enumerate}
The GSFE $\gamma$ was computed from the formula
\begin{equation}
\tag{1}\gamma(u)=\frac{E(u)-E(u=0)}{A}-\left[L(u)-L(u=0)\right]\sigma_{n}.\label{eq:gsfe}
\end{equation}
Here, $E(u=0)$ and $L(u=0)$ represent the energy and length of the
perfect cell under stress $\sigma_{n}$, while $E(u)$ and $L(u)$
are the respective quantities for the faulted cell under the \emph{same}
stress. The second term in the right-hand side of Eq.~(\ref{eq:gsfe})
represents the work (per unit area) done on the system by the external
load. This term is subtracted from the total energy change to obtain
the work of the local forces causing the atomic rearrangements and
bond distortions within the fault region during its creation. This
is the proper thermodynamic definition of the interface/fault energy.
Note that the lattice regions outside the SF are in the same physical
state as in the reference perfect lattice system.

The procedure described above is time-consuming and computationally
expensive in the DFT calculations. Therefore, we applied it only to
obtain the intrinsic SFE ($\gamma_{\mathrm{SF}}$) corresponding to
the displacement $u=\tfrac{a}{6}[11\bar{2}]$. For other shear displacements
$u$, the GSFE was evaluated as 
\begin{equation}
\tag{2}\gamma(u)=\frac{E_{L(u=0)}(u)-E(u=0)}{A},\label{eq:gsfe1}
\end{equation}
where $E_{L(u=0)}(u)$ is the energy of the faulted cell that has
the same length $L(u=0)$ as the perfect cell. Eq.~(\ref{eq:gsfe1})
calculates the energy difference between the perfect and faulted cells
with the same length along the {[}111{]} direction. The normal stress
in the perfect cell meets the target value of $\sigma_{n}$, whereas
$\sigma_{n}$ in the faulted cell does not exactly match the target
value but remains close. In other words, Eq.~(\ref{eq:gsfe1}) compares
the energies of the perfect and faulted cells at fixed cell length,
whereas Eq.~(1) compares them at a fixed stress. Eq.~(\ref{eq:gsfe1})
requires fewer DFT calculations because it bypasses the step of determining
the faulted cell length corresponding to the target stress. We compared
the SFE values predicted by Eqs.~(\ref{eq:gsfe}) and (\ref{eq:gsfe1})
and found that they differed by less than 3\% over the entire stress
range considered in this work, from $-20$ to $20~\text{GPa}$. Thus,
to reduce computational cost, we used Eq.~(\ref{eq:gsfe1}) to calculate
the GSFE and USFE while still using the exact Eq.~(\ref{eq:gsfe})
for the SFE calculations. We note that Eq.~(\ref{eq:gsfe1}) was
frequently employed in previous DFT studies of GSFE~\citep{brandl2007general,branicio2013effect},
in which the outermost layers of the faulted cell were held fixed
during the shear displacement (i.e., under a fixed-strain or clamped-boundary
condition). 

For the shear modulus and Poisson's ratio calculations, the structural
relaxation procedure was the same as in step 1 above. The cell length
along $[111]$ was varied incrementally, the structure was relaxed,
and the normal stress $\sigma_{n}$ was measured after each increment.
This process was repeated until $\sigma_{n}$ matched the target stress
value. To compute the shear modulus, $YZ$ shear strains $\epsilon_{yz}$
ranging from $-0.003$ to $0.003$ were applied to the cell in increments
of $0.001$. The corresponding shear stress $\sigma_{yz}$ was measured
as a function of $\epsilon_{yz}$. The measured stress-strain relation
was fitted with a straight line, and its slope was taken as the shear
modulus $G$. To compute Poisson's ratio, a set of small strains $\epsilon_{xx}$,
$\epsilon_{yy}$, $\epsilon_{zz}$, $\epsilon_{xy}$, $\epsilon_{yz}$,
and $\epsilon_{zx}$ was applied sequentially, and the corresponding
changes in stress were measured to obtain the components of the tangent
stiffness tensor. The compliance tensor $S_{ijkl}$ was then computed
by inverting the stiffness tensor, and Poisson's ratio was calculated
from the equation \citep{Nye-book}
\begin{equation}
\nu_{ij}=-\dfrac{S_{iijj}}{S_{iiii}}.\label{eq:4}
\end{equation}
This ratio represents the lateral strain in the $j$-th direction
in response to axial strain applied in the $i$-th direction. In this
work, we focused on the components $\nu_{31}$ and $\nu_{32}$ corresponding
to the lateral strains in the $X\!\parallel\![1\bar{1}0]$ and $Y\!\parallel\![11\bar{2}]$
directions when tension or compression were applied in the $Z\!\parallel\![111]$
direction. 

\subsection{Calculations with interatomic potentials}

Calculations with interatomic potentials were performed using the
Large-scale Atomic/Molecular Massively Parallel Simulator (LAMMPS)~\citep{thompson2022lammps}.
For consistency, the simulation cells were similar in shape and size
to those in the DFT calculations. The structures were relaxed by energy
minimization until the changes in energy and force fell below $etol=10^{-10}$
and $ftol=10^{-10}\,\mathrm{eV}\,\text{\AA}^{-1}$, respectively.
In contrast to the DFT calculations, it was not necessary to repeatedly
vary the supercell length in the {[}111{]} direction to achieve the
target stress value. LAMMPS provides a direct functionality for controlling
the stress along a specified direction. Consequently, all GSFE calculations
were carried out using Eq.~(\ref{eq:gsfe}), without the need for
the approximate Eq.~(\ref{eq:gsfe1}). In the shear modulus calculations,
shear strains $\epsilon_{yz}=\pm1.0\times10^{-6}$ were applied to
a cell pre-deformed by a set normal stresses $\sigma_{n}$. The shear
modulus was obtained from the relation $G=\delta\sigma_{yz}/\delta\epsilon_{yz}$,
where $\delta\sigma_{yz}$ is the difference in $YZ$ stress between
the two shear strains, and $\delta\epsilon_{yz}=2.0\times10^{-6}$
is the corresponding strain difference. Poisson's ratio was computed
by the same method in the DFT calculations.

Table~\ref{table:table1} lists the interatomic potentials for the
six FCC metals tested in this work. Most of them are in the embedded-atom
method (EAM) \citep{daw1993embedded}, modified EAM (MEAM) \citep{Baskes92},
and angular-dependent potential (ADP) \citep{Mishin05a} formats,
which are most suitable for metallic systems. When we test more than
one EAM potential for the same metal, we distinguish them by the first
author's name.\footnote{The first author's name is only used to label the potentials and does
not imply any credit attribution.} For a broader exploration of the effect of the potential format on
the results, we included modified Tersoff (MT) potentials for Pt \citep{Pt-in-review}
and Al (this work), which are conceptually distinct and lie well outside
the EAM/MEAM/ADP domain. The MT formalism was initially developed
for strongly covalent materials and was recently applied to construct
interatomic potentials for Si \citep{Kumagai:2007ly,Purja-Pun:2017aa}.
The potential includes only interactions with first neighbors and
favors a particular angle between chemical bonds. However, if the
cutoff range is extended well beyond the first coordination shell,
an MT potential becomes a long-range many-body model explicitly capturing
the bond-order effect. As such, it becomes similar in spirit to the
EAM, MEAM, and ADP potentials despite having a different functional
form. Long-range MT potentials can be equally successful in reproducing
a wide range of properties of FCC metals, such as Al and Pt.\footnote{An advantage of MT potentials for metals is that they can be crossed
with MT potentials for covalent elements to describe mixed-bonding
systems, such as Al-Si.} For illustration, some of the properties predicted by the MT Al potential
are summarized in the Supplementary Information file accompanying
this article.

In addition to the classical potentials, we tested several ML potentials
that are currently available for these metals, including a physically-informed
neural network (PINN) potential for Al \citep{pun2020development},
a spectral neighbor analysis potential (SNAP) for Al \citep{li2018quantum},
and moment tensor potentials (MTP) for Cu and Ag \citep{nitol2025evaluating}.
Since more classical and ML potentials are currently available for
Al than for the other five metals studied here, we chose Al as a platform
for the most detailed comparison of different potential models. To
broaden this comparison, we developed a new MTP potential for Al.
The DFT database used for training and the properties predicted by
the MTP Al potential are presented in the Supplementary Information
file. This enabled us to compare three ML potentials for Al in the
PINN, SNAP, and MTP formats. Note that for Au and Pt, only classical
interatomic potentials (EAM, ADP, MT) were tested since, to our knowledge,
no suitable ML potentials are available for them. All potentials studied
in this work are implemented in LAMMPS.

\section{Results}

\subsection{DFT calculations}

\begin{figure}[!ht]
\centering \includegraphics[width=1\textwidth]{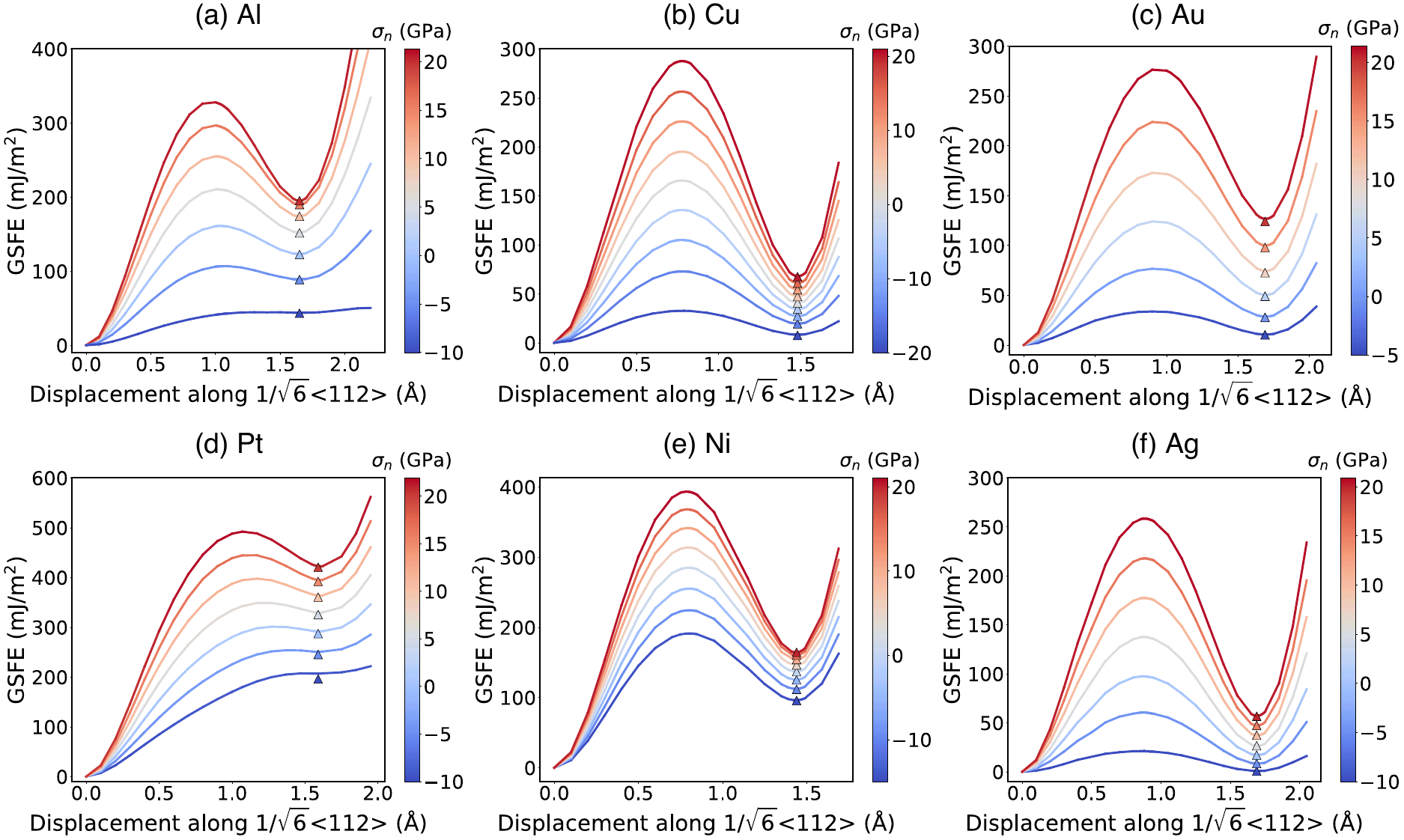}

 \caption{GSFE versus shear displacement under applied normal stress $\sigma_{n}$
obtained by DFT calculations. (a) Al, (b) Cu, (c) Au, (d) Pt, (e)
Ni, and (f) Ag. The curves were computed from Eq.~(\ref{eq:gsfe1}).
The triangular symbols represent SFE calculations from the exact Eq.~(\ref{eq:gsfe}).}
\label{fig:GSFE} 
\end{figure}

Fig.~\ref{fig:GSFE} shows the GSFE plots for the six metals under
different normal stresses $\sigma_{n}$. The results were obtained
by DFT calculations using Eq.~(\ref{eq:gsfe1}). The curves are color-coded
by the value of $\sigma_{n}$. We follow the stress sign convention
adopted in LAMMPS, in which a compressive stress is considered positive
and tensile stress is negative. The USFE values were determined from
the first maximum of the GSFE curve, while the SFE values correspond
to the local minimum. We also show the SFE energies calculated from
Eq.~(\ref{eq:gsfe}) (without the fixed-cell length approximation),
which are represented by the triangular markers. As mentioned above,
the difference between the SFE values obtained from the two equations
is small. 

For all six metals, we applied the normal stresses ranging from $-20$~GPa
(tension) to $20$~GPa (compression). All metals remain stable up
to 20 GPa but some become mechanically unstable under large tensile
stresses. Accordingly, the stress ranges considered in this work are
as follows: Al ($-10$ to $20$~GPa), Cu ($-20$ to $2$0~GPa),
Au ($-5$ to $20$~GPa), Pt ($-10$ to $20$~GPa), Ni ($-15$ to
$20$~GPa), and Ag ($-10$ to $20$~GPa).

\begin{figure}
\includegraphics[width=1\textwidth]{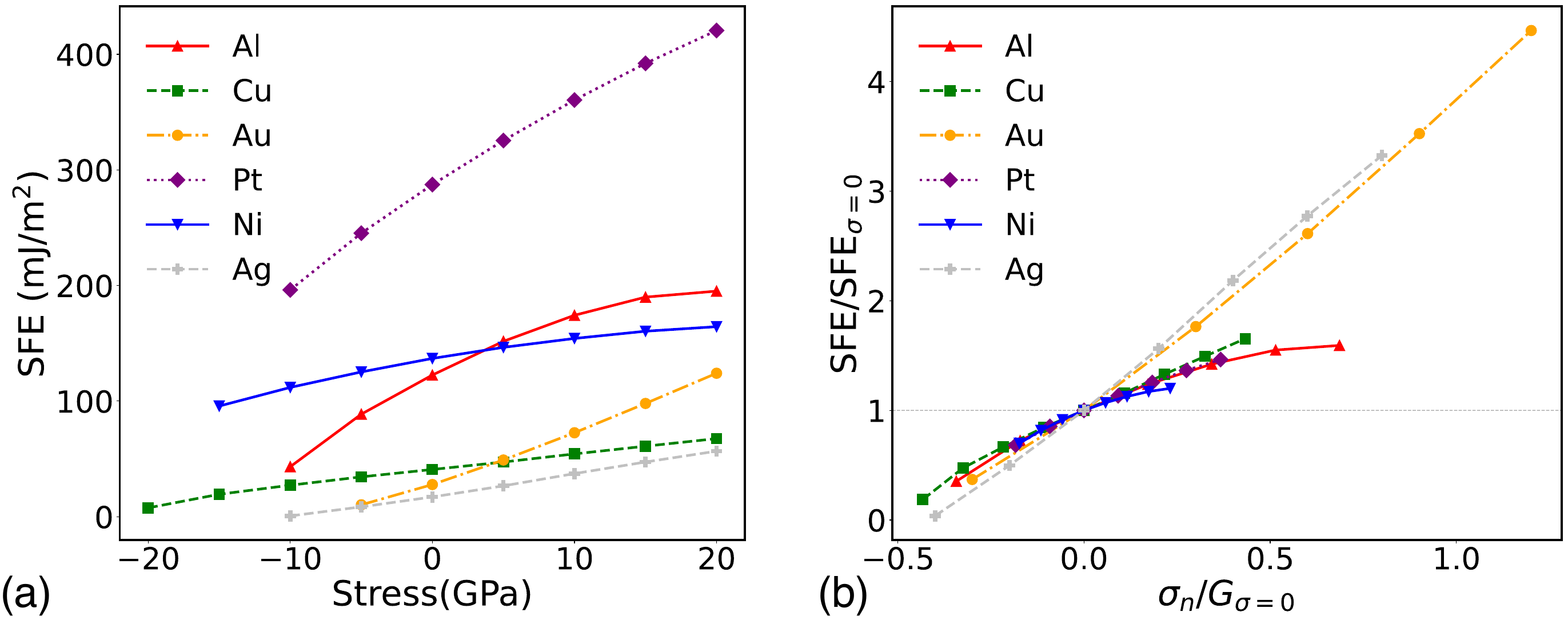}

\caption{SFE in six FCC metals obtained by DFT calculations. (a) Physical coordinates.
(b) Normalized coordinates.}\label{fig:SFE-1}

\end{figure}

\begin{figure}
\includegraphics[width=1\textwidth]{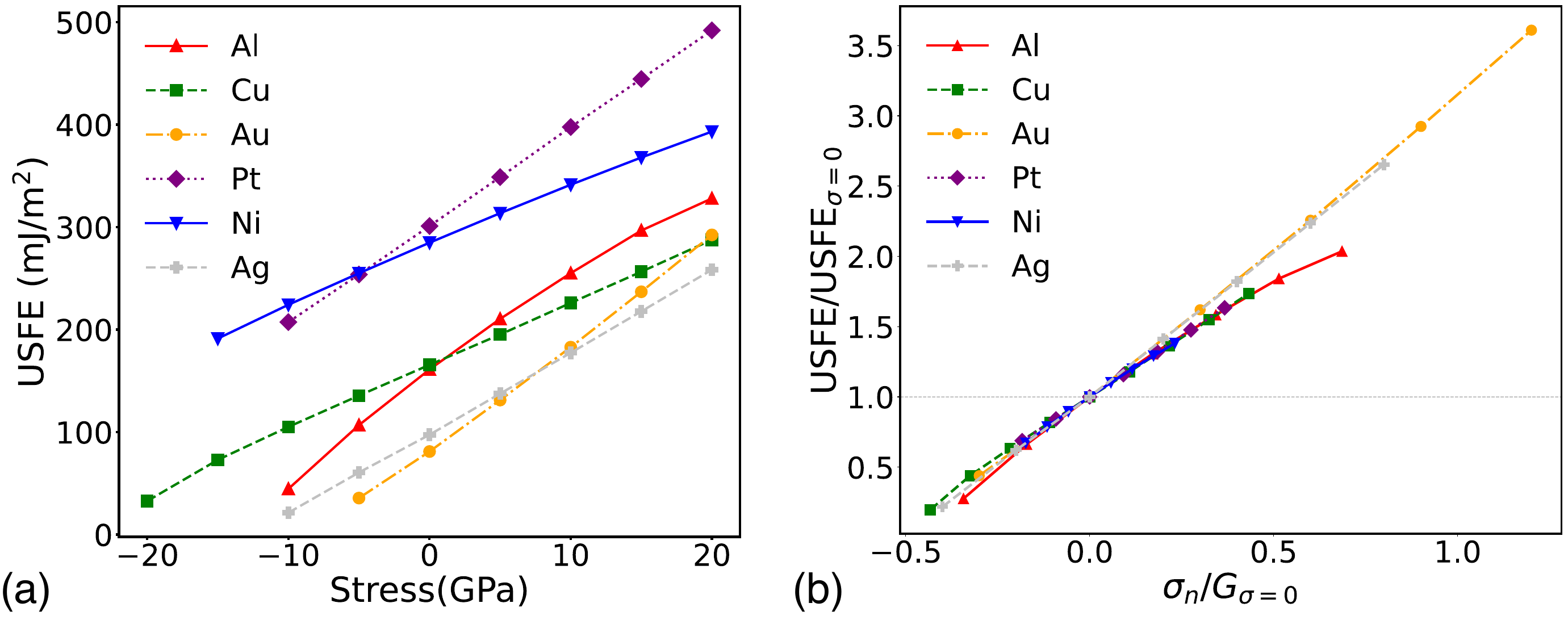}

\caption{USFE in six FCC metals as a function of normal stress obtained by
DFT calculations. (a) Physical coordinates. (b) Normalized coordinates.}\label{fig:USFE-1}
\end{figure}

\begin{figure}[th]
\centering \includegraphics[width=0.5\textwidth]{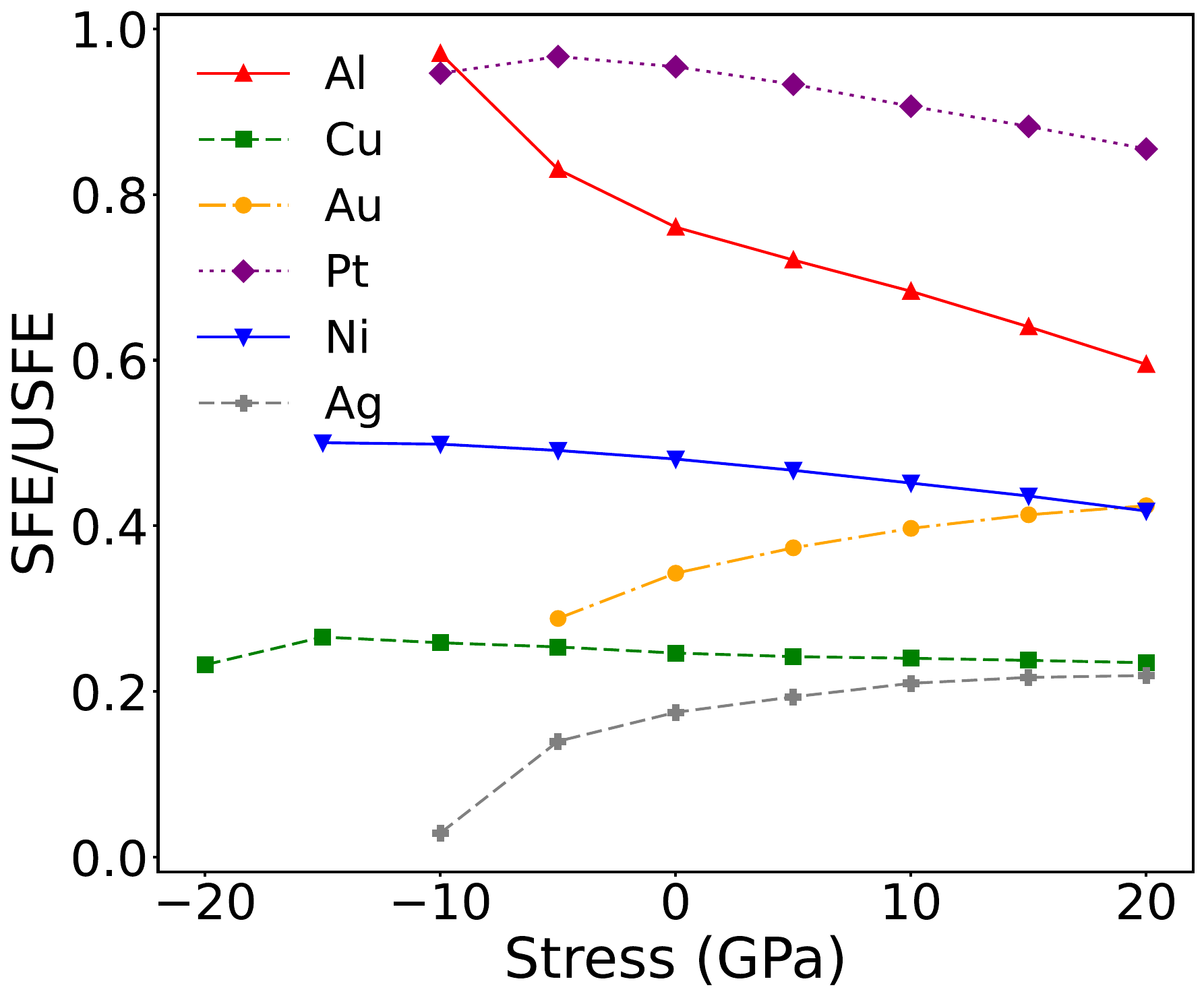}
\caption{The ratio of the SFE and the USFE as a function of normal stress for
six FCC metals obtained by DFT calculations.}
\label{fig:ratio_SFE_over_USFE}
\end{figure}

\begin{figure}
\begin{centering}
\includegraphics[width=0.5\textwidth]{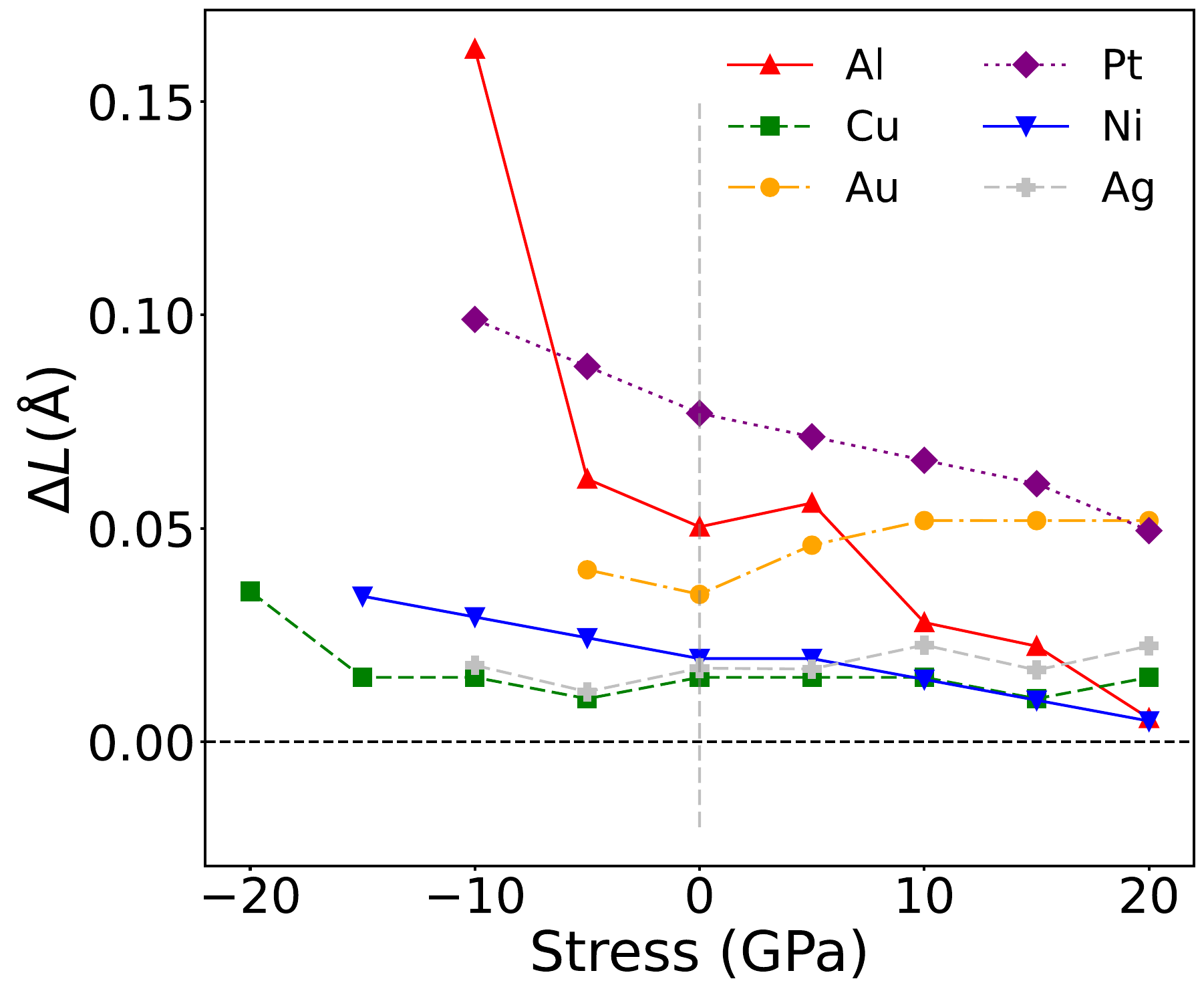}
\par\end{centering}
\caption{SF formation volume as a function of normal stress for six FCC metals
obtained by DFT calculations. }\label{fig:SF-free-volume}

\end{figure}

\begin{figure}
\begin{centering}
\includegraphics[width=1\textwidth]{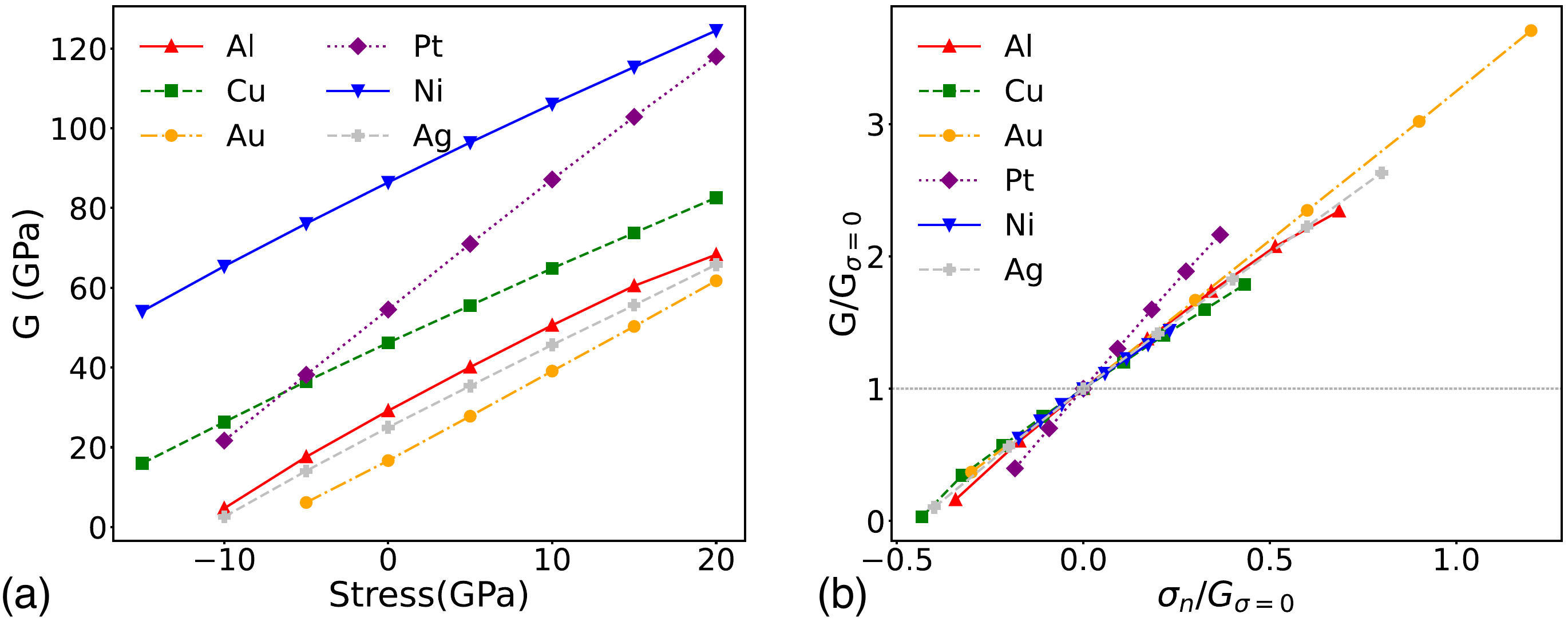}
\par\end{centering}
\caption{The shear modulus $G$ as a function of normal stress for six FCC
metals obtained by DFT calculations. (a) Physical coordinates. (b)
Normalized coordinates.}\label{fig:G-1}

\end{figure}

The DFT results for the intrinsic SFE ($\gamma_{\mathrm{SF}}$), USFE,
and the shear modulus are summarized in Figures \ref{fig:SFE-1} to
\ref{fig:G-1}. In addition to the raw results, we present normalized
plots in which the fault energies and the elastic modulus are normalized
by their stress-free values. All three properties (SFE, USFE, and
$G$) increase monotonically with the normal stress. Thus, normal
compression increases the fault energies and simultaneously makes
the metal stiffer, whereas normal tension produces the opposite effect.
For Al, Cu, and Ni, these trends agree with previous DFT calculations
by Andric et al.~\citep{andric2019stress}.\footnote{Andric et al.~\citep{andric2019stress} considered only tensile loads. }
Branicio et al.~\citep{branicio2013effect} reported similar DFT
results for SFE in Cu under both tension and compression (see their
Fig.~3a).

Note the intersections between some of the curves, which signify changes
in the ranking of the six metals. For example, Al has a smaller SFE
than Ni in the stress-free state and under tension (Fig.~\ref{fig:SFE-1}(a)).
However, at a few GPa of compression, the SFE of Al exceeds that of
Ni. Note also that under sufficiently high tension, Al becomes a low-SFE
metal. Similarly, the SFE of Au is low in the stress-free state and
under tension but increases and eventually exceeds the SFE of Cu under
compression. In the normalized format (Fig.~\ref{fig:SFE-1}(b)),
the SFE plots split into two groups. The plots for Ag and Au merge
together into a straight line, whereas the plots for the remaining
metals merge into a different curve. 

In Fig.~\ref{fig:USFE-1}(a), we plot the USFE as a function of normal
stress for the six metals. Pt and Ni exhibit the highest USFE, while
Ag and Au exhibit the lowest. Other than this, the USFEs do not consistently
follow any particular ranking among the six metals. Some of the plots
cross each other as $\sigma_{n}$ varies within the stress interval
studied here. In normalized coordinates (Fig.~\ref{fig:USFE-1}(b)),
the curves again split into two groups, with Ag and Au forming one
group and the remaining metals the other. This time, however, the
groups are closer together and their plots are more linear. The correlation
within each group is much tighter, suggesting a possible scaling relation
between USFE and normal stress across FCC metals. 

Fig.~\ref{fig:ratio_SFE_over_USFE} shows the stress dependence of
the SFE/USFE ratio. This ratio is slightly below unity for Pt and
Al and relatively small for Au, Cu, and especially Ag. Note that for
Ag and Au, the SFE/USFE ratio increases under compression, whereas
for Cu, Al, Ni, and Pt it decreases.

The atomic density within the SF region is generally different from
that in the perfect FCC lattice. As a result, the formation of an
SF is generally accompanied by inelastic deformation  along the {[}111{]}
direction. This deformation is measured by the difference
\begin{equation}
\Delta L(\sigma_{n})=L(\sigma_{n})-L_{0}(\sigma_{n})\label{eq:SF-volume}
\end{equation}
between the cell dimensions along {[}111{]} with ($L(\sigma_{n})$)
and without ($L_{0}(\sigma_{n})$) the SF under the same normal stress
$\sigma_{n}$. Note that $\Delta L$ coincides with the term in the
square brackets in Eq.~(\ref{eq:gsfe}) when $u=a/\sqrt{6}$. Since
the SF formation does not change the cross-sectional area of the cell,
$\Delta L$ can be interpreted as the formation volume of the SF per
unit area. The DFT calculations indicate that $\Delta L$ is positive
for all six metals under all stresses studied in this work (Fig.~\ref{fig:SF-free-volume}).
In all cases, the SF region has a lower atomic density than the perfect
FCC lattice, leading to the expansion of the system during the SF
formation. Andric et al.~\citep{andric2019stress} arrived at the
same conclusion for Cu and Al under a tensile stress. 

Fig.~\ref{fig:SF-free-volume} shows that for Ni, Al, and Pt, $\Delta L$
decreases with increasing stress. The derivative $-d\Delta L/d\sigma_{n}$
is a measure of compressibility of the SF region, which for these
three metals is positive. For the remaining three metals, the slopes
of the plots are small and could not be reliably resolved due to the
scatter of the points.

The shear modulus $G$ is a fairly linear function of stress with
a positive slope (Fig.~\ref{fig:G-1}(a)). The six metals follow
the ranking Ni $>$ Pt $>$ Cu $>$ Al $>$ Ag $>$ Au, except for
the recrossing of the Cu and Pt plots under a tension of about $-5$
GPa. The normalized plot (Fig.~\ref{fig:G-1}(b)) shows that these
metals generally follow a linear scaling relation with only Pt displaying
a marked deviation. 

\begin{figure}
\begin{centering}
\includegraphics[width=1\textwidth]{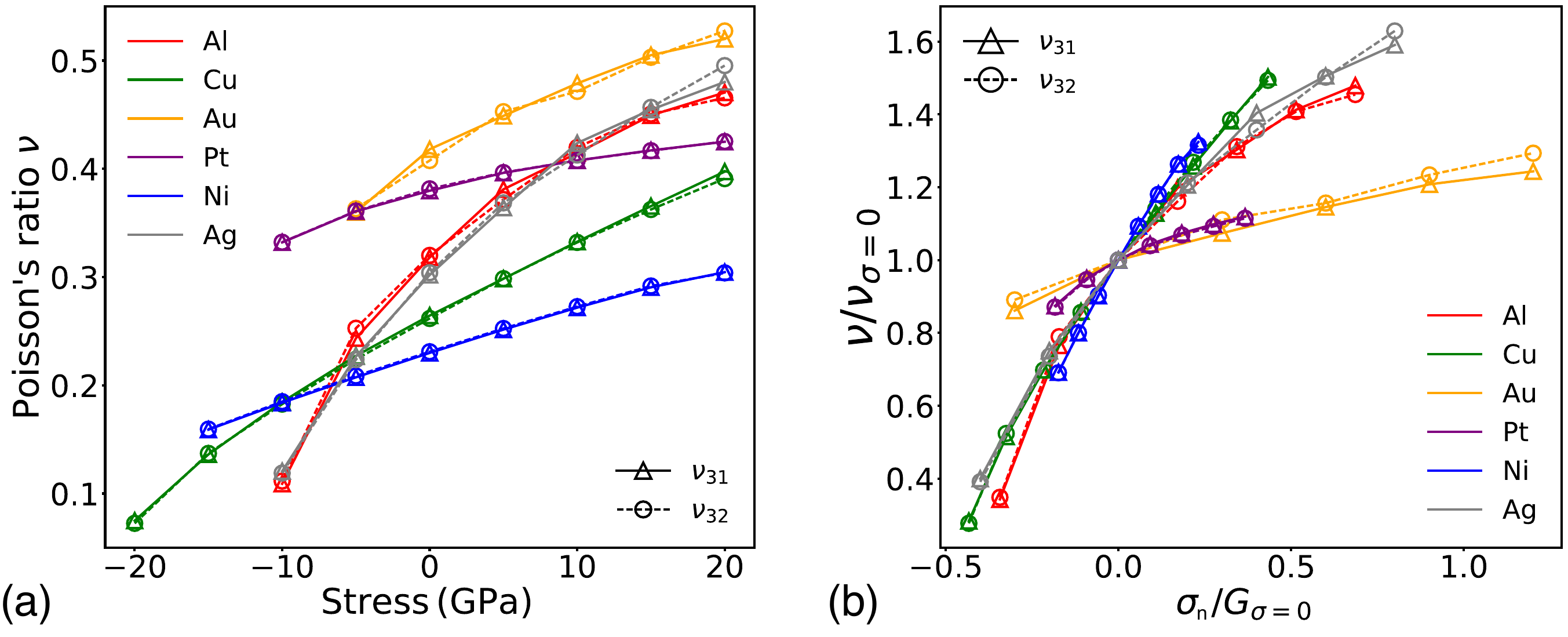}
\par\end{centering}
\caption{Poisson's ratios $\nu_{31}$ and $\nu_{32}$ as a function of normal
stress for six FCC metals obtained by DFT calculations. (a) Physical
coordinates. (b) Normalized coordinates.}\label{fig:Poisson}
\end{figure}

Fig.~\ref{fig:Poisson}(a) presents the DFT results for Poisson's
ratio of the six metals. As expected from symmetry, the two components
$\nu_{31}$ and $\nu_{32}$ nearly coincide within a small margin
($\nu_{31}=\nu_{32}\equiv\nu$). In all cases, $\nu$ increases with
the applied stress, a behavior which is similar to that of the shear
modulus. The $\nu$ values span a wide interval from slightly below
0.1 for Cu under strong tension to nearly 0.5 for Au under strong
compression. No particular ranking is followed because the plots for
individual metals cross each other. In the normalized format (Fig.~\ref{fig:Poisson}(b)),
the plots collapse into two groups: Al, Cu, Ag, and Ni form one group
and Au and Pt form another.

\subsection{Calculations with interatomic potentials}

\begin{figure}[!ht]
\centering \includegraphics[width=1\linewidth]{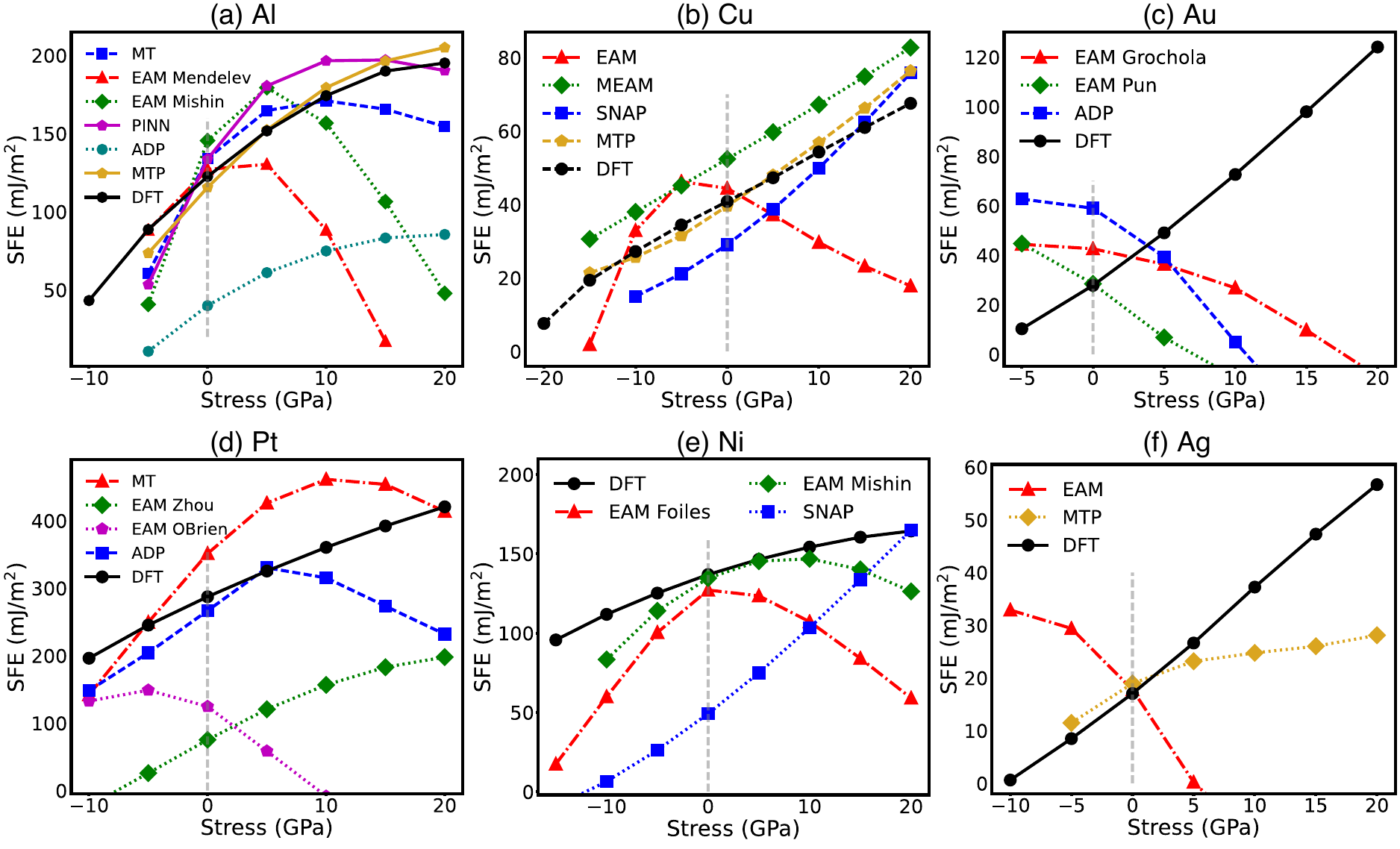} 
 \caption{SFE versus applied normal stress obtained by DFT calculations and
with interatomic potentials. (a) Al, (b) Cu, (c) Au, (d) Pt, (e) Ni,
and (f) Ag. The vertical dashed line corresponds to zero stress and
serves as a visual guide.}
\label{fig:SFE} 
\end{figure}

Figure~\ref{fig:SFE} summarizes the results of SFE calculations
using the interatomic potentials for six metals. The results are compared
with DFT calculations. 

We first examine the comparison for Al (Fig.~\ref{fig:SFE}(a)).
At zero stress, the potentials reproduce the SFE in reasonable agreement
with DFT calculations. The only exception is the ADP potential, which
grossly underestimates the DFT value. Under a tensile stress, all
potentials under-predict the DFT values by up to a factor of 2. Under
compression, only the MTP potential demonstrates good agreement with
DFT. The PINN and MT potentials follow the correct trend but reach
a spurious maximum and start decreasing when $\sigma_{n}$ exceeds
10 GPa. The EAM potentials perform poorly. When the compression reaches
a few GPa, they sharply deviate from the DFT points and develop a
negative slope. As compression increases, the SFE predicted by the
EAM potentials falls below the zero-stress value. At a compression
of 15 to 20 GPa, the EAM potentials predict that Al becomes a low-SFE
metal, which is opposite to the DFT predictions. The ADP potential
correctly reproduces the positive slope of the curve but the SFE remains
significantly below the DFT values. 

For Cu (Fig.~\ref{fig:SFE}(b)), the MTP potential demonstrates excellent
agreement with DFT, except under a tensile stress below $-15$ GPa
when the SF becomes unstable. The MEAM potential correctly predicts
a positive slope of the plot but significantly overestimates the SFE
at all stresses. In addition, with this potential, the SF loses stability
at $\sigma_{n}<-15$ GPa. The SNAP potential also predicts the correct
positive slope but under-predicts the SFE at all stresses below 15
GPa. In addition, with the SNAP potential, the SF loses stability
below $-10$ GPa. The EAM potential performs worst under both tension
and compression. Although it accurately predicts the SFE in the stress-free
state, it drastically underestimates the SFE under tension, eventually
driving it to zero at about $-15$ GPa. Under compression, the EAM
curve has a negative slope instead of a positive one, leading to the
incorrect prediction that compressive stresses reduce the SFE of Cu.
The latter deficiency of this potential was previously noted in \citep{branicio2013effect}.

None of the Ni potentials tested in this work gives satisfactory results
(Fig.~\ref{fig:SFE}(e)). The EAM Mishin potential agrees well with
DFT in the stress-free state and under stresses $\pm5$ GPa but develops
significant downward deviations outside this interval, including the
loss of stability below $-15$ GPa. The EAM Foiles potential deviates
strongly from DFT at all stresses. Both EAM potentials predict a spurious
maximum of the SFE followed by a negative slope at larger compressions,
incorrectly indicating that compression decreases the SFE. 

For Pt (Fig.~\ref{fig:SFE}(d)), the ADP potential approximately
follows the DFT calculations at stresses $\pm5$ GPa but strongly
deviates from them outside this interval. Additionally, the ADP curve
develops a negative slope at $\sigma_{n}>10$ GPa. The curve computed
with the MT potential has a similar shape and strongly overestimates
the DFT values under compression. Both EAM potentials tested here
perform poorly by grossly underestimating the SFE at all stresses. 

Ag and Au pose the greatest challenge to potentials (Figs.~\ref{fig:SFE}(c,f)).
The MTP potential for Ag at least has a positive slope, but significantly
underestimates the SFE under compression and predicts the loss of
stability below $-15$ GPa. The remaining potentials predict the wrong
slope of the plot at \emph{all} stresses. Furthermore, under a sufficiently
strong compression, the SFE becomes zero instead of increasing with
stress and eventually reaching 3 to 4 times the stress-free value. 

\begin{figure}[!ht]
\centering \includegraphics[width=1\linewidth]{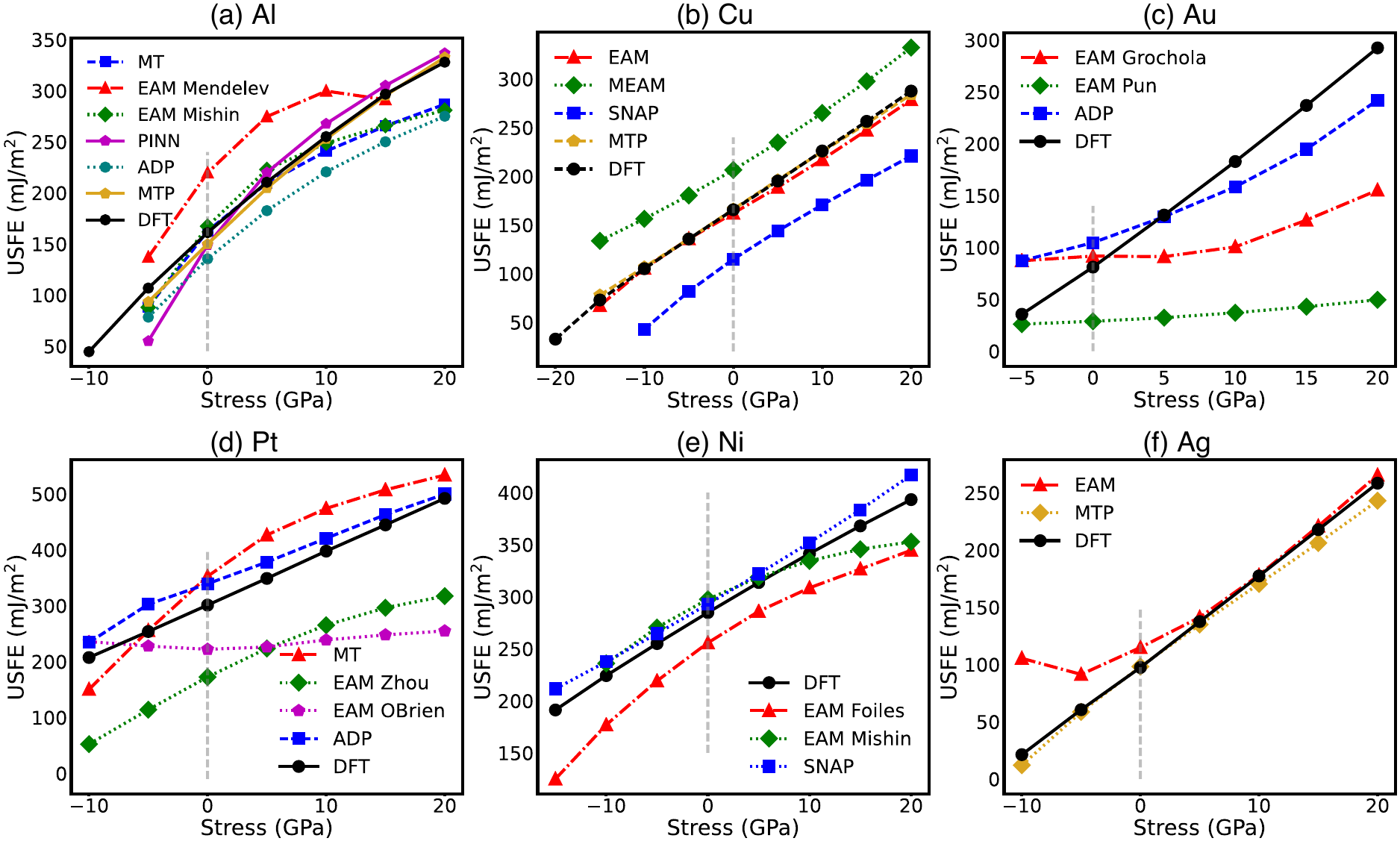} 
 \caption{USFE versus applied normal stress obtained by DFT calculations and
with interatomic potentials. (a) Al, (b) Cu, (c) Au, (d) Pt, (e) Ni,
and (f) Ag. The vertical dashed line corresponds to zero stress and
serves as a visual guide.}
\label{fig:USFE}
\end{figure}

We next consider the USFE calculations for the six metals (Fig.~\ref{fig:USFE}).
For Al, the MTP and PINN potentials accurately reproduce the DFT calculations
under compression but under-predict the USFE under tension (Fig.~\ref{fig:USFE}(a)).
The MT and EAM Mishin potentials match the DFT value at zero stress
but underestimate it under both tension and compression. However,
the general trend is reproduced correctly. The ADP potential reproduces
the correct slope of the plot but is otherwise significantly less
accurate. The EAM Mendelev potential shows the largest deviations
from DFT and predicts a spurious maximum at about 10 GPa. Fig.~\ref{fig:USFE}(a)
does not include an USFE value for the EAM Mendelev potential at 20
GPa because the SFE becomes negative at this stress. For Cu, both
the MTP and EAM potentials are in excellent agreement with DFT calculations,
whereas the MEAM and SNAP potentials overestimate and underestimate
the DFT, respectively (Fig.~\ref{fig:USFE}(b)). However, all four
potentials correctly reproduce the positive slope of the curve. For
Ni, the potentials reproduce the correct increasing trend (Fig.~\ref{fig:USFE}(e)),
with the EAM Foiles potential being least accurate at all stresses.
For Pt (Fig.~\ref{fig:USFE}(d)), the ADP and MT potentials agree
with the DFT calculations reasonably well, but the EAM potential displays
strong deviations. The EAM O'Brien potential fails to reproduce the
correct behavior even qualitatively. For Ag, the MTP potential performs
very well, whereas the EAM potential strongly deviates from DFT under
tension (Fig.~\ref{fig:USFE}(f)). Finally, all Au potentials perform
poorly (Fig.~\ref{fig:USFE}(c)). While the ADP potential predicts
the correct trend, the EAM potentials fail to reproduce the DFT results
even qualitatively. 

Fig.~\ref{fig:G} compares the stress dependencies of the shear modulus
$G$ for the six metals obtained by DFT calculations and predicted
by the potentials. Reasonable agreement is observed for the MEAM potential
for Cu, the ADP potential for Pt, the SNAP potential for Ni, and the
MTP potential for Ag. In all other cases, the agreement is poorer.
In a few cases, the potentials predict an incorrect slope of the curve
under tensile stresses. This includes the EAM Mishin potential for
Al (Fig.~\ref{fig:G}(a)) and the EAM Grachola and ADP potentials
for Au (Fig.~\ref{fig:G}(c)). In addition, the EAM O'Brien potential
for Pt changes the slope of the curve to negative at about 10 GPa
(Fig.~\ref{fig:G}(d)). Several potentials display discontinuous
behavior of $G$ as a function of stress. 

Next, we compare Poisson's ratio $\nu$ (average of $\nu_{31}$ and
$\nu_{32}$) predicted by the potentials and by DFT calculations (Fig.~\ref{fig:nu}).
As with the shear modulus, the most accurate agreement is displayed
by the MTP potentials for Al, Ag, and Cu (in the latter case, a discontinuity
is observed under tension). Incorrect slopes are exhibited by the
EAM Al and EAM O'Brien Pt potentials under tension, as well as the
EAM Mendelev potential for Al, the EAM O'Brien potential for Pt, and
both EAM potentials for Au under compression. In addition, the EAM
Foiles potential for Ni undergoes a discontinuous drop at about 10
GPa of compression.

\begin{figure}[!ht]
\centering \includegraphics[width=1\linewidth]{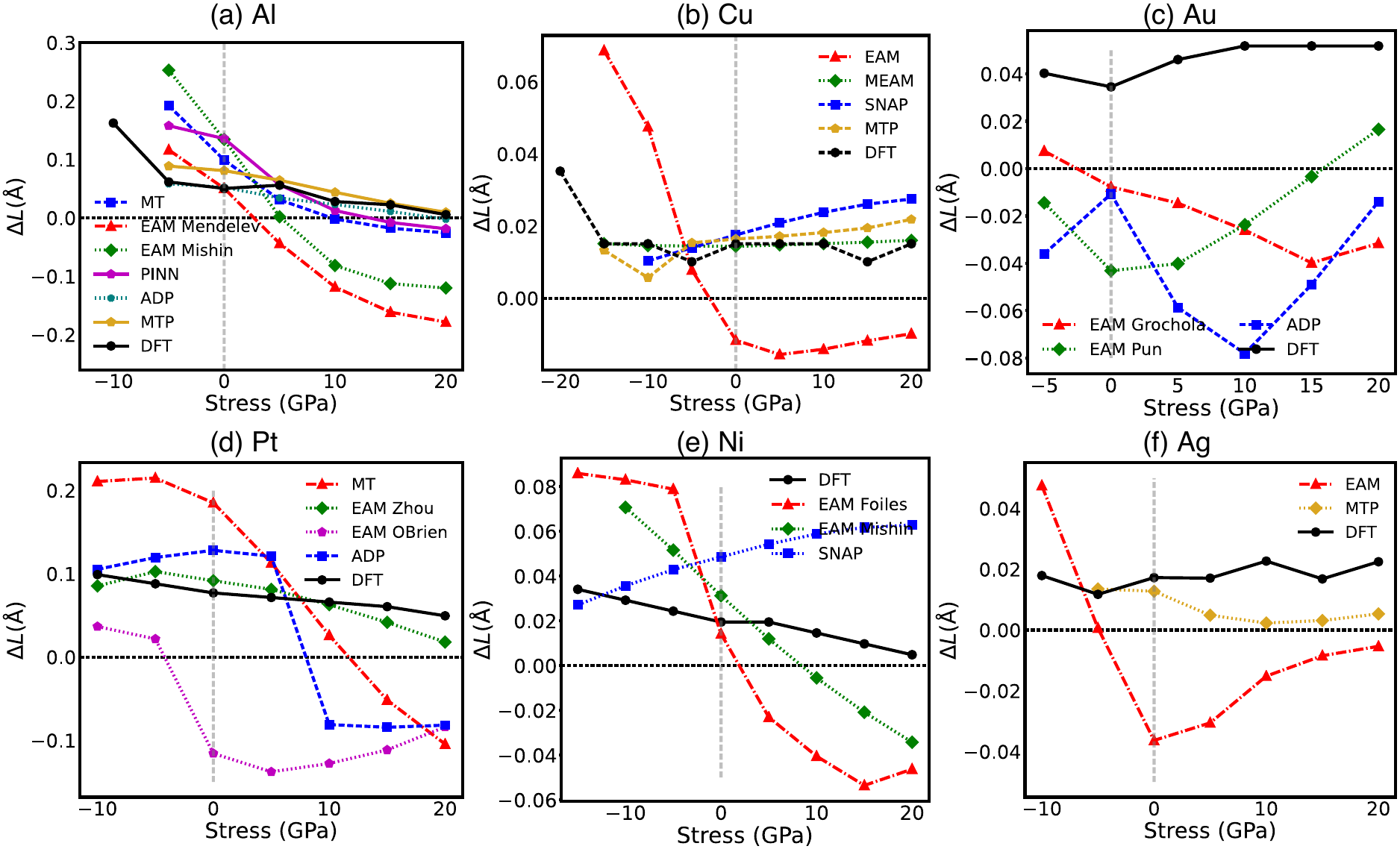} 
 \caption{SF formation volume as a function of normal stress obtained by DFT
calculations and with interatomic potentials. (a) Al, (b) Cu, (c)
Au, (d) Pt, (e) Ni, and (f) Ag. The vertical dashed line corresponds
to zero stress and serves as a visual guide.}
\label{fig:L}
\end{figure}

\begin{figure}[!ht]
\centering \includegraphics[width=1\linewidth]{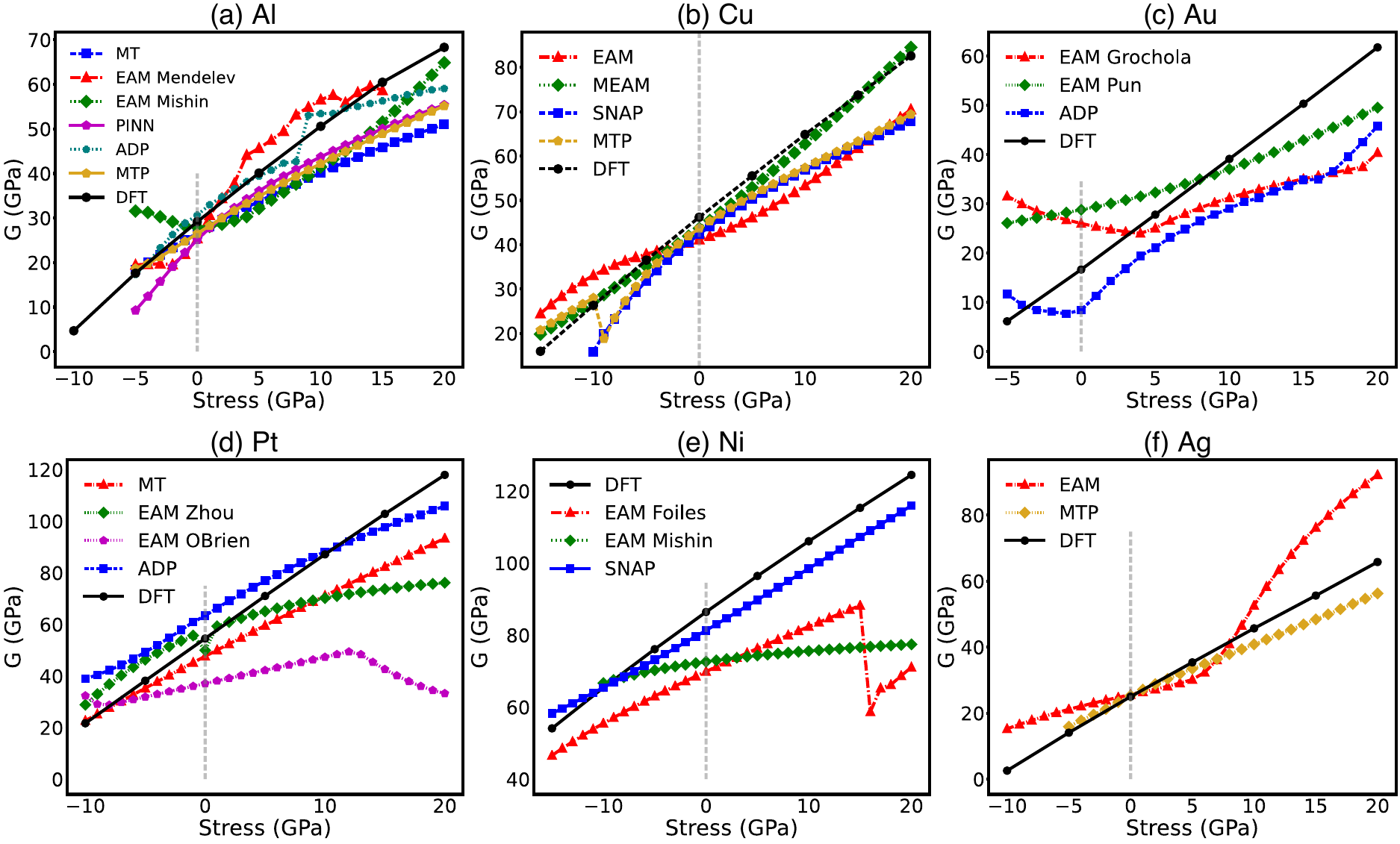} 
 \caption{The shear modulus on the $\{111\}$ plane along the $\langle112\rangle$
direction obtained by DFT calculations and with different interatomic
potentials. (a) Al, (b) Cu, (c) Au, (d) Pt, (e) Ni, and (f) Ag. The
vertical dashed line corresponds to zero stress and serves as a visual
guide. }
\label{fig:G}
\end{figure}

\begin{figure}[!ht]
\centering \includegraphics[width=1\linewidth]{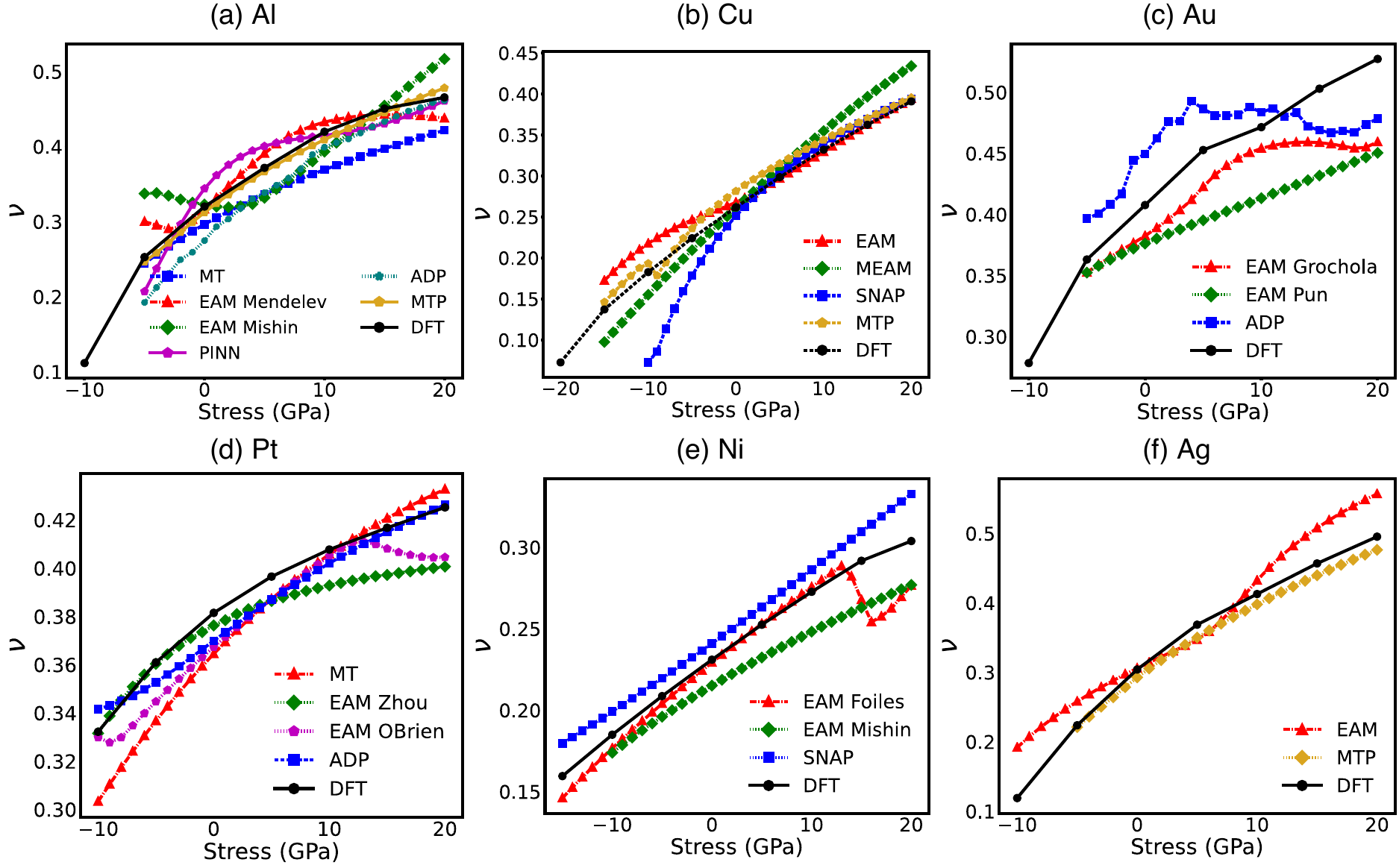} 
 \caption{Poisson's ratio obtained by DFT calculations and with interatomic
potentials. (a) Al, (b) Cu, (c) Au, (d) Pt, (e) Ni, and (f) Ag. }
\label{fig:nu}
\end{figure}

As mentioned above, the DFT calculations predict that the SF formation
volume $\Delta L$ is positive for all six metals. Some potentials
correctly reproduce the positive sign of $\Delta L$ while others
predict a negative sign (Fig.~\ref{fig:L}). For Al, only the MTP
potential yields $\Delta L>0$ at all stresses tested here (Fig.~\ref{fig:L}(a)).
The MT and PINN potentials predict that $\Delta L$ reaches zero at
10 GPa of compression and becomes negative (alreilt small in magnitude)
under further compression. According to the EAM potentials, $\Delta L$
changes sign under a smaller compression and becomes negative and
large in magnitude as $\sigma_{n}$ increases. As a result, the SF
formation is accompanied by a large inelastic contraction instead
of the small expansion predicted by DFT. For Cu (Fig.~\ref{fig:L}(b)),
all potentials reproduce the DFT results well except for the EAM potential.
The latter predicts negative $\Delta L$ values already in the stress-free
state and even under a tension up to a few GPa. Although the magnitude
of the negative $\Delta L$ remains small ($<2\times10^{-2}$ $\textrm{\AA}$),
this is an obvious flaw of the potential. For Pt (Fig.~\ref{fig:L}(d)),
the EAM Zhou potential performs well, while the MT and ADP potentials
incorrectly predict negative $\Delta L$ values when compression exceeds
10 GPa. The EAM O'Brien potential is the least accurate. It predicts
negative $\Delta L$ values with or without applied stresses except
under a strong tension when the sign finally reverses. For Ni (Fig.~\ref{fig:L}(e)),
only the SNAP potential keeps the SF volume positive at all stresses,
although the magnitude of $\Delta L$ is inaccurate and the slope
of the curve is positive. The EAM potentials reverse the sign of $\Delta L$
under a compression of about 5 GPa. For Ag (Fig.~\ref{fig:L}(f)),
only the MTP potential reproduces the correct sign of $\Delta L$
at all stresses. The EAM potential incorrectly predicts negative $\Delta L$
values at all stresses except under a tension of $\sigma_{n}<-5$
GPa. Finally, for Au, all potentials tested here perform poorly. They
predict negative $\Delta L$ values even without applied stresses.
The ADP potential predicts negative $\Delta L$ at all stresses. The
EAM potential reverses the sign to positive only under a strong compression
or tension, but even then, the magnitude of $\Delta L$ remains significantly
below the DFT values.

\section{Discussion}

\subsection{DFT calculations}

The DFT calculations performed in this work indicate that tensile
and compressive stresses $\sigma_{n}$ applied normal to the (111)
plane strongly impact the stable and unstable stacking fault energies
in FCC metals. For all six FCC metals studied here, both SFE and USFE
increase monotonically with applied stress. Normal tension ($\sigma_{n}<0$)
reduces the SFE and USFE while normal compression ($\sigma_{n}>0$)
increases them. The impact of the stress can be very significant,
reaching a factor of four in some cases. In particular, strong enough
tension can suppress the SFE of Cu, Ag, and Au to nearly zero, while
strong compression increases the SFE of Pt to about 0.4 J/m$^{2}$
(Figures \ref{fig:SFE-1} and \ref{fig:USFE-1}). When replotted in
normalized coordinates, the SFE versus stress curves for the six metals
nearly collapse into a single master curve, suggesting the existence
of a scaling law. The USFE versus stress curves also follow a scaling
relation. The nature of this scaling requires further investigation. 

The DFT calculations show that the SF formation is accompanied by
local inelastic expansion of the SF core region in the direction normal
to the SF plane. In other words, the SF formation volume $\Delta L$
(per unit area) is positive, at least for the six metals tested here.
The magnitude of the expansion is small (around 0.02 $\textrm{\AA}$)
for Cu, Ag, and Ni and larger (e.g., 0.05 to 0.1 $\textrm{\AA}$)
for the remaining metals. For Ni, Al, and Pt, $\Delta L$ decreases
with increasing stress. In the remaining cases, the trend could not
be ascertained because of the scatter of the points. 

It has long been assumed, based on the local atomic packing, that
the SFE in FCC metals correlates with the energy difference $\Delta E=E_{\mathrm{HCP}}-E_{\mathrm{FCC}}$
between the HCP and FCC phases per unit area $A$ on the (111) plane;
namely \citep{fanourgakis2003phase}, 
\begin{equation}
\gamma_{\mathrm{SF}}=2\Delta E/A.\label{eq:hcp-fcc}
\end{equation}
To test this hypothesis and determine whether this correlation persists
under a normal stress, we first calculated $\Delta E$ as a function
of atomic volume (Fig.~\ref{fig:HCP-FCC}). We kept the atomic volumes
of the two phases equal because the HCP-like layer within the SF structure
is forced to coherency with the surrounding FCC lattice. This layer
slightly expands in the normal direction, changing its atomic volume.
However, as discussed above, this expansion is small and can be neglected
for this comparison. The triangular symbols in Fig.~\ref{fig:HCP-FCC}(a)
mark the $\Delta E$ values corresponding to the equilibrium FCC volume.
Note that normal tension (i.e., higher atomic volume) decreases the
energy difference between the two structures while normal compression
(i.e., lower atomic volume) increases it. For all six metals, $\Delta E$
remains positive both at the equilibrium atomic volume and under the
tested stresses.

Fig.~\ref{fig:HCP-FCC}(b)compares the SFEs obtained by the DFT calculations
with predictions from Eq.(\ref{eq:hcp-fcc}) at the equilibrium atomic
volume. The bisecting dashed line corresponds to perfect agreement.
Although the data points display a strong positive correlation (Pearson's
correlation factor 0.997), Eq.(\ref{eq:hcp-fcc}) over-predicts the
SFE values for Pt, and to a lesser extent, for Al and Ni. 

The DFT calculations have shown that the shear modulus $G$ and Poisson's
ratio $\nu$ strongly depend on the applied normal stress. Both $G$
and $\nu$ decrease under tension and increase under compression.
In the stress interval examined here, their variation can reach a
factor of four. Similar to the SF energies, both $G$ and $\nu$ follow
scaling relations across the six metals. 

As noted in section \ref{sec:Introduction}, the stable and unstable
SF energies play an important role in plastic deformation of FCC metals.
In particular, the dissociation width of a full dislocation into Shockley
partials plays a critical role in cross-slip and in dislocation nucleation
at surfaces, grain boundaries, and other interfaces. The equilibrium
dissociation width $d$ is dictated by the force balance between the
elastic repulsion of the partials and their attraction due to the
excess energy of the SF ribbon. As demonstrated by our DFT calculations,
a compressive stress applied normal to the SF simultaneously increases
the shear modulus and the SFE, thereby increasing both the elastic
repulsion of the partials and their capillary attraction. Respectively,
a tensile stress decreases both forces. It is not obvious \emph{a
priori} which factor (elasticity or SFE) will dominate the stress
response of the dissociation width. 

To glean a tentative answer to this question, we use the following
equation for the dissociation width of an edge dislocation \citep{anderson2017theory}:
\begin{equation}
d=\dfrac{Gb_{p}^{2}}{8\pi\gamma_{\mathrm{SF}}}\dfrac{2+\nu}{1-\nu},\label{eq:d}
\end{equation}
where $b_{p}=a/\sqrt{6}$ is the magnitude of the Burgers vector of
a Shockley partial dislocation. We emphasize that Eq.(\ref{eq:d})
relies on continuum, isotropic linear elasticity, whereas some of
the metals considered here are elastically anisotropic. In addition,
the dissociation width is in some cases comparable to the lattice
constant $a$, making the continuum approximation questionable. Despite
these approximations, Eq.(\ref{eq:d}) can be useful for examining
trends across the metals and for comparing the DFT calculations with
predictions from interatomic potentials (see below). As evident from
Fig.~\ref{fig:d}(a), the dissociation width computed from Eq.(\ref{eq:d})
is a monotonically increasing function of the normal stress for all
six metals. The rate of increase is moderate to small, but persists
across the entire stress interval studied here. The increasing trend
indicates that the applied stress has a stronger effect on the elastic
repulsion between the partials than on their attraction due to the
SFE. The relatively low rate of increase is explained by the partial
compensation between the stress-induced changes in the repulsive and
attractive forces. Repotting in normalized coordinates reveals a scaling
relation followed by Al, Ni, Cu, and Pt, with Ag and Au splitting
off into a separate group (Fig.~\ref{fig:d}(b)).

\subsection{Comparison with interatomic potentials}

The DFT calculations performed in this work were compared with predictions
of several representative interatomic potentials for the six metals.
In most cases, the potentials fail to reproduce the SF properties
and elastic coefficients under strong tension and compression. In
some cases, the agreement with DFT is reasonable under relatively
small stresses (e.g., $\left|\sigma_{n}\right|<1$ GPa) but rapidly
deteriorates as $\left|\sigma_{n}\right|$ increases. In many cases,
the potential predictions are diametrically opposite to DFT calculations.
For example, many potentials predict that compressive stresses reduce
the SF energy, whereas the DFT calculations predict the opposite trend.
Many potentials predict a negative SF formation volume under compression
or even at zero stress, while according to the DFT calculations, the
SF formation volume is positive under all tensile and compressive
stresses tested in this work. The agreement with DFT is somewhat better
for USFE and elastic properties ($G$ and $\nu$), but some potentials
still predict incorrect trends. 

The classical potentials for Ag and Au perform worst. The MTP potentials
for Al, Cu, and Ag and the PINN potential for Al demonstrate the best
predictive capabilities. Generally, ML potentials can be more reliable
at predicting high-stress behaviors. Their training database usually
contains highly deformed structures informing the potential of the
correct behavior under high-stress conditions. In contrast, classical
potentials are fitted predominantly to equilibrium or near-equilibrium
properties and are not sufficiently exposed to extreme deformations.
Even when they are, reproducing both equilibrium and high-stress behaviors
with a relatively small number of free parameters could be challenging.
The existing classical potentials are most suitable for simulations
under relatively small stresses. Great care should be taken when using
them in simulations involving stresses on the level of 10 GPa and
higher. Improving the performance of classical potentials in high-stress
simulations might be possible but would require developing more general
functional forms with a larger number of free parameters. 

One way to improve classical potentials is through a stricter control
of the HCP-FCC energy difference $\Delta E$ as a function of atomic
volume. As discussed previously, $\Delta E$ correlates positively
with SFE (Fig.~\ref{fig:HCP-FCC}). It can be shown that this correlation
persists under positive and negative normal stresses (Fig.~\ref{fig:SFE-vs-HCP-FCC}).
Thus, the ability of a potential to reproduce the correct $\Delta E$
values and their volume dependence is a predictor of the potential's
reliability in reproducing the stress effect on the SFE. The existing
potentials struggle with reproducing the correct behavior of $\Delta E$
under elastic deformations. As noted above, the DFT values of $\Delta E$
remain positive over a wide range of atomic volumes around the equilibrium
state for all six metals. Fig.~\ref{fig:hcp_fcc_compare} compares
the $\Delta E$ versus volume plots predicted by DFT and computed
with the potentials. As in the DFT calculations, the HCP structure
was constrained such that its in-plane atomic density matched that
of the FCC structure. While the DFT plots exhibit a smooth monotonic
decrease of $\Delta E$ with atomic volume, the $\Delta E$ values
predicted by some of the classical potentials display non-monotonic
behaviors and change the sign from positive to negative. The MTP and
PINN potentials still do not reproduce the DFT curves accurately but
yield the correct positive sign of $\Delta E$ and its monotonic decrease
with atomic volume. The MT potentials for Al and Pt and the MEAM potential
for Cu also perform reasonably well. The classical potentials for
Ag and Au display the poorest performance. The non-monotonic and oscillatory
behavior appears to be a common feature of the classical potentials.
Future research may help elucidate the origin of this behavior and
explore ways to suppress them. 

\subsection{Application to dislocation dissociation }\label{subsec:Application-to-dislocation}

We previously mentioned that, due to a compensation effect, the dislocation
dissociation width $d$ is less sensitive to the applied stress than
the elastic coefficients and the SFE separately. This prediction was
made using Eq.(\ref{eq:d}), which relies on several approximations.
It was interesting to test this prediction by direct simulations of
the dislocation dissociation. As the test material, we chose Cu described
by the MTP and EAM potentials. The MTP Cu potential is fairly accurate
and can be considered a proxy for DFT calculations. As shown in Fig.~\ref{fig:Cu-separation},
the MTP potential accurately reproduces the DFT values of $d$ in
the stress-free state and at moderate stresses of $\pm10$ GPa. Deviations
are observed under higher tension and especially under stronger compression.
However, the predicted dissociation width (between 2.5 and 3 nm) is
in the correct ballpark and displays the slow increase with stress
in agreement with DFT calculations. The EAM potential slightly underestimates
the DFT values at moderate stresses ($\pm10$ GPa) but grossly overestimates
them at higher stresses.

To compare the above predictions with direct simulations, we constructed
a rectangular FCC model with the dimensions of $0.9\times122\times100$
nm$^{3}$. The edges of the simulation block were oriented with $X\!\parallel\![211]$,
$Y\!\parallel\![01\bar{1}]$, and $Z\!\parallel\![\bar{1}11]$. The
boundary conditions were periodic in the $X$ direction and fixed
in the $Y$ and $Z$ directions. Using the $\mathsf{atomman}$ toolkit
\citep{Atomman}, an edge dislocation was created at the center of
the simulation block with the dislocation line parallel to the $X$
axis. The construction used the lattice parameter and elastic constants
corresponding to the tested potential. Atoms within a 0.5-nm-thick
surface layer were frozen while all other atoms remained dynamic.
The system was subjected to energy minimization followed by an MD
simulation in the NVE ensemble and then another energy minimization.
As a result, the dislocation dissociated into Shockley partials with
the dissociation width corresponding to the zero-stress condition.
To apply uniaxial stress, a separate simulation block was created
with the same dimensions and number of atoms and with all-periodic
boundary conditions. An NPT MD simulation was performed at 1 K and
with the imposed stress of $20$ GPa (compression) or $-10$ GPa (tension)
parallel to the $Z$ axis while keeping the $X$ and $Y$ dimensions
fixed. The measured uniaxial strain was applied to the system containing
the dislocation. The strain resulted in approximately the same stress
parallel to {[}111{]} as in the perfect-lattice system. The equilibration
procedure described above was reapplied, resulting in a new dissociation
width corresponding to the applied stress. 

The simulations yielded dislocation dissociation widths under stresses
of $20$ GPa, $0$ GPa, and $-10$ GPa with the two interatomic potentials.
The results are summarized in the Supplementary Table \ref{table:Cu_partial}
and illustrated in Fig.~\ref{fig:MD-separation-distances}. With
the MTP potential, the observed $d$ values decrease from 4.8 nm to
2.6 nm as the stress increases from $-10$ GPa to $20$ GPa. This
is contrary to Eq.(\ref{eq:d}), which predicts a mild increase from
approximately 2.5 nm to 3 nm. With the EAM potential, the observed
$d$ values also decrease from 4.7 nm to 3.6 nm as the stress changes
from $-10$ GPa to zero. This trend also contradicts Eq.(\ref{eq:d})
but is in reasonable agreement with the MTP potential. However, compression
causes a drastic increase in the dissociation width, which reaches
10.7 nm at 20 GPa. This increase is predicted by Eq.(\ref{eq:d})
(cf.~Fig.~\ref{fig:Cu-separation}) and is in stark contrast with
the MTP calculations. The unrealistically wide dislocation dissociation
is an artifact of the EAM potential and may lead to unphysical behaviors
of dislocations during high-stress deformation.

\section{Conclusions}

DFT calculations performed in this work demonstrate that uniaxial
tensile and compressive stresses applied normal to (111) plane in
FCC metals have a strong impact on the energies of the intrinsic and
unstable SFs. Both energies increase under compression and decrease
under tension. Normal stress also shifts the metal's shear modulus
$G$ and Poisson's ratio $\nu$ in the same direction as the SF energies.
Both the SF energies and the elastic coefficients follow scaling relations
across the six FCC metals tested in this work. The DFT calculations
show that the SF formation is accompanied by inelastic expansion of
the SF core region in the normal direction.

Stress-induced changes in SF energies can affect many dislocation-controlled
processes in FCC metals, such as dislocation nucleation at surfaces
and interfaces in nanoscale objects and at crack tips. The magnitude
of the stress effect varies. For example, the dislocation nucleation
barrier depends on the USFE. The latter strongly depends on the magnitude
and sign of the stress component normal to the slip plane, especially
when it reaches a 10 GPa level. On the other hand, the width of the
dislocation dissociation into Shockley partials is less sensitive
to the normal stress because the latter changes the elastic repulsion
force and the capillary attraction force between the partials in the
same direction.

Atomistic simulations of deformation and fracture require spatial
and temporal scales beyond those accessible to DFT calculations, and
therefore rely on interatomic potentials. To assess the ability of
potentials to reproduce the stress effect on SF energies, we tested
several representative classical and ML potentials for the six metals.
In addition to potentials taken from the literature, we included several
classical and ML potentials developed in this work. Most of the potentials
tested here perform poorly under large stresses. ML potentials in
the MTP and PINN formats demonstrate markedly better agreement with
DFT calculations than the classical potentials. Many classical potentials
perform well when the normal stress is relatively small (say, below
1 GPa) but display significant deviations from the DFT calculations
under larger positive and negative stresses (e.g., $\pm10$ GPa and
higher). Under high-stress conditions, some classical potentials produce
artifacts, such as decrease in SFE under strong compression and unrealistically
wide dislocation dissociation.

Classical interatomic potentials remain the workhorse of large-scale
atomistic simulations of the mechanical behavior of materials. Although
less accurate than ML potentials, they are computationally faster
and often display more stable behaviors. This work suggests that they
should be carefully tested before use in simulations involving high-stress
conditions, such as shock deformation and nucleation-controlled plasticity
in defect-free nanowires and nanoparticles. The development of new
classical potentials for FCC metals should include training on larger
deformations and fitting to the stress dependence of the HCP-FCC energy
difference, or fitting to the SFE and USFE directly under normal tension
and compression. 

\bigskip{}

\bigskip{}

\noindent\textbf{Acknowledgments}

\noindent This research was supported by the U.S.~Department of Energy,
Office of Basic Energy Sciences, Division of Materials Sciences and
Engineering, under Award \# DE-SC0023102.


\begin{thebibliography}{91}
    \expandafter\ifx\csname natexlab\endcsname\relax\def\natexlab#1{#1}\fi
    \providecommand{\url}[1]{\texttt{#1}}
    \providecommand{\href}[2]{#2}
    \providecommand{\path}[1]{#1}
    \providecommand{\DOIprefix}{doi:}
    \providecommand{\ArXivprefix}{arXiv:}
    \providecommand{\URLprefix}{URL: }
    \providecommand{\Pubmedprefix}{pmid:}
    \providecommand{\doi}[1]{\href{http://dx.doi.org/#1}{\path{#1}}}
    \providecommand{\Pubmed}[1]{\href{pmid:#1}{\path{#1}}}
    \providecommand{\bibinfo}[2]{#2}
    \ifx\xfnm\relax \def\xfnm[#1]{\unskip,\space#1}\fi
    \bibitem[{Anderson et~al.(2017)Anderson, Hirth, and Lothe}]{anderson2017theory}
    \bibinfo{author}{P.~M. Anderson}, \bibinfo{author}{J.~P. Hirth},
    \bibinfo{author}{J.~Lothe}, \bibinfo{title}{Theory of dislocations},
    \bibinfo{publisher}{Cambridge University Press}, \bibinfo{year}{2017}.
    \bibitem[{Zimmerman et~al.(2000)Zimmerman, Gao, and
        Abraham}]{zimmerman2000generalized}
    \bibinfo{author}{J.~A. Zimmerman}, \bibinfo{author}{H.~Gao},
    \bibinfo{author}{F.~F. Abraham},
    \newblock \bibinfo{title}{Generalized stacking fault energies for embedded atom
        fcc metals},
    \newblock \bibinfo{journal}{Modelling and Simulation in Materials Science and
        Engineering} \bibinfo{volume}{8} (\bibinfo{year}{2000}) \bibinfo{pages}{103}.
    \bibitem[{Lu et~al.(2000)Lu, Kioussis, Bulatov, and Kaxiras}]{Lu00a}
    \bibinfo{author}{G.~Lu}, \bibinfo{author}{N.~Kioussis}, \bibinfo{author}{V.~V.
        Bulatov}, \bibinfo{author}{E.~Kaxiras},
    \newblock \bibinfo{title}{Generalized-stacking-fault energy surface and
        dislocation properties of aluminum},
    \newblock \bibinfo{journal}{Phys. Rev. {\rm B}} \bibinfo{volume}{62}
    (\bibinfo{year}{2000}) \bibinfo{pages}{3099--3108}.
    \bibitem[{Yan et~al.(2004)Yan, Wang, and Wang}]{Yan04}
    \bibinfo{author}{J.~A. Yan}, \bibinfo{author}{C.~Y. Wang},
    \bibinfo{author}{S.~Y. Wang},
    \newblock \bibinfo{title}{Generalized stacking fault energy and dislocation
        properties in bcc {Fe}: {A} first-principles study},
    \newblock \bibinfo{journal}{Phys. Rev. {\rm B}} \bibinfo{volume}{70}
    (\bibinfo{year}{2004}) \bibinfo{pages}{174105}.
    \bibitem[{Wang et~al.(2015)Wang, Wang, Huang, Xue, Que, and
        Jiang}]{Wang:2015aa}
    \bibinfo{author}{C.~Wang}, \bibinfo{author}{H.~Wang},
    \bibinfo{author}{T.~Huang}, \bibinfo{author}{X.~Xue},
    \bibinfo{author}{F.~Que}, \bibinfo{author}{Q.~Jiang},
    \newblock \bibinfo{title}{Generalized-stacking-fault energy and twin-boundary
        energy of hexagonal close-packed {Au}: {A} first-principles calculation},
    \newblock \bibinfo{journal}{Scientific Reports} \bibinfo{volume}{5}
    (\bibinfo{year}{2015}) \bibinfo{pages}{10213}.
    \bibitem[{Bulatov et~al.(1998)Bulatov, Abraham, Kubin, Devincre, and
        Yip}]{bulatov1998connecting}
    \bibinfo{author}{V.~Bulatov}, \bibinfo{author}{F.~F. Abraham},
    \bibinfo{author}{L.~Kubin}, \bibinfo{author}{B.~Devincre},
    \bibinfo{author}{S.~Yip},
    \newblock \bibinfo{title}{Connecting atomistic and mesoscale simulations of
        crystal plasticity},
    \newblock \bibinfo{journal}{Nature} \bibinfo{volume}{391}
    (\bibinfo{year}{1998}) \bibinfo{pages}{669--672}.
    \bibitem[{Rice(1992)}]{Rice1992}
    \bibinfo{author}{J.~R. Rice},
    \newblock \bibinfo{title}{Dislocation nucleation from a crack tip: an analysis
        based on the peierls concept},
    \newblock \bibinfo{journal}{J. Mech. Phys. Solids} \bibinfo{volume}{40}
    (\bibinfo{year}{1992}) \bibinfo{pages}{239--271}.
    \bibitem[{Liu et~al.(2017)Liu, Cheng, Wang, Chen, and Shen}]{Liu2017}
    \bibinfo{author}{G.~S. Liu}, \bibinfo{author}{X.~Cheng},
    \bibinfo{author}{J.~Wang}, \bibinfo{author}{K.~G. Chen},
    \bibinfo{author}{Y.~Shen},
    \newblock \bibinfo{title}{Improvement of nonlocal {Peierls-Nabarro} models},
    \newblock \bibinfo{journal}{Comp. Mater. Sci.} \bibinfo{volume}{131}
    (\bibinfo{year}{2017}) \bibinfo{pages}{69--77}.
    \bibitem[{Lu et~al.(2000)Lu, Kioussis, Bulatov, and Kaxiras}]{Lu2000b}
    \bibinfo{author}{G.~Lu}, \bibinfo{author}{N.~Kioussis}, \bibinfo{author}{V.~V.
        Bulatov}, \bibinfo{author}{E.~Kaxiras},
    \newblock \bibinfo{title}{The {Peierls-Nabarro} model revisited},
    \newblock \bibinfo{journal}{Philos. Mag. Lett.} \bibinfo{volume}{80}
    (\bibinfo{year}{2000}) \bibinfo{pages}{675--682}.
    \bibitem[{Schoeck(2001)}]{Schoeck01}
    \bibinfo{author}{G.~Schoeck},
    \newblock \bibinfo{title}{The core structure of dislocations. {Peierls} model
        vs. atomic simulations in {Pd}},
    \newblock \bibinfo{journal}{Compu. Mater. Sci.} \bibinfo{volume}{21}
    (\bibinfo{year}{2001}) \bibinfo{pages}{124--134}.
    \bibitem[{Lu(2005)}]{Lu:2005aa}
    \bibinfo{author}{G.~Lu}, \bibinfo{title}{The Peierls---Nabarro model of
        dislocations: {A} venerable theory and its current development},
    \bibinfo{publisher}{Springer Netherlands}, \bibinfo{address}{Dordrecht},
    \bibinfo{year}{2005}, pp. \bibinfo{pages}{793--811}.
    \bibitem[{Beyerlein and Hunter(2016)}]{beyerlein2016understanding}
    \bibinfo{author}{I.~Beyerlein}, \bibinfo{author}{A.~Hunter},
    \newblock \bibinfo{title}{Understanding dislocation mechanics at the mesoscale
        using phase field dislocation dynamics},
    \newblock \bibinfo{journal}{Philosophical Transactions of the Royal Society A:
        Mathematical, Physical and Engineering Sciences} \bibinfo{volume}{374}
    (\bibinfo{year}{2016}) \bibinfo{pages}{20150166}.
    \bibitem[{Kim et~al.(2021)Kim, Mathew, Luscher, and Hunter}]{Kim:2021aa}
    \bibinfo{author}{H.~Kim}, \bibinfo{author}{N.~Mathew}, \bibinfo{author}{D.~J.
        Luscher}, \bibinfo{author}{A.~Hunter},
    \newblock \bibinfo{title}{Phase field dislocation dynamics ({PFDD}) modeling of
        non-{Schmid} behavior in {BCC} metals informed by atomistic simulations},
    \newblock \bibinfo{journal}{Journal of the Mechanics and Physics of Solids}
    \bibinfo{volume}{152} (\bibinfo{year}{2021}) \bibinfo{pages}{104460}.
    \bibitem[{Hunter et~al.(2011)Hunter, Beyerlein, Germann, and
        Koslowski}]{hunter2011influence}
    \bibinfo{author}{A.~Hunter}, \bibinfo{author}{I.~J. Beyerlein},
    \bibinfo{author}{T.~C. Germann}, \bibinfo{author}{M.~Koslowski},
    \newblock \bibinfo{title}{Influence of the stacking fault energy surface on
        partial dislocations in fcc metals with a three-dimensional phase field
        dislocations dynamics model},
    \newblock \bibinfo{journal}{Physical Review B---Condensed Matter and Materials
        Physics} \bibinfo{volume}{84} (\bibinfo{year}{2011}) \bibinfo{pages}{144108}.
    \bibitem[{Borovikov et~al.(2016)Borovikov, Mendelev, and
        King}]{borovikov2016effects}
    \bibinfo{author}{V.~Borovikov}, \bibinfo{author}{M.~I. Mendelev},
    \bibinfo{author}{A.~H. King},
    \newblock \bibinfo{title}{Effects of stable and unstable stacking fault energy
        on dislocation nucleation in nano-crystalline metals},
    \newblock \bibinfo{journal}{Modelling and Simulation in Materials Science and
        Engineering} \bibinfo{volume}{24} (\bibinfo{year}{2016})
    \bibinfo{pages}{085017}.
    \bibitem[{Anderson and Rice(1986)}]{AndersonR86}
    \bibinfo{author}{P.~M. Anderson}, \bibinfo{author}{J.~R. Rice},
    \newblock \bibinfo{title}{Dislocation emission from cracks in crystals or along
        crystal interfaces},
    \newblock \bibinfo{journal}{Scr Metall} \bibinfo{volume}{20}
    (\bibinfo{year}{1986}) \bibinfo{pages}{1467--1472}.
    \bibitem[{Xu and Zhang(2003)}]{Xu:2003aa}
    \bibinfo{author}{G.~Xu}, \bibinfo{author}{C.~Zhang},
    \newblock \bibinfo{title}{Analysis of dislocation nucleation from a crystal
        surface based on the {Peierls--Nabarro} dislocation model},
    \newblock \bibinfo{journal}{Journal of the Mechanics and Physics of Solids}
    \bibinfo{volume}{51} (\bibinfo{year}{2003}) \bibinfo{pages}{1371--1394}.
    \bibitem[{Cao and Deng(2015)}]{Cao2015}
    \bibinfo{author}{R.~Cao}, \bibinfo{author}{C.~Deng},
    \newblock \bibinfo{title}{The ultra-small strongest grain size in
        nanocrystalline {Ni} nanowires},
    \newblock \bibinfo{journal}{Scripta Mater.} \bibinfo{volume}{94}
    (\bibinfo{year}{2015}) \bibinfo{pages}{9--12}.
    \bibitem[{Li et~al.(2018)Li, Xu, Hara, Li, and Ma}]{Li:2018ab}
    \bibinfo{author}{Q.-J. Li}, \bibinfo{author}{B.~Xu}, \bibinfo{author}{S.~Hara},
    \bibinfo{author}{J.~Li}, \bibinfo{author}{E.~Ma},
    \newblock \bibinfo{title}{Sample-size-dependent surface dislocation nucleation
        in nanoscale crystals},
    \newblock \bibinfo{journal}{Acta Materialia} \bibinfo{volume}{145}
    (\bibinfo{year}{2018}) \bibinfo{pages}{19--29}.
    \bibitem[{Mordehai et~al.(2018)Mordehai, David, and Kositski}]{Mordehai:2018aa}
    \bibinfo{author}{D.~Mordehai}, \bibinfo{author}{O.~David},
    \bibinfo{author}{R.~Kositski},
    \newblock \bibinfo{title}{Nucleation-controlled plasticity of metallic
        nanowires and nanoparticles},
    \newblock \bibinfo{journal}{Advanced Materials} \bibinfo{volume}{30}
    (\bibinfo{year}{2018}) \bibinfo{pages}{1706710}.
    \bibitem[{Amodeo and Lizoul(2017)}]{amodeo2017mechanical}
    \bibinfo{author}{J.~Amodeo}, \bibinfo{author}{K.~Lizoul},
    \newblock \bibinfo{title}{Mechanical properties and dislocation nucleation in
        nanocrystals with blunt edges},
    \newblock \bibinfo{journal}{Materials \& Design} \bibinfo{volume}{135}
    (\bibinfo{year}{2017}) \bibinfo{pages}{223--231}.
    \bibitem[{Sharma et~al.(2018)Sharma, Hickman, Gazit, Rabkin, and
        Mishin}]{Sharma:2018aa}
    \bibinfo{author}{A.~Sharma}, \bibinfo{author}{J.~Hickman},
    \bibinfo{author}{N.~Gazit}, \bibinfo{author}{E.~Rabkin},
    \bibinfo{author}{Y.~Mishin},
    \newblock \bibinfo{title}{Nickel nanoparticles set a new record of strength},
    \newblock \bibinfo{journal}{Nature Communications} \bibinfo{volume}{9}
    (\bibinfo{year}{2018}) \bibinfo{pages}{4102}.
    \bibitem[{Liang et~al.(2024)Liang, Magar, Koju, Chesser, Zimmerman, Mishin, and
        Rabkin}]{Liang:2024aa}
    \bibinfo{author}{Z.~Liang}, \bibinfo{author}{N.~T. Magar},
    \bibinfo{author}{R.~K. Koju}, \bibinfo{author}{I.~Chesser},
    \bibinfo{author}{J.~Zimmerman}, \bibinfo{author}{Y.~Mishin},
    \bibinfo{author}{E.~Rabkin},
    \newblock \bibinfo{title}{Ultimate compressive strength and severe plastic
        deformation of equilibrated single-crystalline copper nanoparticles},
    \newblock \bibinfo{journal}{Acta Materialia} \bibinfo{volume}{276}
    (\bibinfo{year}{2024}) \bibinfo{pages}{120101}.
    \bibitem[{Yan et~al.(2021)Yan, Zhang, Yu, Li, Hu, Yang, and
        Zhang}]{yan2021effects}
    \bibinfo{author}{J.~Yan}, \bibinfo{author}{Z.~Zhang}, \bibinfo{author}{H.~Yu},
    \bibinfo{author}{K.~Li}, \bibinfo{author}{Q.~Hu}, \bibinfo{author}{J.~Yang},
    \bibinfo{author}{Z.~Zhang},
    \newblock \bibinfo{title}{Effects of pressure on the generalized stacking fault
        energy and twinning propensity of face-centered cubic metals},
    \newblock \bibinfo{journal}{Journal of Alloys and Compounds}
    \bibinfo{volume}{866} (\bibinfo{year}{2021}) \bibinfo{pages}{158869}.
    \bibitem[{Linda et~al.(2022)Linda, Tripathi, Nagar, and
        Bhowmick}]{linda2022effect}
    \bibinfo{author}{A.~Linda}, \bibinfo{author}{P.~K. Tripathi},
    \bibinfo{author}{S.~Nagar}, \bibinfo{author}{S.~Bhowmick},
    \newblock \bibinfo{title}{Effect of pressure on stacking fault energy and
        deformation behavior of face-centered cubic metals},
    \newblock \bibinfo{journal}{Materialia} \bibinfo{volume}{26}
    (\bibinfo{year}{2022}) \bibinfo{pages}{101598}.
    \bibitem[{Brandl et~al.(2007)Brandl, Derlet, and
        Van~Swygenhoven}]{brandl2007general}
    \bibinfo{author}{C.~Brandl}, \bibinfo{author}{P.~Derlet},
    \bibinfo{author}{H.~Van~Swygenhoven},
    \newblock \bibinfo{title}{General-stacking-fault energies in highly strained
        metallic environments: Ab initio calculations},
    \newblock \bibinfo{journal}{Physical Review B---Condensed Matter and Materials
        Physics} \bibinfo{volume}{76} (\bibinfo{year}{2007}) \bibinfo{pages}{054124}.
    \bibitem[{Branicio et~al.(2013)Branicio, Zhang, and
        Srolovitz}]{branicio2013effect}
    \bibinfo{author}{P.~Branicio}, \bibinfo{author}{J.~Zhang},
    \bibinfo{author}{D.~Srolovitz},
    \newblock \bibinfo{title}{Effect of strain on the stacking fault energy of
        copper: a first-principles study},
    \newblock \bibinfo{journal}{Physical Review B---Condensed Matter and Materials
        Physics} \bibinfo{volume}{88} (\bibinfo{year}{2013}) \bibinfo{pages}{064104}.
    \bibitem[{Andric et~al.(2019)Andric, Yin, and Curtin}]{andric2019stress}
    \bibinfo{author}{P.~Andric}, \bibinfo{author}{B.~Yin},
    \bibinfo{author}{W.~Curtin},
    \newblock \bibinfo{title}{Stress-dependence of generalized stacking fault
        energies},
    \newblock \bibinfo{journal}{Journal of the Mechanics and Physics of Solids}
    \bibinfo{volume}{122} (\bibinfo{year}{2019}) \bibinfo{pages}{262--279}.
    \bibitem[{Tschopp and McDowell(2008)}]{tschopp2008influence}
    \bibinfo{author}{M.~Tschopp}, \bibinfo{author}{D.~McDowell},
    \newblock \bibinfo{title}{Influence of single crystal orientation on
        homogeneous dislocation nucleation under uniaxial loading},
    \newblock \bibinfo{journal}{Journal of the Mechanics and Physics of Solids}
    \bibinfo{volume}{56} (\bibinfo{year}{2008}) \bibinfo{pages}{1806--1830}.
    \bibitem[{Zhang et~al.(2015)Zhang, Cheng, Zhao, and Pei}]{zhang2015molecular}
    \bibinfo{author}{L.~Zhang}, \bibinfo{author}{L.~Cheng},
    \bibinfo{author}{X.~Zhao}, \bibinfo{author}{L.-Q. Pei},
    \newblock \bibinfo{title}{Molecular dynamics simulation on generalized stacking
        fault energies of fcc metals under preloading stress},
    \newblock \bibinfo{journal}{Chinese Physics B} \bibinfo{volume}{24}
    (\bibinfo{year}{2015}) \bibinfo{pages}{088106}.
    \bibitem[{Mishin(2021)}]{Mishin:2021uc}
    \bibinfo{author}{Y.~Mishin},
    \newblock \bibinfo{title}{Machine-learning interatomic potentials for materials
        science},
    \newblock \bibinfo{journal}{Acta Mater.} \bibinfo{volume}{214}
    (\bibinfo{year}{2021}) \bibinfo{pages}{116980}.
    \bibitem[{Zhang et~al.(2025)Zhang, Sorkin, Aitken, Politano, Behler,
        P~Thompson, Ko, Ong, Chalykh, Korogod, Podryabinkin, Shapeev, Li, Mishin,
        Pei, Liu, Kim, Park, Hwang, Han, Sheriff, Cao, and Freitas}]{Zhang:2025aa}
    \bibinfo{author}{Y.-W. Zhang}, \bibinfo{author}{V.~Sorkin},
    \bibinfo{author}{Z.~H. Aitken}, \bibinfo{author}{A.~Politano},
    \bibinfo{author}{J.~Behler}, \bibinfo{author}{A.~P~Thompson},
    \bibinfo{author}{T.~W. Ko}, \bibinfo{author}{S.~P. Ong},
    \bibinfo{author}{O.~Chalykh}, \bibinfo{author}{D.~Korogod},
    \bibinfo{author}{E.~Podryabinkin}, \bibinfo{author}{A.~Shapeev},
    \bibinfo{author}{J.~Li}, \bibinfo{author}{Y.~Mishin},
    \bibinfo{author}{Z.~Pei}, \bibinfo{author}{X.~Liu}, \bibinfo{author}{J.~Kim},
    \bibinfo{author}{Y.~Park}, \bibinfo{author}{S.~Hwang},
    \bibinfo{author}{S.~Han}, \bibinfo{author}{K.~Sheriff},
    \bibinfo{author}{Y.~Cao}, \bibinfo{author}{R.~Freitas},
    \newblock \bibinfo{title}{Roadmap for the development of machine learning-based
        interatomic potentials},
    \newblock \bibinfo{journal}{Modelling and Simulation in Materials Science and
        Engineering} \bibinfo{volume}{33} (\bibinfo{year}{2025})
    \bibinfo{pages}{023301}.
    \bibitem[{Jhon et~al.(2005)Jhon, Glaeser, and Chrzan}]{jhon2005computational}
    \bibinfo{author}{M.~Jhon}, \bibinfo{author}{A.~Glaeser},
    \bibinfo{author}{D.~Chrzan},
    \newblock \bibinfo{title}{Computational study of stacking faults in sapphire
        using total energy methods},
    \newblock \bibinfo{journal}{Physical Review B---Condensed Matter and Materials
        Physics} \bibinfo{volume}{71} (\bibinfo{year}{2005}) \bibinfo{pages}{214101}.
    \bibitem[{Kibey et~al.(2006)Kibey, Liu, Johnson, and
        Sehitoglu}]{kibey2006generalized}
    \bibinfo{author}{S.~Kibey}, \bibinfo{author}{J.~Liu}, \bibinfo{author}{D.~D.
        Johnson}, \bibinfo{author}{H.~Sehitoglu},
    \newblock \bibinfo{title}{Generalized planar fault energies and twinning in
        cu--al alloys},
    \newblock \bibinfo{journal}{Applied physics letters} \bibinfo{volume}{89}
    (\bibinfo{year}{2006}).
    \bibitem[{Kresse and
        Furthm{\"u}ller(1996{\natexlab{a}})}]{kresse1996efficiency}
    \bibinfo{author}{G.~Kresse}, \bibinfo{author}{J.~Furthm{\"u}ller},
    \newblock \bibinfo{title}{Efficiency of ab-initio total energy calculations for
        metals and semiconductors using a plane-wave basis set},
    \newblock \bibinfo{journal}{Computational materials science}
    \bibinfo{volume}{6} (\bibinfo{year}{1996}{\natexlab{a}})
    \bibinfo{pages}{15--50}.
    \bibitem[{Kresse and Furthm{\"u}ller(1996{\natexlab{b}})}]{kresse1996efficient}
    \bibinfo{author}{G.~Kresse}, \bibinfo{author}{J.~Furthm{\"u}ller},
    \newblock \bibinfo{title}{Efficient iterative schemes for ab initio
        total-energy calculations using a plane-wave basis set},
    \newblock \bibinfo{journal}{Physical review B} \bibinfo{volume}{54}
    (\bibinfo{year}{1996}{\natexlab{b}}) \bibinfo{pages}{11169}.
    \bibitem[{Bl{\"o}chl(1994)}]{blochl1994projector}
    \bibinfo{author}{P.~E. Bl{\"o}chl},
    \newblock \bibinfo{title}{Projector augmented-wave method},
    \newblock \bibinfo{journal}{Physical review B} \bibinfo{volume}{50}
    (\bibinfo{year}{1994}) \bibinfo{pages}{17953}.
    \bibitem[{Perdew et~al.(1996)Perdew, Burke, and
        Ernzerhof}]{perdew1996generalized}
    \bibinfo{author}{J.~P. Perdew}, \bibinfo{author}{K.~Burke},
    \bibinfo{author}{M.~Ernzerhof},
    \newblock \bibinfo{title}{Generalized gradient approximation made simple},
    \newblock \bibinfo{journal}{Physical review letters} \bibinfo{volume}{77}
    (\bibinfo{year}{1996}) \bibinfo{pages}{3865}.
    \bibitem[{Nye(1985)}]{Nye-book}
    \bibinfo{author}{J.~F. Nye}, \bibinfo{title}{Physical Properties of Crystals},
    \bibinfo{publisher}{Clarendon Press}, \bibinfo{address}{Oxford},
    \bibinfo{year}{1985}.
    \bibitem[{Thompson et~al.(2022)Thompson, Aktulga, Berger, Bolintineanu, Brown,
        Crozier, In't~Veld, Kohlmeyer, Moore, Nguyen et~al.}]{thompson2022lammps}
    \bibinfo{author}{A.~P. Thompson}, \bibinfo{author}{H.~M. Aktulga},
    \bibinfo{author}{R.~Berger}, \bibinfo{author}{D.~S. Bolintineanu},
    \bibinfo{author}{W.~M. Brown}, \bibinfo{author}{P.~S. Crozier},
    \bibinfo{author}{P.~J. In't~Veld}, \bibinfo{author}{A.~Kohlmeyer},
    \bibinfo{author}{S.~G. Moore}, \bibinfo{author}{T.~D. Nguyen}, et~al.,
    \newblock \bibinfo{title}{Lammps-a flexible simulation tool for particle-based
        materials modeling at the atomic, meso, and continuum scales},
    \newblock \bibinfo{journal}{Computer physics communications}
    \bibinfo{volume}{271} (\bibinfo{year}{2022}) \bibinfo{pages}{108171}.
    \bibitem[{Daw et~al.(1993)Daw, Foiles, and Baskes}]{daw1993embedded}
    \bibinfo{author}{M.~S. Daw}, \bibinfo{author}{S.~M. Foiles},
    \bibinfo{author}{M.~I. Baskes},
    \newblock \bibinfo{title}{The embedded-atom method: a review of theory and
        applications},
    \newblock \bibinfo{journal}{Materials Science Reports} \bibinfo{volume}{9}
    (\bibinfo{year}{1993}) \bibinfo{pages}{251--310}.
    \bibitem[{Baskes(1992)}]{Baskes92}
    \bibinfo{author}{M.~I. Baskes},
    \newblock \bibinfo{title}{Modified embedded-atom potentials for cubic metals
        and impurities},
    \newblock \bibinfo{journal}{Phys. Rev. {\rm B}} \bibinfo{volume}{46}
    (\bibinfo{year}{1992}) \bibinfo{pages}{2727--2742}.
    \bibitem[{Mishin et~al.(2005)Mishin, Mehl, and Papaconstantopoulos}]{Mishin05a}
    \bibinfo{author}{Y.~Mishin}, \bibinfo{author}{M.~J. Mehl},
    \bibinfo{author}{D.~A. Papaconstantopoulos},
    \newblock \bibinfo{title}{Phase stability in the {Fe-Ni} system: Investigation
        by first-principles calculations and atomistic simulations},
    \newblock \bibinfo{journal}{Acta Mater.} \bibinfo{volume}{53}
    (\bibinfo{year}{2005}) \bibinfo{pages}{4029--4041}.
    \bibitem[{Koju et~al.(2005)Koju, Li, and Mishin}]{Pt-in-review}
    \bibinfo{author}{R.~K. Koju}, \bibinfo{author}{Y.~Li},
    \bibinfo{author}{Y.~Mishin}, \bibinfo{title}{Comparison of interatomic
        potentials for platinum}, \bibinfo{howpublished}{In preparation},
    \bibinfo{year}{2005}.
    \bibitem[{Kumagai et~al.(2007)Kumagai, Izumi, Hara, and Sakai}]{Kumagai:2007ly}
    \bibinfo{author}{T.~Kumagai}, \bibinfo{author}{S.~Izumi},
    \bibinfo{author}{S.~Hara}, \bibinfo{author}{S.~Sakai},
    \newblock \bibinfo{title}{Development of bond-order potentials that can
        reproduce the elastic constants and melting point of silicon for classical
        molecular dynamics simulation},
    \newblock \bibinfo{journal}{Comp. Mater. Sci.} \bibinfo{volume}{39}
    (\bibinfo{year}{2007}) \bibinfo{pages}{457--464}.
    \bibitem[{{Purja Pun} and Mishin(2017)}]{Purja-Pun:2017aa}
    \bibinfo{author}{G.~P. {Purja Pun}}, \bibinfo{author}{Y.~Mishin},
    \newblock \bibinfo{title}{Optimized interatomic potential for silicon and its
        application to thermal stability of silicene},
    \newblock \bibinfo{journal}{Phys. Rev. {\rm B}} \bibinfo{volume}{95}
    (\bibinfo{year}{2017}) \bibinfo{pages}{224103}.
    \bibitem[{Pun et~al.(2020)Pun, Yamakov, Hickman, Glaessgen, and
        Mishin}]{pun2020development}
    \bibinfo{author}{G.~P. Pun}, \bibinfo{author}{V.~Yamakov},
    \bibinfo{author}{J.~Hickman}, \bibinfo{author}{E.~Glaessgen},
    \bibinfo{author}{Y.~Mishin},
    \newblock \bibinfo{title}{Development of a general-purpose machine-learning
        interatomic potential for aluminum by the physically informed neural network
        method},
    \newblock \bibinfo{journal}{Physical Review Materials} \bibinfo{volume}{4}
    (\bibinfo{year}{2020}) \bibinfo{pages}{113807}.
    \bibitem[{Li et~al.(2018)Li, Hu, Chen, Deng, Luo, and Ong}]{li2018quantum}
    \bibinfo{author}{X.-G. Li}, \bibinfo{author}{C.~Hu}, \bibinfo{author}{C.~Chen},
    \bibinfo{author}{Z.~Deng}, \bibinfo{author}{J.~Luo}, \bibinfo{author}{S.~P.
        Ong},
    \newblock \bibinfo{title}{Quantum-accurate spectral neighbor analysis potential
        models for {Ni-Mo} binary alloys and fcc metals},
    \newblock \bibinfo{journal}{Physical Review B} \bibinfo{volume}{98}
    (\bibinfo{year}{2018}) \bibinfo{pages}{094104}.
    \bibitem[{Nitol et~al.(2025)Nitol, Iriarte, Dickel, and
        Fensin}]{nitol2025evaluating}
    \bibinfo{author}{M.~S. Nitol}, \bibinfo{author}{M.~J.~E. Iriarte},
    \bibinfo{author}{D.~E. Dickel}, \bibinfo{author}{S.~J. Fensin},
    \newblock \bibinfo{title}{Evaluating moment tensor potential in {Ag-Cu} alloy:
        {Accuracy}, transferability, and phase diagram fidelity},
    \newblock \bibinfo{journal}{arXiv preprint arXiv:2508.18129}
    (\bibinfo{year}{2025}).
    \bibitem[{Fanourgakis et~al.(2003)Fanourgakis, Pontikis, and
        Z{\'e}rah}]{fanourgakis2003phase}
    \bibinfo{author}{G.~S. Fanourgakis}, \bibinfo{author}{V.~Pontikis},
    \bibinfo{author}{G.~Z{\'e}rah},
    \newblock \bibinfo{title}{Phase stability and intrinsic stacking faults in
        aluminum under pressure},
    \newblock \bibinfo{journal}{Physical Review B} \bibinfo{volume}{67}
    (\bibinfo{year}{2003}) \bibinfo{pages}{094102}.
    \bibitem[{Hale et~al.(2023)Hale, Trautt, and Becker}]{Atomman}
    \bibinfo{author}{L.~M. Hale}, \bibinfo{author}{Z.~T. Trautt},
    \bibinfo{author}{C.~A. Becker}, \bibinfo{title}{Atomistic manipulation
        toolkit (atomman)}, \bibinfo{howpublished}{National Institute of Standards
        and Technology}, \bibinfo{year}{2023}.
    \bibitem[{Mishin et~al.(2001)Mishin, Mehl, Papaconstantopoulos, Voter, and
        Kress}]{mishin2001structural}
    \bibinfo{author}{Y.~Mishin}, \bibinfo{author}{M.~J. Mehl},
    \bibinfo{author}{D.~A. Papaconstantopoulos}, \bibinfo{author}{A.~F. Voter},
    \bibinfo{author}{J.~D. Kress},
    \newblock \bibinfo{title}{Structural stability and lattice defects in copper:
        Ab initio, tight-binding, and embedded-atom calculations},
    \newblock \bibinfo{journal}{Physical Review B} \bibinfo{volume}{63}
    (\bibinfo{year}{2001}) \bibinfo{pages}{224106}.
    \bibitem[{Mendelev et~al.(2008)Mendelev, Kramer, Becker, and
        Asta}]{mendelev2008analysis}
    \bibinfo{author}{M.~Mendelev}, \bibinfo{author}{M.~Kramer},
    \bibinfo{author}{C.~A. Becker}, \bibinfo{author}{M.~Asta},
    \newblock \bibinfo{title}{Analysis of semi-empirical interatomic potentials
        appropriate for simulation of crystalline and liquid {Al} and {Cu}},
    \newblock \bibinfo{journal}{Philosophical Magazine} \bibinfo{volume}{88}
    (\bibinfo{year}{2008}) \bibinfo{pages}{1723--1750}.
    \bibitem[{Mishin et~al.(1999)Mishin, Farkas, Mehl, and
        Papaconstantopoulos}]{mishin1999interatomic}
    \bibinfo{author}{Y.~Mishin}, \bibinfo{author}{D.~Farkas},
    \bibinfo{author}{M.~Mehl}, \bibinfo{author}{D.~Papaconstantopoulos},
    \newblock \bibinfo{title}{Interatomic potentials for monoatomic metals from
        experimental data and ab initio calculations},
    \newblock \bibinfo{journal}{Physical Review B} \bibinfo{volume}{59}
    (\bibinfo{year}{1999}) \bibinfo{pages}{3393}.
    \bibitem[{Starikov et~al.(2020)Starikov, Gordeev, Lysogorskiy, Kolotova, and
        Makarov}]{starikov2020optimized}
    \bibinfo{author}{S.~Starikov}, \bibinfo{author}{I.~Gordeev},
    \bibinfo{author}{Y.~Lysogorskiy}, \bibinfo{author}{L.~Kolotova},
    \bibinfo{author}{S.~Makarov},
    \newblock \bibinfo{title}{Optimized interatomic potential for study of
        structure and phase transitions in {Si-Au} and {Si-Al} systems},
    \newblock \bibinfo{journal}{Computational Materials Science}
    \bibinfo{volume}{184} (\bibinfo{year}{2020}) \bibinfo{pages}{109891}.
    \bibitem[{Etesami and Asadi(2018)}]{etesami2018molecular}
    \bibinfo{author}{S.~A. Etesami}, \bibinfo{author}{E.~Asadi},
    \newblock \bibinfo{title}{Molecular dynamics for near melting temperatures
        simulations of metals using modified embedded-atom method},
    \newblock \bibinfo{journal}{Journal of Physics and Chemistry of Solids}
    \bibinfo{volume}{112} (\bibinfo{year}{2018}) \bibinfo{pages}{61--72}.
    \bibitem[{Grochola et~al.(2005)Grochola, Russo, and
        Snook}]{grochola2005fitting}
    \bibinfo{author}{G.~Grochola}, \bibinfo{author}{S.~P. Russo},
    \bibinfo{author}{I.~K. Snook},
    \newblock \bibinfo{title}{On fitting a gold embedded atom method potential
        using the force matching method},
    \newblock \bibinfo{journal}{The Journal of chemical physics}
    \bibinfo{volume}{123} (\bibinfo{year}{2005}).
    \bibitem[{Purja~Pun(2017)}]{Pun2017_Au_EAM_IPR}
    \bibinfo{author}{G.~P. Purja~Pun}, \bibinfo{title}{{Au\_GLJ10\_3.eam.alloy
            (unpublished)}}, \bibinfo{howpublished}{NIST Interatomic Potentials
        Repository, 2017--Purja-Pun-G-P--Au}, \bibinfo{year}{2017}. \URLprefix
    \url{https://www.ctcms.nist.gov/potentials/system/Au/}.
    \bibitem[{Zhou et~al.(2004)Zhou, Johnson, and Wadley}]{zhou2004misfit}
    \bibinfo{author}{X.~W. Zhou}, \bibinfo{author}{R.~Johnson},
    \bibinfo{author}{H.~N. Wadley},
    \newblock \bibinfo{title}{Misfit-energy-increasing dislocations in
        vapor-deposited {CoFe/NiFe} multilayers},
    \newblock \bibinfo{journal}{Physical Review B} \bibinfo{volume}{69}
    (\bibinfo{year}{2004}) \bibinfo{pages}{144113}.
    \bibitem[{O'Brien et~al.(2018)O'Brien, Barr, Price, Hattar, and
        Foiles}]{o2018grain}
    \bibinfo{author}{C.~O'Brien}, \bibinfo{author}{C.~Barr},
    \bibinfo{author}{P.~Price}, \bibinfo{author}{K.~Hattar},
    \bibinfo{author}{S.~Foiles},
    \newblock \bibinfo{title}{Grain boundary phase transformations in {PtAu} and
        relevance to thermal stabilization of bulk nanocrystalline metals},
    \newblock \bibinfo{journal}{Journal of Materials Science} \bibinfo{volume}{53}
    (\bibinfo{year}{2018}) \bibinfo{pages}{2911--2927}.
    \bibitem[{Foiles and Hoyt(2006)}]{foiles2006computation}
    \bibinfo{author}{S.~M. Foiles}, \bibinfo{author}{J.~Hoyt},
    \newblock \bibinfo{title}{Computation of grain boundary stiffness and mobility
        from boundary fluctuations},
    \newblock \bibinfo{journal}{Acta Materialia} \bibinfo{volume}{54}
    (\bibinfo{year}{2006}) \bibinfo{pages}{3351--3357}.
    \bibitem[{Mishin(2004)}]{mishin2004atomistic}
    \bibinfo{author}{Y.~Mishin},
    \newblock \bibinfo{title}{Atomistic modeling of the $\gamma$ and
        $\gamma^\prime$-phases of the {Ni-Al} system},
    \newblock \bibinfo{journal}{Acta Materialia} \bibinfo{volume}{52}
    (\bibinfo{year}{2004}) \bibinfo{pages}{1451--1467}.
    \bibitem[{Zuo et~al.(2020)Zuo, Chen, Li, Deng, Chen, Behler, Cs{\'a}nyi,
        Shapeev, Thompson, Wood et~al.}]{zuo2020performance}
    \bibinfo{author}{Y.~Zuo}, \bibinfo{author}{C.~Chen}, \bibinfo{author}{X.~Li},
    \bibinfo{author}{Z.~Deng}, \bibinfo{author}{Y.~Chen},
    \bibinfo{author}{J.~Behler}, \bibinfo{author}{G.~Cs{\'a}nyi},
    \bibinfo{author}{A.~V. Shapeev}, \bibinfo{author}{A.~P. Thompson},
    \bibinfo{author}{M.~A. Wood}, et~al.,
    \newblock \bibinfo{title}{Performance and cost assessment of machine learning
        interatomic potentials},
    \newblock \bibinfo{journal}{The Journal of Physical Chemistry A}
    \bibinfo{volume}{124} (\bibinfo{year}{2020}) \bibinfo{pages}{731--745}.
    \bibitem[{Williams et~al.(2006)Williams, Mishin, and
        Hamilton}]{williams2006embedded}
    \bibinfo{author}{P.~Williams}, \bibinfo{author}{Y.~Mishin},
    \bibinfo{author}{J.~Hamilton},
    \newblock \bibinfo{title}{An embedded-atom potential for the {Cu-Ag} system},
    \newblock \bibinfo{journal}{Modelling and Simulation in Materials Science and
        Engineering} \bibinfo{volume}{14} (\bibinfo{year}{2006})
    \bibinfo{pages}{817}.
    \bibitem[{Shapeev(2016)}]{shapeev2016moment}
    \bibinfo{author}{A.~V. Shapeev},
    \newblock \bibinfo{title}{Moment tensor potentials: A class of systematically
        improvable interatomic potentials},
    \newblock \bibinfo{journal}{Multiscale Modeling \& Simulation}
    \bibinfo{volume}{14} (\bibinfo{year}{2016}) \bibinfo{pages}{1153--1173}.
    \bibitem[{Gubaev et~al.(2018)Gubaev, Podryabinkin, and
        Shapeev}]{gubaev2018machine}
    \bibinfo{author}{K.~Gubaev}, \bibinfo{author}{E.~V. Podryabinkin},
    \bibinfo{author}{A.~V. Shapeev},
    \newblock \bibinfo{title}{Machine learning of molecular properties: Locality
        and active learning},
    \newblock \bibinfo{journal}{The Journal of chemical physics}
    \bibinfo{volume}{148} (\bibinfo{year}{2018}).
    \bibitem[{Vocadlo et~al.(2001)}]{vocadlo2001ab}
    \bibinfo{author}{L.~Vocadlo}, et~al.,
    \newblock \bibinfo{title}{The ab initio melting curve of aluminium},
    \newblock \bibinfo{journal}{arXiv preprint cond-mat/0108460}
    (\bibinfo{year}{2001}).
    \bibitem[{De~Jong et~al.(2015)De~Jong, Chen, Angsten, Jain, Notestine, Gamst,
        Sluiter, Krishna~Ande, Van Der~Zwaag, Plata et~al.}]{de2015charting}
    \bibinfo{author}{M.~De~Jong}, \bibinfo{author}{W.~Chen},
    \bibinfo{author}{T.~Angsten}, \bibinfo{author}{A.~Jain},
    \bibinfo{author}{R.~Notestine}, \bibinfo{author}{A.~Gamst},
    \bibinfo{author}{M.~Sluiter}, \bibinfo{author}{C.~Krishna~Ande},
    \bibinfo{author}{S.~Van Der~Zwaag}, \bibinfo{author}{J.~J. Plata}, et~al.,
    \newblock \bibinfo{title}{Charting the complete elastic properties of inorganic
        crystalline compounds},
    \newblock \bibinfo{journal}{Scientific data} \bibinfo{volume}{2}
    (\bibinfo{year}{2015}) \bibinfo{pages}{1--13}.
    \bibitem[{Tran et~al.(2016)Tran, Xu, Radhakrishnan, Winston, Sun, Persson, and
        Ong}]{tran2016surface}
    \bibinfo{author}{R.~Tran}, \bibinfo{author}{Z.~Xu},
    \bibinfo{author}{B.~Radhakrishnan}, \bibinfo{author}{D.~Winston},
    \bibinfo{author}{W.~Sun}, \bibinfo{author}{K.~A. Persson},
    \bibinfo{author}{S.~P. Ong},
    \newblock \bibinfo{title}{Surface energies of elemental crystals},
    \newblock \bibinfo{journal}{Scientific data} \bibinfo{volume}{3}
    (\bibinfo{year}{2016}) \bibinfo{pages}{1--13}.
    \bibitem[{Qiu et~al.(2017)Qiu, Lu, Ao, Huang, Tang, and
        Chen}]{qiu2017energetics}
    \bibinfo{author}{R.~Qiu}, \bibinfo{author}{H.~Lu}, \bibinfo{author}{B.~Ao},
    \bibinfo{author}{L.~Huang}, \bibinfo{author}{T.~Tang},
    \bibinfo{author}{P.~Chen},
    \newblock \bibinfo{title}{Energetics of intrinsic point defects in aluminium
        via orbital-free density functional theory},
    \newblock \bibinfo{journal}{Philosophical Magazine} \bibinfo{volume}{97}
    (\bibinfo{year}{2017}) \bibinfo{pages}{2164--2181}.
    \bibitem[{Zhuang et~al.(2016)Zhuang, Chen, and Carter}]{zhuang2016elastic}
    \bibinfo{author}{H.~Zhuang}, \bibinfo{author}{M.~Chen}, \bibinfo{author}{E.~A.
        Carter},
    \newblock \bibinfo{title}{Elastic and thermodynamic properties of complex mg-al
        intermetallic compounds via orbital-free density functional theory},
    \newblock \bibinfo{journal}{Physical Review Applied} \bibinfo{volume}{5}
    (\bibinfo{year}{2016}) \bibinfo{pages}{064021}.
    \bibitem[{Iyer et~al.(2014)Iyer, Gavini, and Pollock}]{iyer2014energetics}
    \bibinfo{author}{M.~Iyer}, \bibinfo{author}{V.~Gavini}, \bibinfo{author}{T.~M.
        Pollock},
    \newblock \bibinfo{title}{Energetics and nucleation of point defects in
        aluminum under extreme tensile hydrostatic stresses},
    \newblock \bibinfo{journal}{Physical Review B} \bibinfo{volume}{89}
    (\bibinfo{year}{2014}) \bibinfo{pages}{014108}.
    \bibitem[{Sjostrom et~al.(2016)Sjostrom, Crockett, and
        Rudin}]{sjostrom2016multiphase}
    \bibinfo{author}{T.~Sjostrom}, \bibinfo{author}{S.~Crockett},
    \bibinfo{author}{S.~Rudin},
    \newblock \bibinfo{title}{Multiphase aluminum equations of state via density
        functional theory},
    \newblock \bibinfo{journal}{Physical Review B} \bibinfo{volume}{94}
    (\bibinfo{year}{2016}) \bibinfo{pages}{144101}.
    \bibitem[{Devlin(1974)}]{devlin1974stacking}
    \bibinfo{author}{J.~F. Devlin},
    \newblock \bibinfo{title}{Stacking fault energies of be, mg, al, cu, ag, and
        au},
    \newblock \bibinfo{journal}{Journal of Physics F: Metal Physics}
    \bibinfo{volume}{4} (\bibinfo{year}{1974}) \bibinfo{pages}{1865}.
    \bibitem[{Ogata et~al.(2002)Ogata, Li, and Yip}]{ogata2002ideal}
    \bibinfo{author}{S.~Ogata}, \bibinfo{author}{J.~Li}, \bibinfo{author}{S.~Yip},
    \newblock \bibinfo{title}{Ideal pure shear strength of aluminum and copper},
    \newblock \bibinfo{journal}{Science} \bibinfo{volume}{298}
    (\bibinfo{year}{2002}) \bibinfo{pages}{807--811}.
    \bibitem[{Jahn{\'a}tek et~al.(2009)Jahn{\'a}tek, Hafner, and
        Kraj{\v{c}}{\'\i}}]{jahnatek2009shear}
    \bibinfo{author}{M.~Jahn{\'a}tek}, \bibinfo{author}{J.~Hafner},
    \bibinfo{author}{M.~Kraj{\v{c}}{\'\i}},
    \newblock \bibinfo{title}{Shear deformation, ideal strength, and stacking fault
        formation of fcc metals: A density-functional study of al and cu},
    \newblock \bibinfo{journal}{Physical Review B---Condensed Matter and Materials
        Physics} \bibinfo{volume}{79} (\bibinfo{year}{2009}) \bibinfo{pages}{224103}.
    \bibitem[{Kibey et~al.(2007)Kibey, Liu, Johnson, and
        Sehitoglu}]{kibey2007predicting}
    \bibinfo{author}{S.~Kibey}, \bibinfo{author}{J.~Liu},
    \bibinfo{author}{D.~Johnson}, \bibinfo{author}{H.~Sehitoglu},
    \newblock \bibinfo{title}{Predicting twinning stress in fcc metals: Linking
        twin-energy pathways to twin nucleation},
    \newblock \bibinfo{journal}{Acta materialia} \bibinfo{volume}{55}
    (\bibinfo{year}{2007}) \bibinfo{pages}{6843--6851}.
    \bibitem[{Desai(1987)}]{desai1987thermodynamic}
    \bibinfo{author}{P.~Desai},
    \newblock \bibinfo{title}{Thermodynamic properties of aluminum},
    \newblock \bibinfo{journal}{International journal of thermophysics}
    \bibinfo{volume}{8} (\bibinfo{year}{1987}) \bibinfo{pages}{621--638}.
    \bibitem[{Kozyrev and Gordeev(2022)}]{kozyrev2022thermodynamic}
    \bibinfo{author}{N.~V. Kozyrev}, \bibinfo{author}{V.~V. Gordeev},
    \newblock \bibinfo{title}{Thermodynamic properties and equation of state for
        solid and liquid aluminum},
    \newblock \bibinfo{journal}{Metals} \bibinfo{volume}{12} (\bibinfo{year}{2022})
    \bibinfo{pages}{1346}.
    \bibitem[{Ning and Zhang(2023)}]{ning2023ab}
    \bibinfo{author}{B.-Y. Ning}, \bibinfo{author}{L.-Y. Zhang},
    \newblock \bibinfo{title}{An ab initio study of structural phase transitions of
        crystalline aluminium under ultrahigh pressures based on ensemble theory},
    \newblock \bibinfo{journal}{Computational Materials Science}
    \bibinfo{volume}{218} (\bibinfo{year}{2023}) \bibinfo{pages}{111960}.
    \bibitem[{Dewaele et~al.(2004)Dewaele, Loubeyre, and
        Mezouar}]{dewaele2004equations}
    \bibinfo{author}{A.~Dewaele}, \bibinfo{author}{P.~Loubeyre},
    \bibinfo{author}{M.~Mezouar},
    \newblock \bibinfo{title}{Equations of state of six metals above 94 gpa},
    \newblock \bibinfo{journal}{Physical Review B---Condensed Matter and Materials
        Physics} \bibinfo{volume}{70} (\bibinfo{year}{2004}) \bibinfo{pages}{094112}.
    \bibitem[{Akahama et~al.(2006)Akahama, Nishimura, Kinoshita, Kawamura, and
        Ohishi}]{akahama2006evidence}
    \bibinfo{author}{Y.~Akahama}, \bibinfo{author}{M.~Nishimura},
    \bibinfo{author}{K.~Kinoshita}, \bibinfo{author}{H.~Kawamura},
    \bibinfo{author}{Y.~Ohishi},
    \newblock \bibinfo{title}{Evidence of a fcc-hcp transition in aluminum at
        multimegabar pressure},
    \newblock \bibinfo{journal}{Physical review letters} \bibinfo{volume}{96}
    (\bibinfo{year}{2006}) \bibinfo{pages}{045505}.
    \bibitem[{Stedman and Nilsson(1966)}]{stedman1966dispersion}
    \bibinfo{author}{R.~t. Stedman}, \bibinfo{author}{G.~Nilsson},
    \newblock \bibinfo{title}{Dispersion relations for phonons in aluminum at 80
        and 300 k},
    \newblock \bibinfo{journal}{Physical Review} \bibinfo{volume}{145}
    (\bibinfo{year}{1966}) \bibinfo{pages}{492}.
    \bibitem[{Ziegler and Biersack(1985)}]{ziegler1985stopping}
    \bibinfo{author}{J.~F. Ziegler}, \bibinfo{author}{J.~P. Biersack},
    \newblock \bibinfo{title}{The stopping and range of ions in matter},
    \newblock in: \bibinfo{booktitle}{Treatise on heavy-ion science: volume 6:
        astrophysics, chemistry, and condensed matter},
    \bibinfo{publisher}{Springer}, \bibinfo{year}{1985}, pp.
    \bibinfo{pages}{93--129}.
    \bibitem[{Tago and Tonaka(2015)}]{phonopy}
    \bibinfo{author}{A.~Tago}, \bibinfo{author}{I.~Tonaka},
    \newblock \bibinfo{title}{First principles phonon calculations in materials
        science},
    \newblock \bibinfo{journal}{Scripta Mater.} \bibinfo{volume}{108}
    (\bibinfo{year}{2015}) \bibinfo{pages}{1--5}.
    \bibitem[{Morris et~al.(1994)Morris, Wang, Ho, and Chan}]{Morris94}
    \bibinfo{author}{J.~Morris}, \bibinfo{author}{C.~Wang},
    \bibinfo{author}{K.~Ho}, \bibinfo{author}{C.~Chan},
    \newblock \bibinfo{title}{Melting line of aluminum from simulations of
        coexisting phases},
    \newblock \bibinfo{journal}{Phys. Rev. B} \bibinfo{volume}{49}
    (\bibinfo{year}{1994}) \bibinfo{pages}{3109--3115}.
    \bibitem[{Morris and Song(2002)}]{Morris02}
    \bibinfo{author}{J.~Morris}, \bibinfo{author}{X.~Song},
    \newblock \bibinfo{title}{The melting lines of model systems calculated from
        coexistence simulations},
    \newblock \bibinfo{journal}{J. Chem. Phys.} \bibinfo{volume}{116}
    (\bibinfo{year}{2002}) \bibinfo{pages}{9352--9358}.
    \bibitem[{Howells and Mishin(2018)}]{Howells:2018aa}
    \bibinfo{author}{C.~A. Howells}, \bibinfo{author}{Y.~Mishin},
    \newblock \bibinfo{title}{Angular-dependent interatomic potential for the
        binary {Ni-C}r system},
    \newblock \bibinfo{journal}{Model. Simul. Mater. Sci. Eng.}
    \bibinfo{volume}{26} (\bibinfo{year}{2018}) \bibinfo{pages}{085008}.
    \bibitem[{Tersoff(1988{\natexlab{a}})}]{Tersoff88}
    \bibinfo{author}{J.~Tersoff},
    \newblock \bibinfo{title}{New empirical approach for the structure and energy
        of covalent systems},
    \newblock \bibinfo{journal}{Phys. Rev. {\rm B}} \bibinfo{volume}{37}
    (\bibinfo{year}{1988}{\natexlab{a}}) \bibinfo{pages}{6991--7000}.
    \bibitem[{Tersoff(1988{\natexlab{b}})}]{Tersoff:1988dn}
    \bibinfo{author}{J.~Tersoff},
    \newblock \bibinfo{title}{Empirical interatomic potential for silicon with
        improved elastic properties},
    \newblock \bibinfo{journal}{Phys. Rev. {\rm B}} \bibinfo{volume}{38}
    (\bibinfo{year}{1988}{\natexlab{b}}) \bibinfo{pages}{9902--9905}.
    \bibitem[{Tersoff(1989)}]{Tersoff:1989wj}
    \bibinfo{author}{J.~Tersoff},
    \newblock \bibinfo{title}{Modeling solid-state chemistry: Interatomic
        potentials for multicomponent systems},
    \newblock \bibinfo{journal}{Phys. Rev. {\rm B}} \bibinfo{volume}{39}
    (\bibinfo{year}{1989}) \bibinfo{pages}{5566--5568}.
    
\end{thebibliography}

\newpage{}

\begin{figure}
\begin{centering}
\includegraphics[width=1\textwidth]{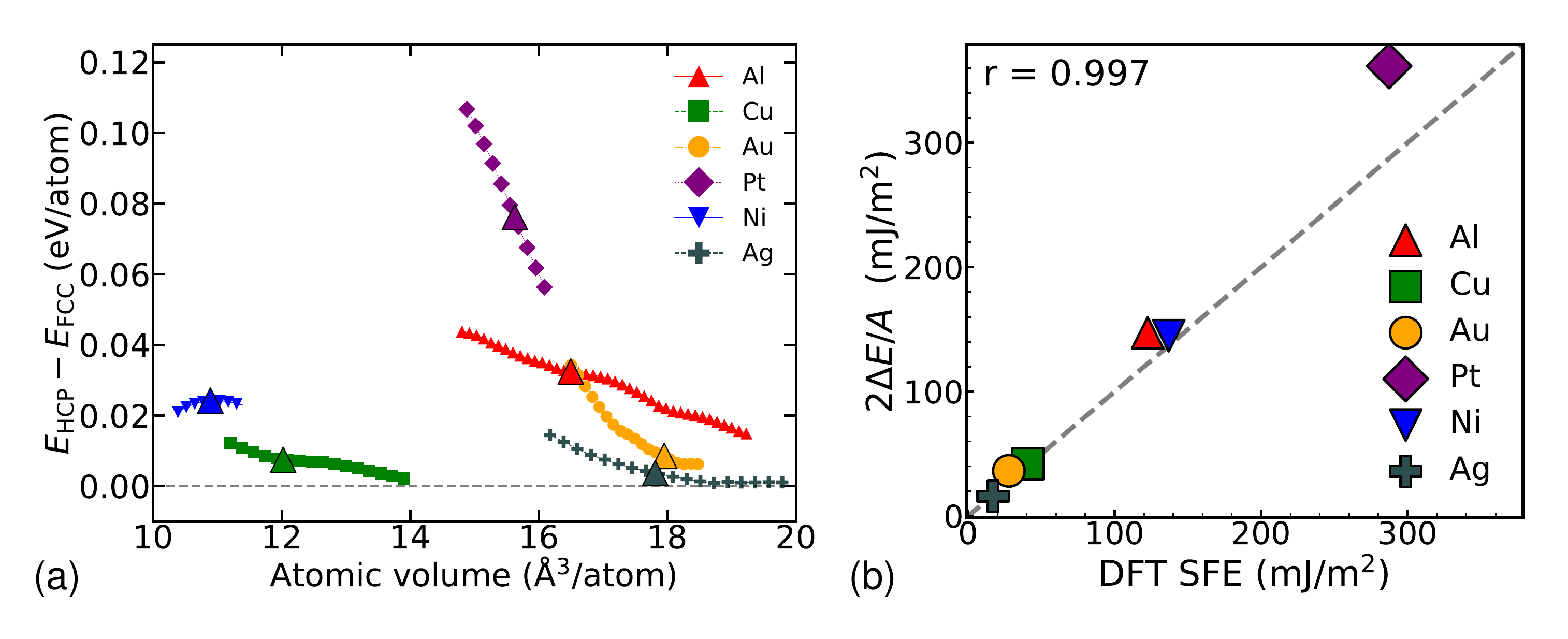}
\par\end{centering}
\caption{(a) HCP-FCC energy difference for six FCC metals obtained by DFT calculations
as a function of atomic volume. Only the atomic volumes for which
the SF energy remains stable are included. The triangular symbols
mark the data for the equilibrium FCC volume. (b) Comparison the SFE
with predictions of Eq.(\ref{eq:hcp-fcc}). The dashed line represents
perfect agreement. }\label{fig:HCP-FCC}
\end{figure}

\begin{figure}
\includegraphics[width=1\textwidth]{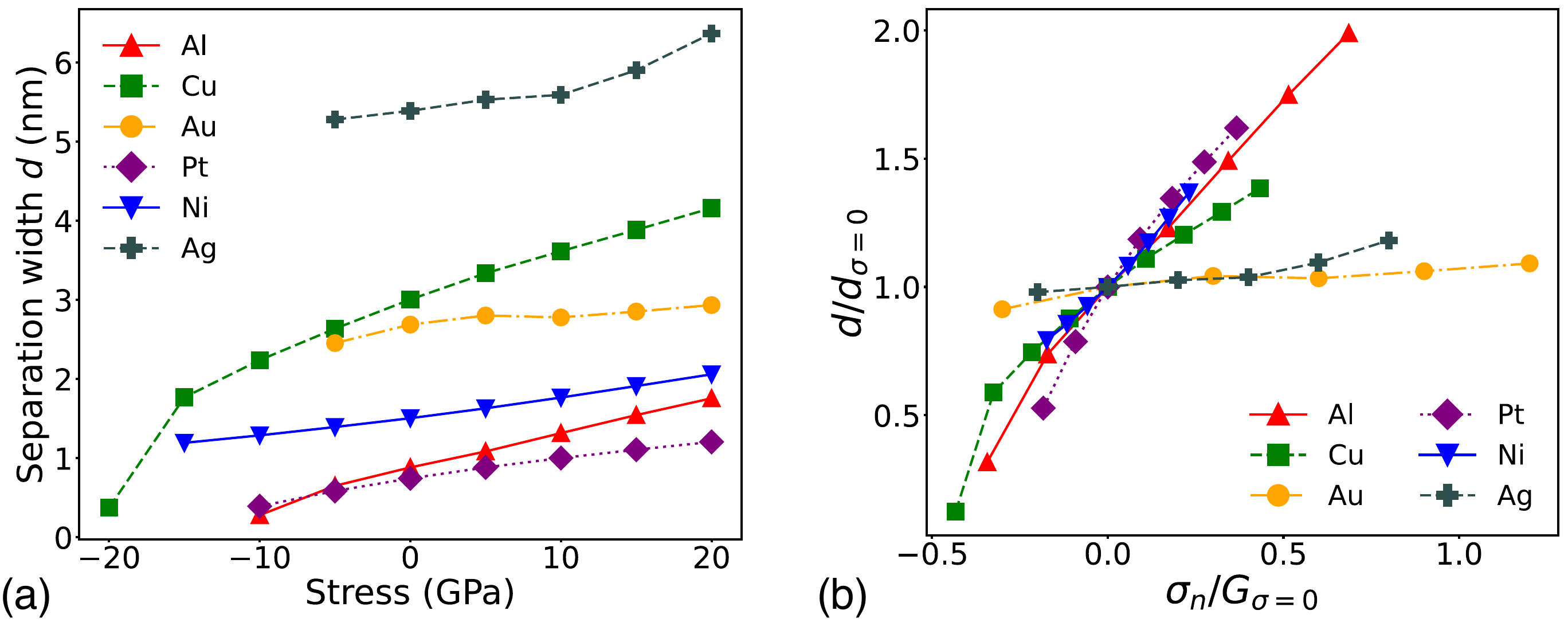}

\caption{Dissociation width of an edge dislocation as a function of normal
stress in six FCC metals obtained from Eq.(\ref{eq:d}) with input
from DFT calculations. (a) Physical coordinates. (b) Normalized coordinates.}\label{fig:d}
\end{figure}

\begin{figure}[!ht]
\centering \includegraphics[width=1\linewidth]{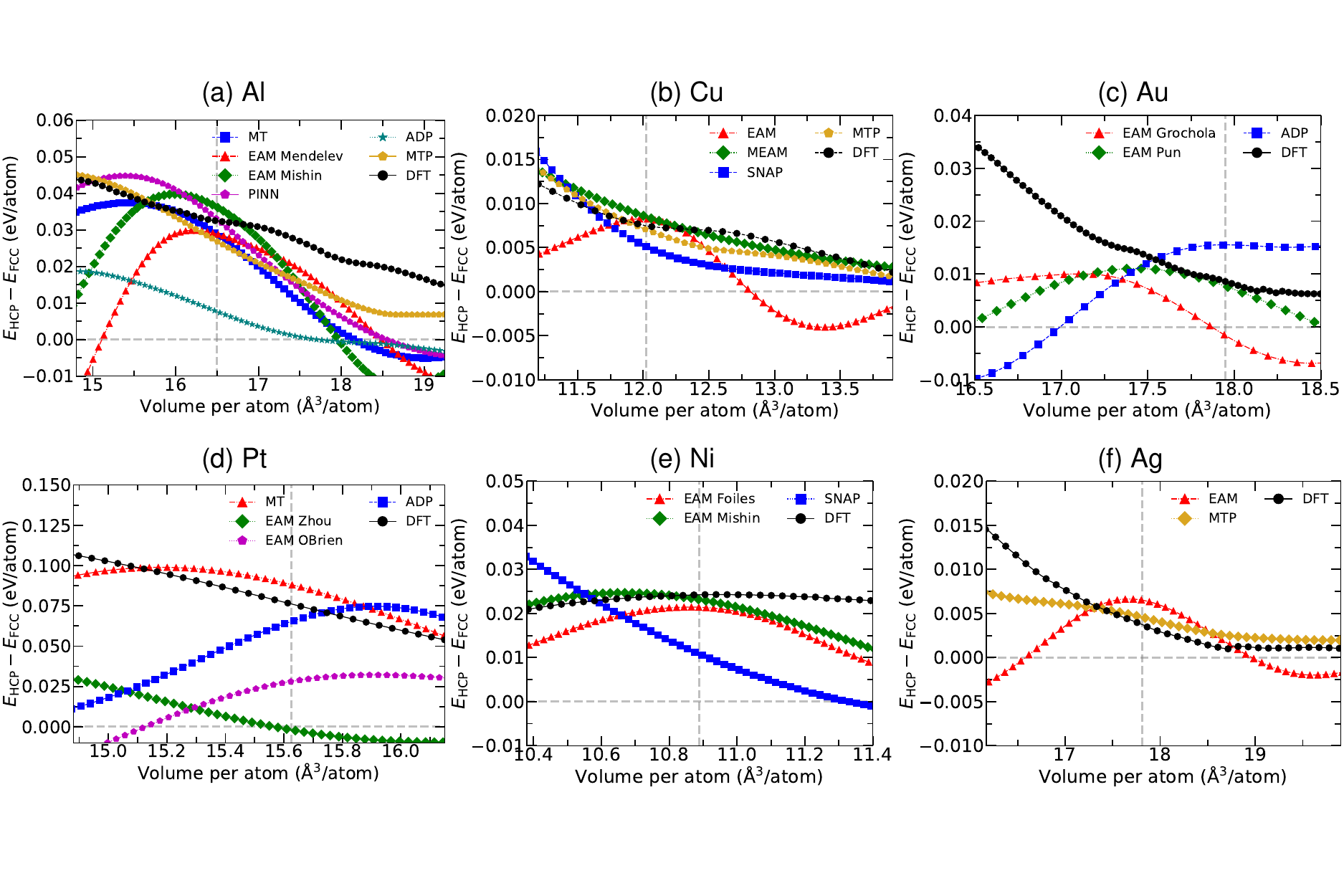}
 \caption{HCP-FCC energy difference for six metals as a function of atomic volume
obtained by DFT calculations in comparison with interatomic potentials.
Only the atomic volumes for which the SF energy remains stable are
included. (a) Al, (b) Cu, (c) Au, (d) Pt, (e) Ni, and (f) Ag. The
vertical dashed line corresponds to zero stress and serves as a visual
guide.}
\label{fig:hcp_fcc_compare}
\end{figure}

\begin{figure}
\begin{centering}
\includegraphics[width=0.5\textwidth]{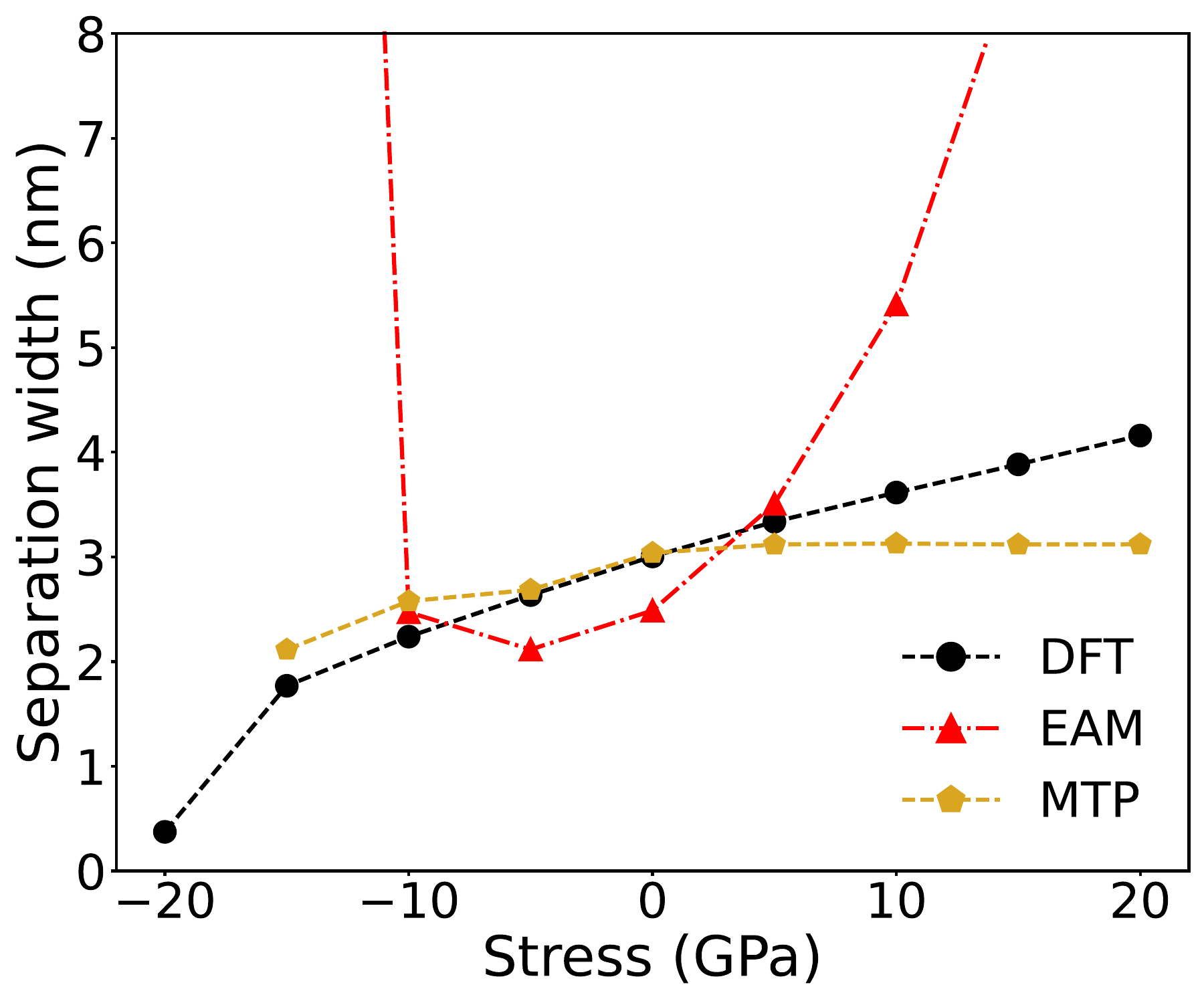}
\par\end{centering}
\caption{Equilibrium separation between partial dislocations in Cu as a function
of applied stresses predicted by Eq.(\ref{eq:d}) with input from
DFT calculations and from the MTP \citep{nitol2025evaluating} and
EAM \citep{mishin2001structural} potentials.}\label{fig:Cu-separation}

\end{figure}

\begin{figure}[!ht]
\centering \includegraphics[width=1\linewidth]{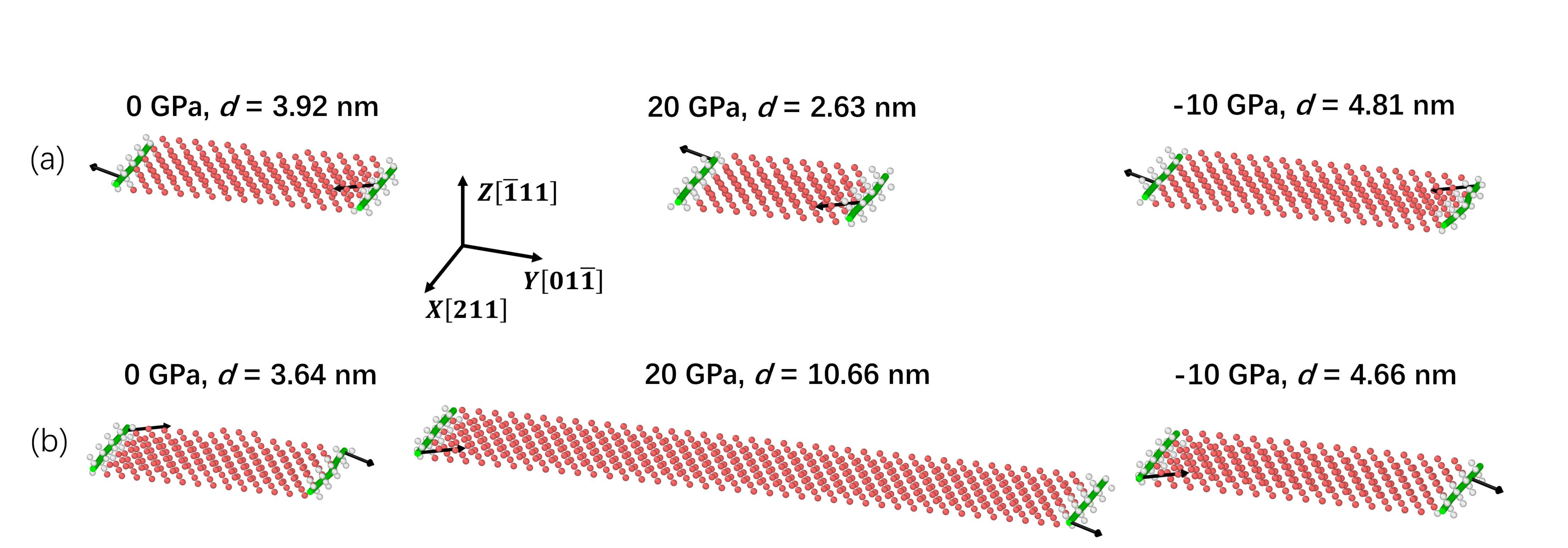}
 \caption{Equilibrium separation distance $d$ of partial dislocations in Cu
subjected to normal stresses of $0$, $20$, and $-10$ GPa (positive
values indicate compression, negative values indicate tension). (a)
Results obtained with the MTP potential \citep{nitol2025evaluating}.
(b) Results obtained with the EAM potential \citep{mishin2001structural}.
Green lines mark partial dislocations, and black arrows indicate the
Burgers vectors. Atoms within the stacking fault are shown in red.
Atoms surrounding the dislocations shown in grey do not belong to
any crystalline structure.}\label{fig:MD-separation-distances}
\end{figure}

\begin{table}[h]
\centering \caption{Interatomic potentials tested in this work.}
\begin{tabular}{lccc}
\hline 
Label  & Potential model  & Cite  & Fitting/Training data \tabularnewline
\hline 
\multicolumn{4}{c}{Al}\tabularnewline
MT  & Modified Tersoff potential  & This work  & DFT \tabularnewline
EAM Mendelev  & Embedded atom method  & \citep{mendelev2008analysis}  & DFT + Experiment \tabularnewline
EAM Mishin  & Embedded atom method  & \citep{mishin1999interatomic}  & DFT + Experiment \tabularnewline
PINN  & Physically-informed neural network  & \citep{pun2020development}  & DFT \tabularnewline
ADP  & Angular-dependent potential  & \citep{starikov2020optimized}  & DFT \tabularnewline
MTP  & Moment Tensor potential  & This work  & DFT \tabularnewline
\hline 
\multicolumn{4}{c}{Cu}\tabularnewline
EAM  & Embedded atom method  & \citep{mishin2001structural}  & DFT + Experiment \tabularnewline
MEAM  & Modified embedded-atom method  & \citep{etesami2018molecular}  & Experiment \tabularnewline
SNAP  & Spectral neighbor analysis potential  & \citep{li2018quantum}  & DFT + ab initio MD \tabularnewline
MTP  & Moment Tensor potential  & \citep{nitol2025evaluating}  & DFT \tabularnewline
\hline 
\multicolumn{4}{c}{Au}\tabularnewline
EAM Grochola  & Embedded atom method  & \citep{grochola2005fitting}  & DFT + Experiment \tabularnewline
EAM Pun  & Embedded atom method  & \citep{Pun2017_Au_EAM_IPR}  & DFT \tabularnewline
ADP  & Angular-dependent potential  & \citep{starikov2020optimized}  & DFT \tabularnewline
\hline 
\multicolumn{4}{c}{Pt}\tabularnewline
MT  & Modified Tersoff potential  & \citep{Pt-in-review}  & DFT \tabularnewline
EAM Zhou  & Embedded atom method  & \citep{zhou2004misfit}  & Experiment \tabularnewline
EAM O'Brien  & Embedded atom method  & \citep{o2018grain}  & DFT \tabularnewline
ADP  & Angular-dependent potential  & \citep{Pt-in-review}  & DFT \tabularnewline
\hline 
\multicolumn{4}{c}{Ni}\tabularnewline
EAM Foiles  & Embedded atom method  & \citep{daw1993embedded,foiles2006computation}  & Experiment \tabularnewline
EAM Mishin  & Embedded atom method  & \citep{mishin2004atomistic}  & DFT + Experiment \tabularnewline
SNAP  & Spectral neighbor analysis potential  & \citep{zuo2020performance}  & DFT \tabularnewline
\hline 
\multicolumn{4}{c}{Ag}\tabularnewline
EAM  & Embedded atom method  & \citep{williams2006embedded}  & DFT + Experiment \tabularnewline
MTP  & Moment Tensor potential  & \citep{nitol2025evaluating}  & DFT \tabularnewline
\hline 
\end{tabular}\vspace{0.3em}
 \label{table:table1} 
\end{table}

\clearpage{}

\appendix

\section*{SUPPLEMENTARY INFORMATION}

\setcounter{figure}{0} \setcounter{table}{0} 
\global\long\def\thefigure{S\arabic{figure}}%
 
\global\long\def\thetable{S\arabic{table}}%

\begin{center}
{\Large\textbf{The effect of normal stress on stacking fault energy
in face-centered cubic metals}}{\Large\par}
\par\end{center}

\begin{center}
{\large Yang Li and Yuri Mishin}\\
\bigskip{}
{\large Department of Physics and Astronomy, MSN 3F3,}\\
{\large{} George Mason University, Fairfax, VA 22030, USA}{\large\par}
\par\end{center}

\bigskip{}

\section{Summary of DFT results}

The tables below summarize the main results of the DFT calculations
performed in this work. The dash symbol represents cases when the
simulation cell was unstable at the applied stress. 

\begin{table}[h]
\centering \caption{DFT SF energy (mJ/m$^{2}$) at different normal stresses applied to
the fault plane.}
\begin{tabular}{lccccccccc}
\hline 
 & $-20$ GPa & $-15$ GPa & $-10$ GPa & $-5$ GPa & $0$ & $5$ GPa & $10$ GPa & $15$ GPa & $20$ GPa\tabularnewline
\hline 
Al & - & - & 43.21 & 88.67 & 122.52 & 151.70 & 174.24 & 189.96 & 195.13\tabularnewline
Cu & 7.59 & 19.37 & 27.20 & 34.42 & 40.82 & 47.27 & 54.29 & 60.96 & 67.54\tabularnewline
Au & - & - & - & 10.26 & 27.78 & 49.02 & 72.59 & 97.94 & 124.08\tabularnewline
Pt & - & - & 196.19 & 245.27 & 287.30 & 325.43 & 360.47 & 392.17 & 420.61\tabularnewline
Ni & - & 95.69 & 111.82 & 125.22 & 136.89 & 146.48 & 154.19 & 160.44 & 164.37\tabularnewline
Ag & - & - & 0.62 & 8.48 & 17.04 & 26.63 & 37.24 & 47.29 & 56.68\tabularnewline
\hline 
\end{tabular}\vspace{0.3em}
 \label{table:DFT_data_SFE}
\end{table}

\begin{table}[h]
\centering \caption{DFT USFE (mJ/m$^{2}$) at different normal stresses applied to the
fault plane.}
\begin{tabular}{lccccccccc}
\hline 
l & $-20$ GPa & $-15$ GPa & $-10$ GPa & $-5$ GPa & $0$ & $5$ GPa & $10$ GPa & $15$ GPa & $20$ GPa\tabularnewline
\hline 
Al & - & - & 44.54 & 106.80 & 161.10 & 210.45 & 255.05 & 296.68 & 327.99\tabularnewline
Cu & 32.68 & 72.87 & 105.04 & 135.58 & 165.62 & 195.16 & 226.02 & 256.48 & 287.59\tabularnewline
Au & - & - & - & 35.59 & 81.02 & 131.23 & 182.85 & 236.91 & 292.41\tabularnewline
Pt & - & - & 207.31 & 253.79 & 301.09 & 348.81 & 397.66 & 444.55 & 491.98\tabularnewline
Ni & - & 191.24 & 224.24 & 255.01 & 284.81 & 313.60 & 341.38 & 367.95 & 393.23\tabularnewline
Ag & - & - & 21.29 & 60.55 & 97.39 & 137.51 & 177.35 & 217.80 & 258.39\tabularnewline
\hline 
\end{tabular}\vspace{0.3em}
 \label{table:DFT_data_USFE}
\end{table}

\clearpage{}

\begin{table}[th]
\centering \caption{DFT shear modulus (GPa) corresponding to shear parallel to $\{111\}$
plane along a $\langle112\rangle$ direction at different normal stresses
applied in the {[}111{]} direction.}
\begin{tabular}{lccccccccc}
\hline 
 & $-20$ GPa & $-15$ GPa & $-10$ GPa & $-5$ GPa & $0$ & $5$ GPa & $10$ GPa & $15$ GPa & $20$ GPa\tabularnewline
\hline 
Al & - & - & 4.68 & 17.60 & 29.16 & 40.09 & 50.59 & 60.47 & 68.30\tabularnewline
Cu & 1.46 & 15.96 & 26.31 & 36.52 & 46.20 & 55.54 & 64.87 & 73.74 & 82.58\tabularnewline
Au & - & - & - & 6.15 & 16.66 & 27.79 & 39.10 & 50.30 & 61.74\tabularnewline
Pt & - & - & 21.66 & 38.17 & 54.52 & 71.01 & 87.14 & 102.84 & 117.95\tabularnewline
Ni & - & 54.10 & 65.39 & 76.09 & 86.42 & 96.42 & 106.06 & 115.33 & 124.48\tabularnewline
Ag & - & - & 2.58 & 14.08 & 25.00 & 35.38 & 45.68 & 55.62 & 65.79\tabularnewline
\hline 
\end{tabular}\vspace{0.3em}
 \label{table:DFT_data_G}
\end{table}

\begin{table}[th]
\centering \caption{DFT Poisson's ratio (average of $\nu_{31}$ and $\nu_{32}$) at different
normal stresses applied in the {[}111{]} direction. $\nu_{31}$ and
$\nu_{32}$ correspond to uniaxial strain along $[111]$ and the transverse
strains along $[1\bar{1}0]$ and $[11\bar{2}]$, respectively.}
\begin{tabular}{lccccccccc}
\hline 
 & $-20$ GPa & $-15$ GPa & $-10$ GPa & $-5$ GPa & $0$ & $5$ GPa & $10$ GPa & $15$ GPa & $20$ GPa\tabularnewline
\hline 
Al & -- & -- & 0.110 & 0.248 & 0.319 & 0.376 & 0.417 & 0.450 & 0.468\tabularnewline
Cu & 0.074 & 0.137 & 0.184 & 0.226 & 0.263 & 0.299 & 0.332 & 0.364 & 0.394\tabularnewline
Au & -- & -- & -- & 0.362 & 0.413 & 0.451 & 0.475 & 0.504 & 0.524\tabularnewline
Pt & -- & -- & 0.332 & 0.361 & 0.381 & 0.396 & 0.408 & 0.417 & 0.425\tabularnewline
Ni & -- & 0.159 & 0.184 & 0.208 & 0.230 & 0.252 & 0.272 & 0.291 & 0.304\tabularnewline
Ag & -- & -- & 0.120 & 0.225 & 0.303 & 0.366 & 0.418 & 0.456 & 0.488\tabularnewline
\hline 
\end{tabular}\label{table:DFT_data_nu}
\end{table}

\section{The moment tensor potential for Al}

In this section, we describe the development of a moment tensor potential
(MTP) for Al. The MTP is a machine-learning interatomic potential
built on a set of local moment tensor descriptors. For each atom $i$,
its neighborhood (within a cutoff range) is encoded by the moment
tensors \citep{shapeev2016moment,gubaev2018machine} 
\[
\mathbf{M}_{\mu,\nu}(i)\;=\;\sum_{j\in\mathcal{N}_{i}}f_{\mu}(r_{ij})\,\underbrace{\mathbf{r}_{ij}\otimes\cdots\otimes\mathbf{r}_{ij}}_{\nu\ \text{times}},
\]
where $\mathbf{r}_{ij}$ is the relative position vector of a neighboring
atom $j$, $r_{ij}=|\mathbf{r}_{ij}|$, $f_{\mu}(r)$ are radial basis
functions, and $\mu,\nu$ index radial and angular orders, respectively.
Scalar basis functions $B_{\alpha}(i)$ are then formed as all possible
contractions of the moment tensors. These functions are invariant
under translations, rotations, and permutations of chemical species.
The atomic energy $E_{i}$ is expressed as a linear expansion in the
basis functions, 
\[
E_{i}\;=\;\sum_{\alpha}c_{\alpha}\,B_{\alpha}(i),
\]
where $\{c_{\alpha}\}$ are trainable coefficients. The total energy
is then obtained by
\[
E_{\text{tot}}\;=\;\sum_{i}E_{i}.
\]
Increasing the maximum tensor orders $(\mu,\nu)$ expands the basis
and raises the effective body order. Training consists of determining
$\{c_{\alpha}\}$ by minimizing a regularized loss function that interpolates
the DFT reference dataset. Further details of MTP training can be
found in Refs.~\citep{shapeev2016moment,gubaev2018machine}.

The reference database was composed of energies, forces, and stresses
for a set of atomic configurations computed within the density-functional
theory (DFT). Calculations were performed with the Vienna \textit{ab
initio} Simulation Package (VASP)~\citep{kresse1996efficiency,kresse1996efficient},
employing the projector-augmented-wave (PAW) method~\citep{blochl1994projector}
and the Perdew--Burke--Ernzerhof (PBE) generalized-gradient approximation
for exchange-correlation~\citep{perdew1996generalized}. The PAW
dataset used was \texttt{PAW\_PBE} Al (04Jan2001). A plane-wave energy
cutoff of 500~eV was applied. Brillouin-zone integrations used Monkhorst--Pack
meshes corresponding to a k-point density of $10{,}000$ k-points
per reciprocal atom (kppa $=10{,}000$). First-order Methfessel--Paxton
smearing with a width of 0.20~eV was employed, and electronic self-consistency
was converged to $10^{-6}$~eV.

\begin{table}[h]
\centering \caption{DFT database used to create the MTP potential for Al. The first column
lists the structure type; the second gives the number of atoms per
structure; the third describes the number of configurations for each
type of structure.}
\begin{tabular}{lccc}
\hline 
Structure & Number of atoms & Number of configurations & \tabularnewline
\hline 
Bulk deformations, FCC & 32 & 202 & \tabularnewline
Bulk shear deformations, FCC & 32 & 33 & \tabularnewline
AIMD, FCC & 32 & 80 & \tabularnewline
Bulk deformations, A15 & 8 & 99 & \tabularnewline
Bulk deformations, BCC & 2 & 120 & \tabularnewline
Bulk deformations, DC & 8 & 109 & \tabularnewline
Bulk deformations, & 32 & 202 & \tabularnewline
Bulk deformations, HEX & 1 & 107 & \tabularnewline
Bulk deformations, SC & 8 & 119 & \tabularnewline
Surface$\{100\}$\  & 144 & 37 & \tabularnewline
Surface$\{110\}$\  & 128 & 40 & \tabularnewline
Surface$\{111\}$\  & 16 & 40 & \tabularnewline
Stacking fault$\{111\}$$\langle112\rangle$ & 30 & 40 & \tabularnewline
Dimer & 2 & 10 & \tabularnewline
Interstitial & 33 & 50 & \tabularnewline
Vacancy & 31, 255 & 120 & \tabularnewline
Liquid phase & 108 & 51 & \tabularnewline
\hline 
\end{tabular}\vspace{0.3em}
 \label{table:mtpdataset}
\end{table}

The DFT database comprised multiple crystalline phases: face-centered
cubic (FCC), body-centered cubic (BCC), hexagonal close-packed (HCP),
simple cubic (SC), simple hexagonal (HEX), A15, and diamond cubic
(DC), as well as liquid, point defects, stacking faults, and free-surface
structures. Details of these structures are summarized in Table~\ref{table:mtpdataset}.
Bulk structures were subjected to isotropic (volumetric) deformations
to generate multiple configurations. In addition, FCC bulk structures
were deformed under applied shear and were also relaxed at 500~K
using ab initio molecular dynamics (AIMD) simulations to generate
additional configurations. Free-surface and point-defect models were
randomized by applying small atomic displacements to sample distinct
local environments. Stacking-fault (SF) configurations were constructed
by imposing shear displacements along $\langle112\rangle$ within
the $\{111\}$ plane. Dimer configurations were generated with varying
separation distances. Liquid-phase Al configurations were obtained
from molecular dynamics trajectories generated with the EAM Mishin
potential \citep{mishin1999interatomic}. Representative snapshots
were subsequently re-evaluated using DFT to obtain energies, forces,
and stresses. The volumetrically deformed FCC and HCP bulk structures
were assigned more configurations. This choice reflects our observation
that the stacking-fault energy (SFE) under normal stress is highly
sensitive to the FCC--HCP energy difference as a function of volumetric
strain. Emphasizing FCC and HCP configurations at different deformation
states improves the MTP’s ability to reproduce the DFT-computed SFE
at high stresses. 

For liquid-phase configurations, we applied a uniform energy shift
of $+\!0.02\ \mathrm{eV/atom}$. This correction is motivated by the
tendency of GGA-based DFT to under-predict the melting temperature
of Al at low pressure \citep{vocadlo2001ab}. This trend was confirmed
by our own MTP simulations, which yields a melting temperature that
was too low. Raising the liquid energies relative to the solid enlarges
the solid--liquid free-energy gap and brings the predicted melting
temperature closer to experiment. The value of $0.02\ \mathrm{eV/atom}$
was selected empirically after testing several shifts and choosing
the one that yielded the best melting temperature. 

\begin{table}[h]
\centering \caption{Comparison of Al properties from DFT, MTP, MT, PINN~\citep{pun2020development},
EAM~\citep{mishin1999interatomic}, and ADP~\citep{starikov2020optimized}
potentials. The properties marked with an asterisk were calculated
in this work and are different from those reported in the original
publication.}
\label{tab:al_properties} %
\begin{tabular}{lcccccc}
\hline 
Property & DFT & MTP$^{\dagger}$ & MT$^{\dagger}$ & PINN & EAM & ADP\tabularnewline
\hline 
$E_{0}$ (eV/atom) & 3.7480$^{a}$ & 3.3555 & 3.3560 & 3.3604 & 3.3600 & 3.3810\tabularnewline
$a_{0}$ (\AA) & 4.039$^{a,d}$; 3.9725--4.0676$^{c}$ & 4.0405 & 4.0284 & 4.0399 & 4.0500 & 4.0254\tabularnewline
$B$ (GPa) & 83$^{a}$; 81$^{f}$ & 75 & 83 & 81 & 79 & 63\tabularnewline
$C_{11}$ (GPa) & 104$^{a}$; 103--106$^{d}$ & 106 & 110 & 112 & 114 & 99{*}\tabularnewline
$C_{12}$ (GPa) & 73$^{a}$; 57--66$^{d}$ & 59 & 70 & 65 & 62 & 46{*}\tabularnewline
$C_{44}$ (GPa) & 32$^{a}$; 28--33$^{d}$ & 32 & 39 & 28 & 32 & 39{*}\tabularnewline
$\gamma_{S}$(100) (J/m$^{2}$) & 0.92$^{b}$ & 0.910 & 0.761 & 0.904 & 0.944 & 0.822\tabularnewline
$\gamma_{S}$(110) (J/m$^{2}$) & 0.98$^{b}$ & 0.951 & 0.818 & 0.954 & 1.006 & 0.858\tabularnewline
$\gamma_{S}$(111) (J/m$^{2}$) & 0.80$^{b}$ & 0.802 & 0.641 & 0.804 & 0.871 & 0.784\tabularnewline
$E_{v}^{f}$ (eV) & 0.6646--1.3458$^{c}$; 0.7$^{e}$ & 0.694 & 0.670 & 0.703 & 0.676 & 0.734\tabularnewline
$E_{v}^{m}$ (eV) & 0.3041--0.6251$^{c}$ & 0.639 & 0.575 & 0.628 & 0.636 & 0.649\tabularnewline
$E_{f}^{T}$ (Td) (eV) & 2.2001--3.2941$^{c}$ & 3.319 & 2.837 & 2.706 & 3.105 & 3.177\tabularnewline
$E_{f}^{T}$ (Oh) (eV) & 2.5313--2.9485$^{c}$ & 2.833 & 2.649 & 2.739 & 2.798 & 2.780\tabularnewline
$E_{f}^{T}$ $<$100$>$ (eV) & 2.2953--2.6073$^{c}$ & 2.721 & 2.342 & 2.517 & 2.600 & 2.575\tabularnewline
$E_{f}^{T}$ $<$110$>$ (eV) & 2.5432--2.9809$^{c}$ & 2.900 & 2.674 & 2.843 & 2.930 & 2.917\tabularnewline
$E_{f}^{T}$ $<$111$>$ (eV) & 2.6793--3.1821$^{c}$ & 3.232 & 2.838 & 2.775 & 3.018 & 3.129\tabularnewline
$\gamma_{SF}$ (mJ/m$^{2}$) & 134$^{i}$; 145.67$^{g}$; 158$^{h}$; 123$^{\dagger}$ & 116 & 134 & 134 & 145 & 40\tabularnewline
$\gamma_{US}$ (mJ/m$^{2}$) & 162$^{j}$; 175$^{h}$; 161$^{\dagger}$ & 150 & 163 & 150 & 167 & 135\tabularnewline
$T_{m}$ (K) & 933 (Experiment)$^{k}$ & 966 & 928 & 975 & 1038 & 855{*}\tabularnewline
\hline 
\end{tabular}
\raggedright{}$^{a}$Ref \citep{de2015charting}, $^{b}$Ref\citep{tran2016surface},
$^{c}$Ref\citep{qiu2017energetics}, $^{d}$Ref \citep{zhuang2016elastic},
$^{e}$Ref\citep{iyer2014energetics}, $^{f}$Ref\citep{sjostrom2016multiphase},
$^{g}$Ref\citep{devlin1974stacking}, $^{h}$Ref\citep{ogata2002ideal},
$^{i}$Ref\citep{jahnatek2009shear},$^{j}$Ref\citep{kibey2007predicting},
$^{k}$Ref\citep{desai1987thermodynamic}. $^{\dagger}$This work.
\end{table}

\begin{figure}[!ht]
\vspace{0cm}
 \centering \includegraphics[height=5cm]{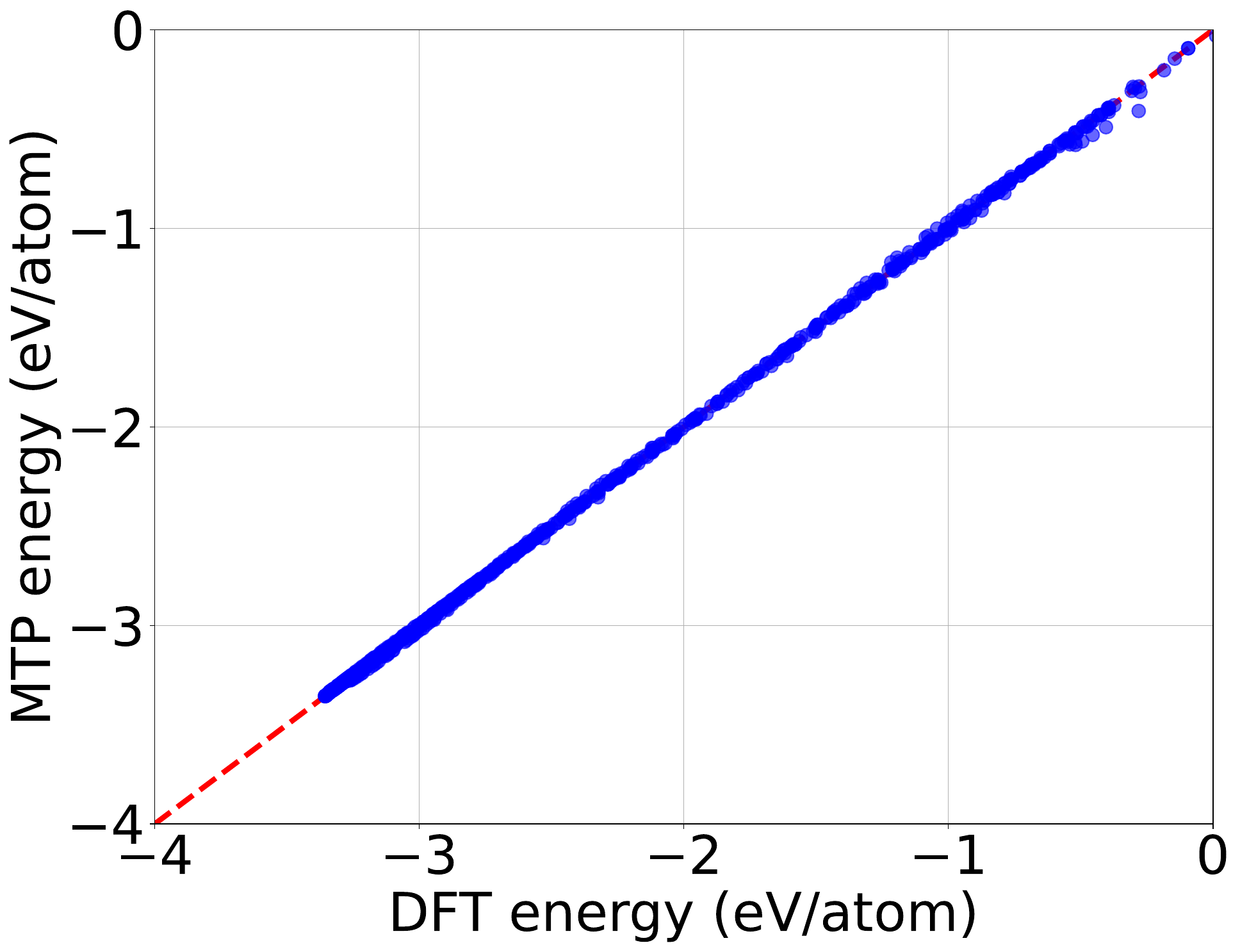}\qquad{}
\includegraphics[height=5cm]{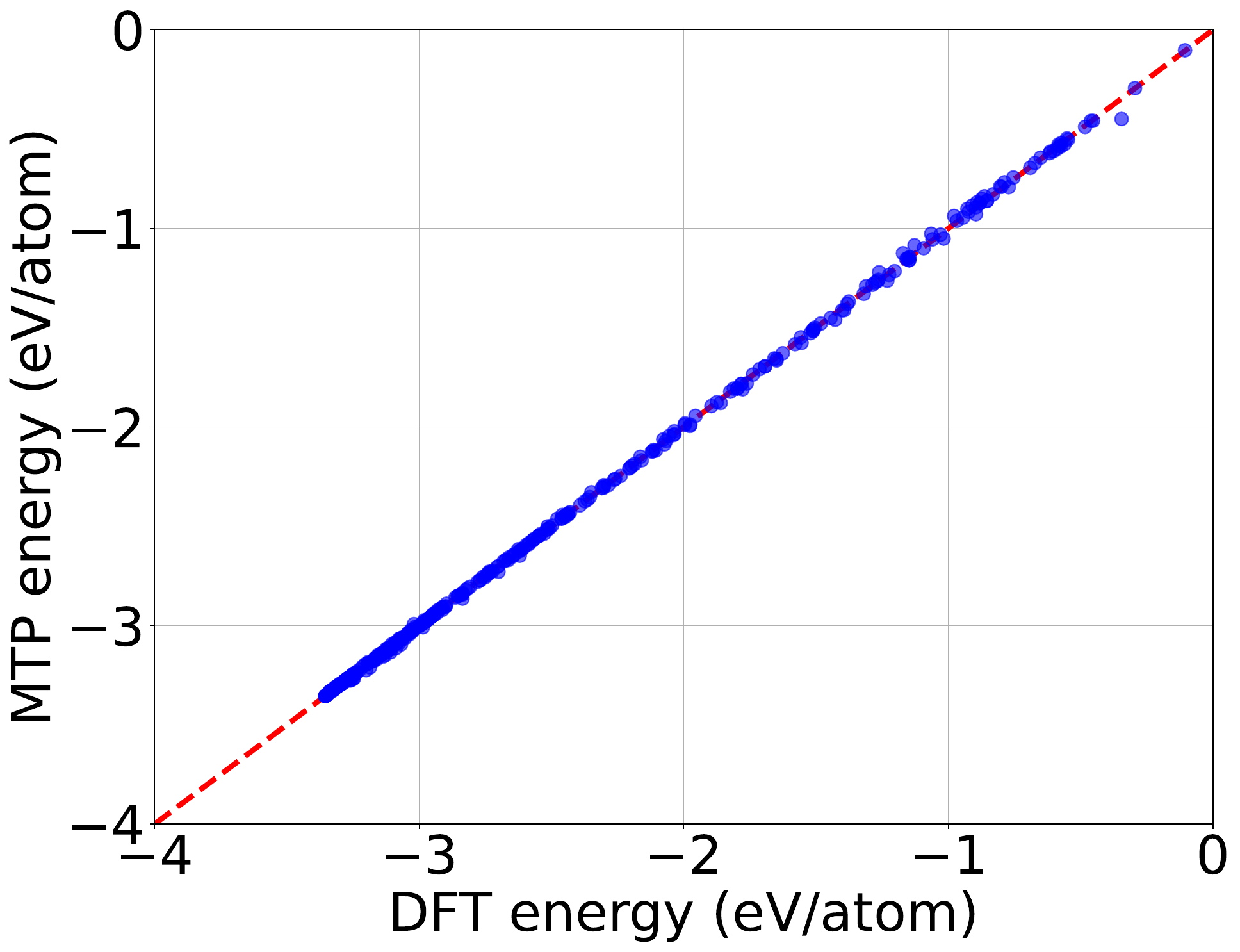} \caption[]{ Comparison of DFT-computed and MTP-predicted energies of Al in the
training set (left) and the validation set (right). The dashed lines
represent that the perfect match.}
\label{fig:MTP_validate}
\end{figure}

The reference dataset was randomly divided into training (70\%) and
validation (30\%) subsets. The training subset was employed to fit
the MTP model. Default weights of 1.0, 0.01, and 0.001 were assigned
to the energy, forces, and stress components, respectively. The cutoff
radius was set to 6.0~\AA, with a minimum cutoff of 0.5~\AA. Among
several cutoff values examined here, this choice provided the most
accurate and stable results. An MTP of level~16 was used. Several
random training/validation splits were performed, and in all cases
the trained MTP accurately reproduced the DFT-calculated energies
of the validation configurations. Results for one representative split,
comparing DFT-calculated energies with MTP-predicted energies for
both the training and validation sets, are shown in Fig.~\ref{fig:MTP_validate}.
Strong correlation is observed between the DFT-computed and MTP-predicted
total energies, confirming that the MTP is well-trained without noticeable
under-fitting or over-fitting. The average absolute differences of
the MTP and DFT-computed energy per atom are 6.2 meV/atom and 6.8
meV/atom.

\begin{figure}[!ht]
\vspace{0cm}
 \centering \includegraphics[height=5cm]{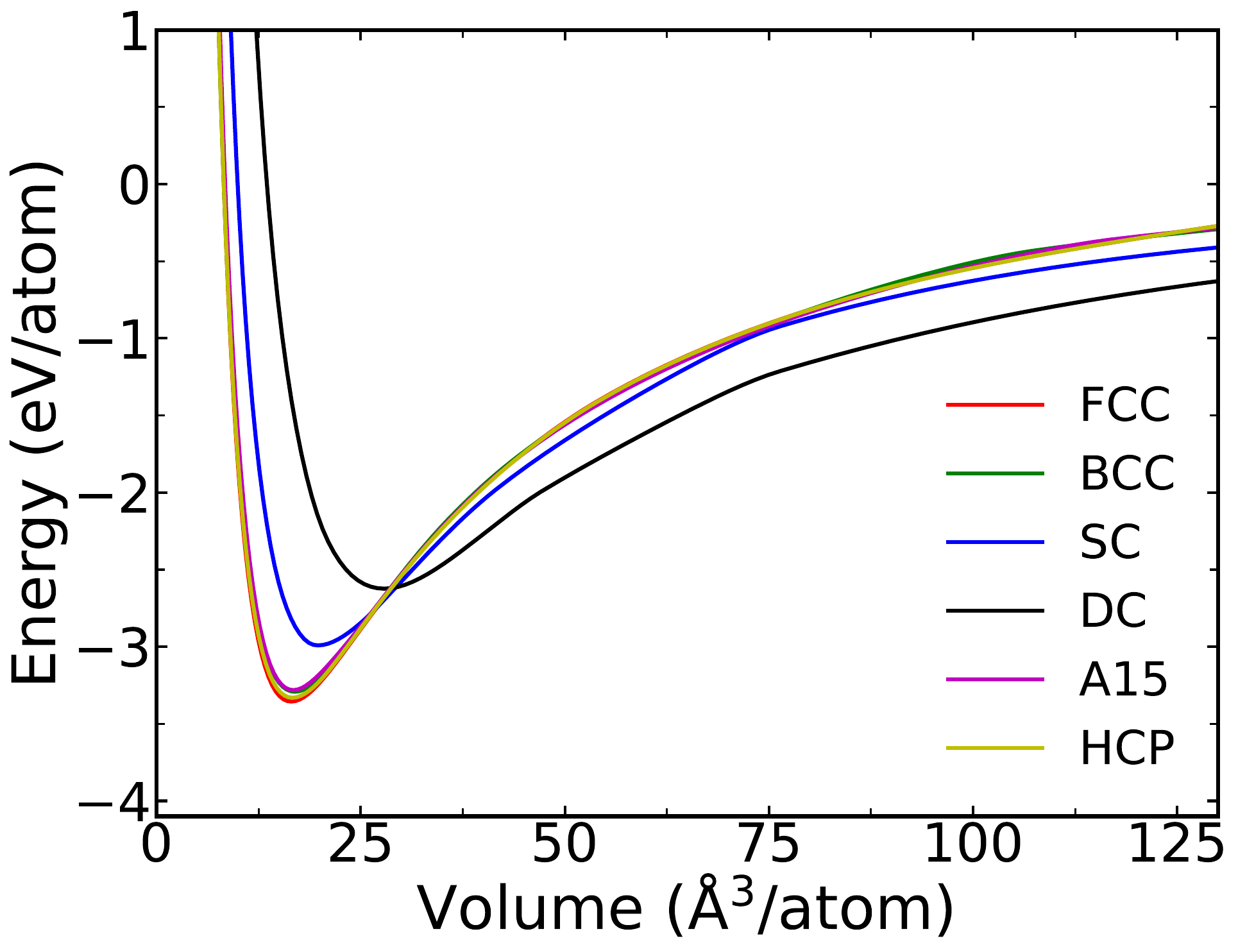}\qquad{} \includegraphics[height=5cm]{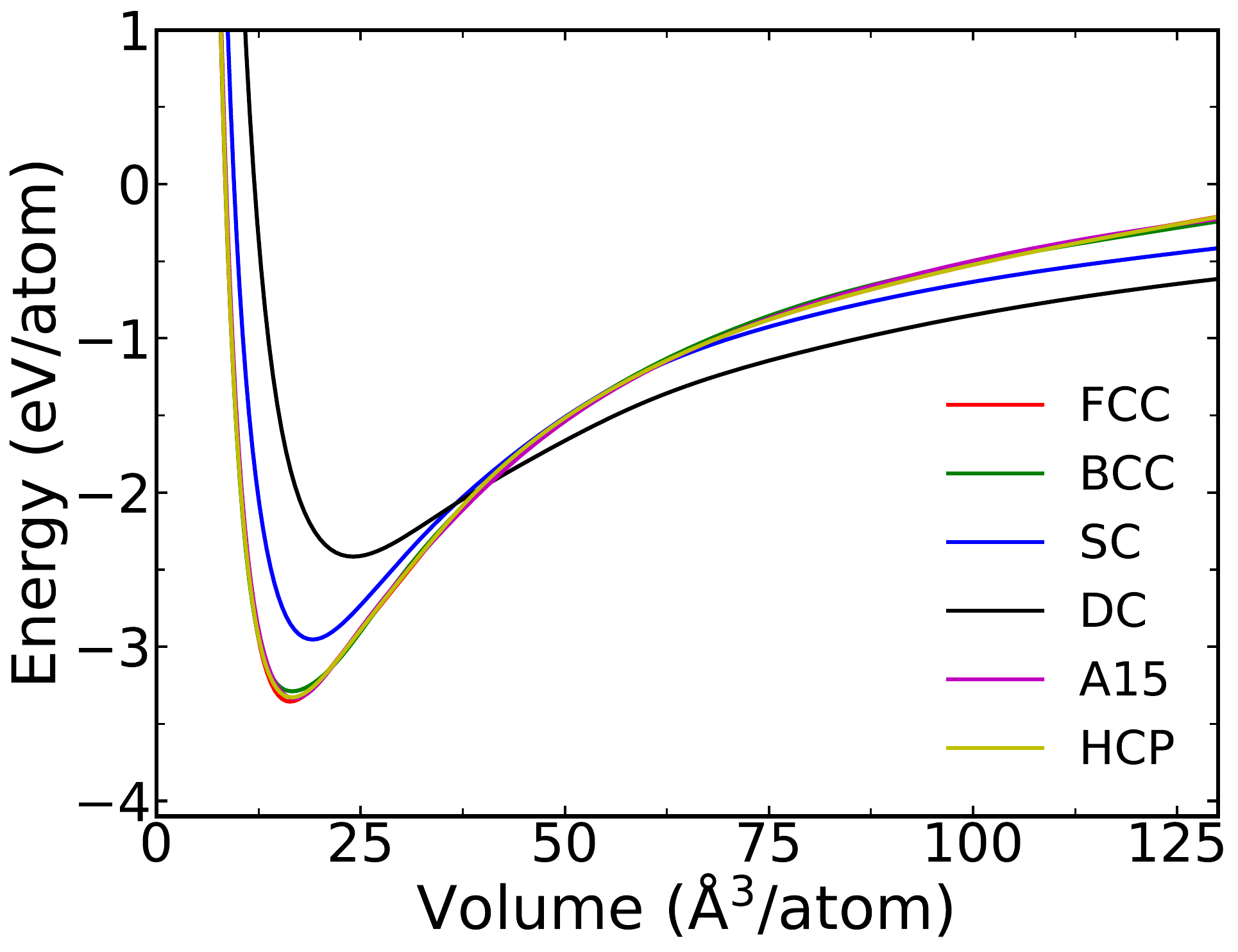}

\caption[]{Equation of states of Al computed with the MTP (left) and MT (right)
potentials developed in this work.}
\label{fig:EOS}
\end{figure}

\begin{figure}[!ht]
\vspace{0cm}
 \centering \includegraphics[height=5cm]{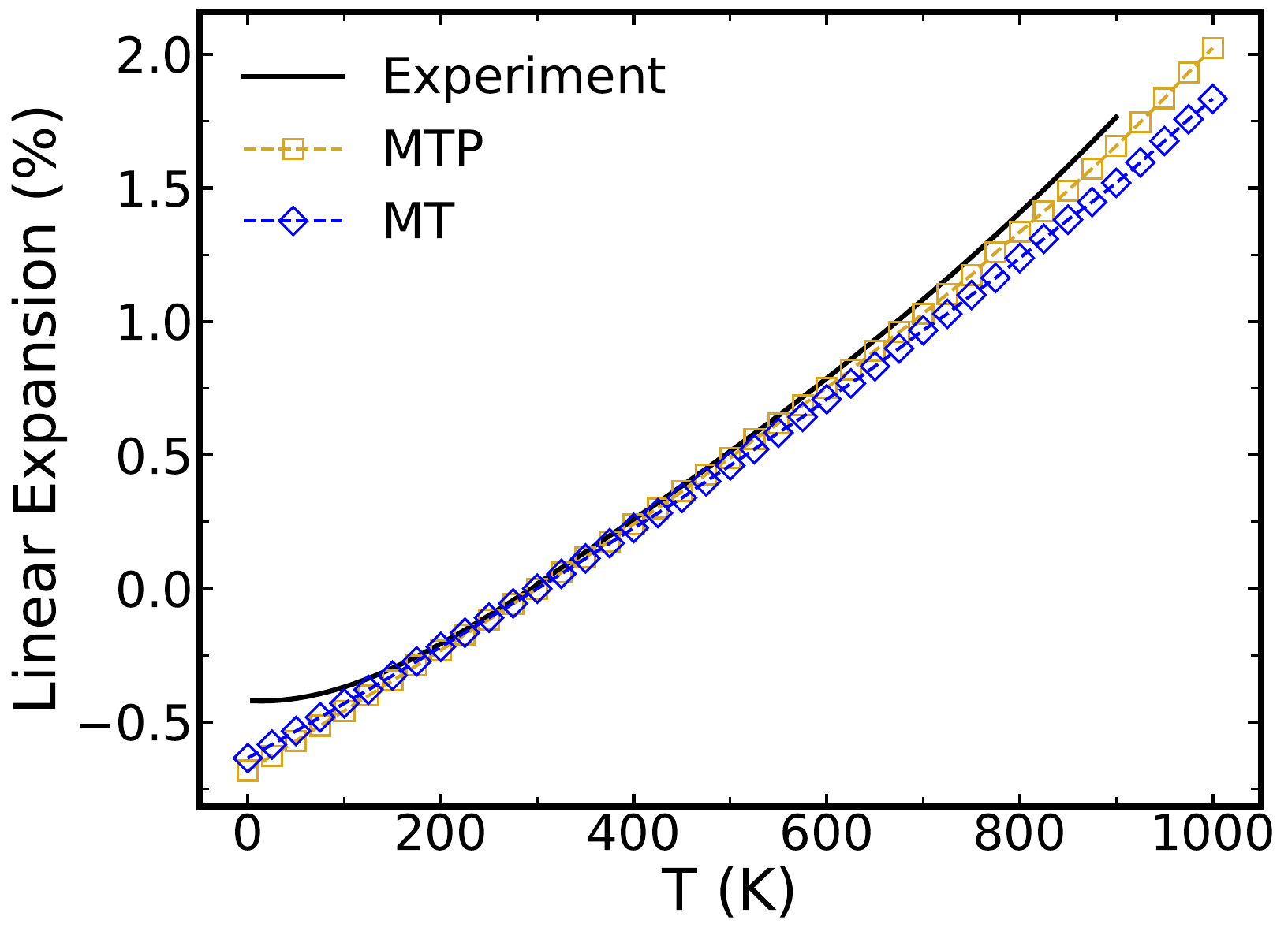}
\caption[]{Linear thermal expansion relative to room temperature (293K) predicted
by the MTP and MT potentials for Al in comparison with experimental
data \citep{kozyrev2022thermodynamic}.}
\label{fig:thermal_expansion}
\end{figure}

\begin{figure}[!ht]
\vspace{0cm}
 \centering \includegraphics[height=5cm]{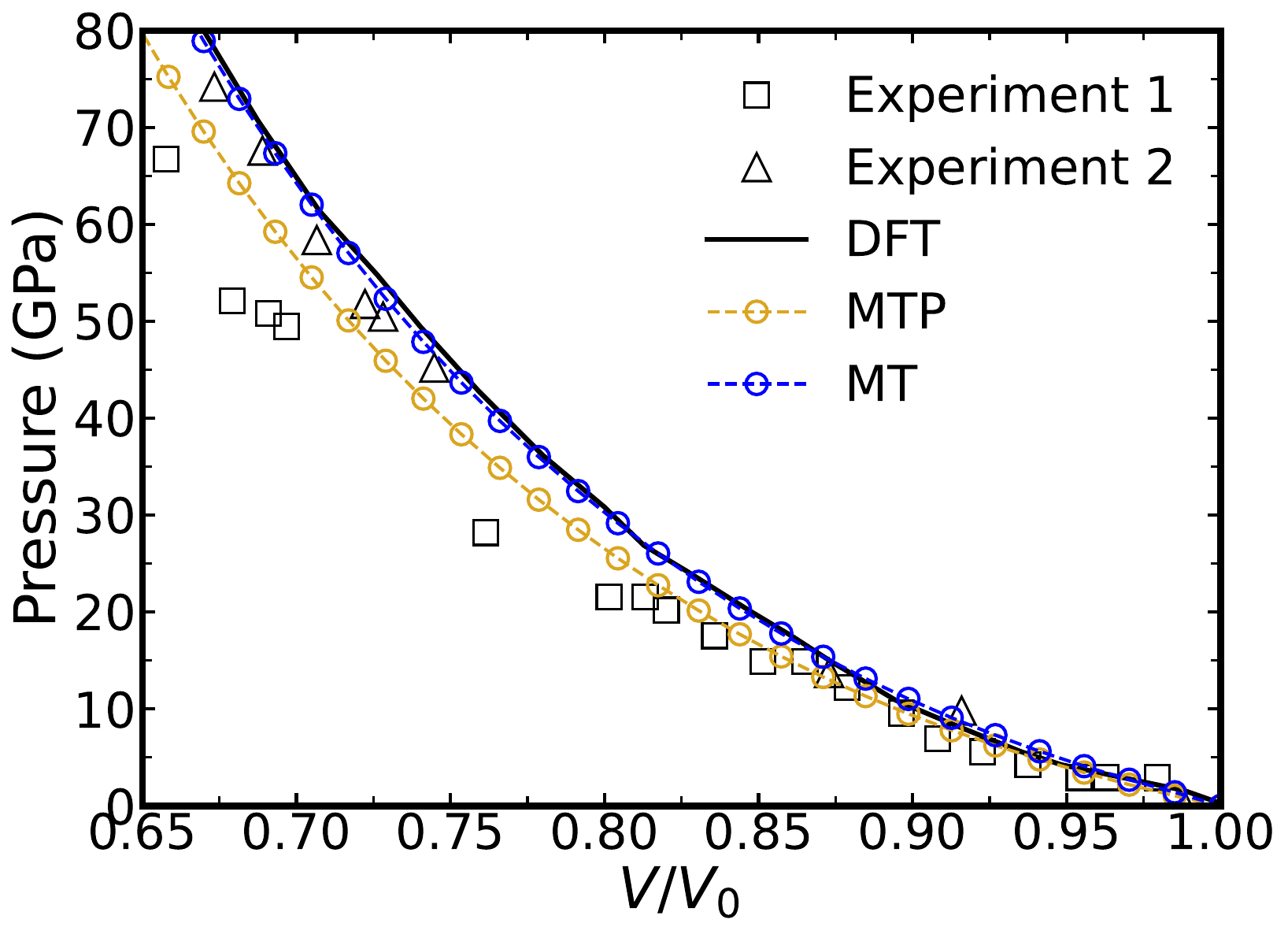} \caption[]{Isotropic compression predicted by the MTP and MT potentials for
Al in comparison with DFT calculations \citep{ning2023ab} and two
sets of experimental data (Experiment 1: Ref.\citep{dewaele2004equations};
Experiment 2: Ref.\citep{akahama2006evidence} ).}
\label{fig:strain_pressure}
\end{figure}

\begin{figure}[!ht]
\centering \includegraphics[height=5cm]{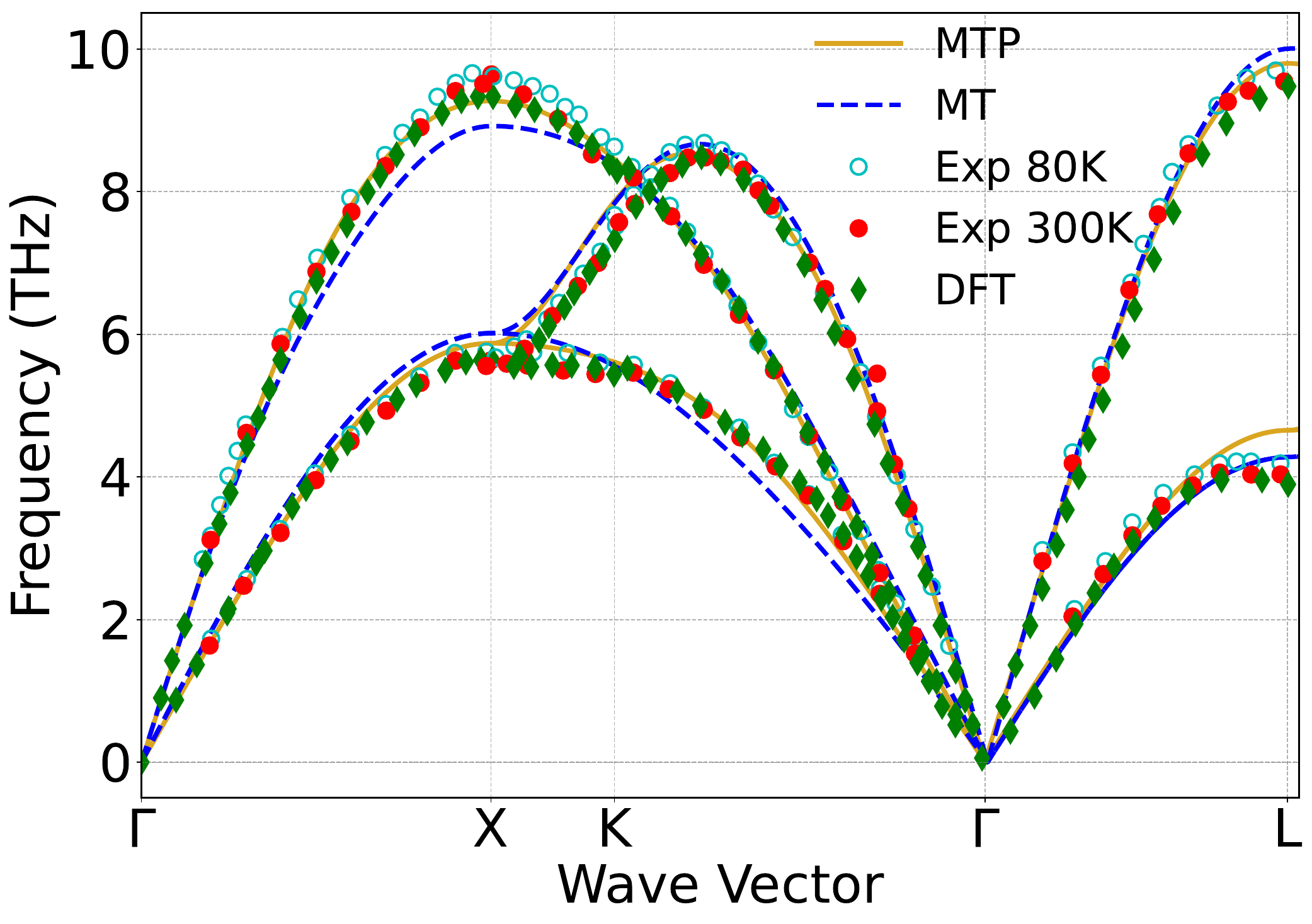} 
 \caption{Phonon dispersion curves of Al computed at 0 K with the MTP and MT
potentials in comparison with DFT~\citep{pun2020development} and
experimental data~\citep{stedman1966dispersion}.}
\label{fig:phonon}
\end{figure}

Since the MTP model employed in this work has a minimum cutoff distance
of 0.5~\AA, unphysical atomic motions may occur if two atoms approach
closer than this limit. To prevent such artifacts, the Ziegler--Biersack--Littmark
(ZBL) screened nuclear repulsion potential~\citep{ziegler1985stopping}
was applied to describe short-range, high-energy atomic collisions.
The cutoff distance for the ZBL interaction was set to 1.0~\AA,
and the switching function was activated at 0.5~\AA. Consequently,
whenever two atoms approach within 1.0~\AA (a regime where the MTP
may lose accuracy due to the absence of such configurations in the
training database) the ZBL potential introduces a strong repulsive
force, effectively preventing unphysical atomic overlap.

The Al properties predicted by the MTP potential developed in this
work are summarized in Table \ref{tab:al_properties}. For comparison,
we also include the properties predicted by the DFT calculations,
the modified Tersoff (MT) potential developed in this work, the PINN
potential \citep{pun2020development}, the EAM potential \citep{mishin1999interatomic},
and the ADP potential \citep{starikov2020optimized}. For the ADP
potential, the results for the elastic constants $C_{11}$, $C_{12}$,
$C_{44}$, and the melting temperature $T_{m}$ calculated in this
work differ from the values reported in the original publication \citep{starikov2020optimized}.
All defect energies correspond to fully relaxed configurations. The
phonon dispersion curves were computed using the \textsf{phonopy}
code \citep{phonopy}. The melting temperature was calculated using
the phase coexistence method \citep{Morris94,Morris02,Howells:2018aa}.
A periodic simulation cell was first equilibrated in the NPT ensemble
at the expected melting temperature and zero pressure. One half of
the cell was then heated to form a liquid phase, while the other half
remained solid. The entire system was relaxed in the NPH (constant
enthalpy) ensemble. During the relaxation, the solid--liquid interface
moved as the volume fractions of the solid and liquid phases adjusted
the equilibrium. In the process, the system temperature evolved until
it converged to a value corresponding to the melting temperature at
the ambient pressure.

The performance of the MTP and MT potentials in reproducing the equation
of states, thermal expansion, pressure-volume relationship, and phonon
dispersion relation, as well as their comparison with DFT and experimental
data, are shown in Figs. \ref{fig:EOS}, \ref{fig:thermal_expansion},
\ref{fig:strain_pressure}, \ref{fig:phonon}.

\section{The modified Tersoff potential for Al}

In this section, we describe the development of a new modified Tersoff
(MT) potential for Al. 

In the MT model, the total energy of a single-component system is
represented in the form 
\[
E=\dfrac{1}{2}\sum_{i\neq j}V_{ij}(r_{ij}),
\]
where $r_{ij}$ is distance between atoms $i$ and $j$. The pair
interaction energy $V_{ij}$ is represented in the form 
\begin{equation}
V_{ij}=f_{c}(r_{ij})\left[A\exp(-\lambda_{1}r_{ij})-b_{ij}B\exp(-\lambda_{2}r_{ij})+c_{0}\right],\label{eq:MT_1}
\end{equation}
where the bond order factor $b_{ij}$ is given by 
\begin{equation}
b_{ij}=\left(1+\xi_{ij}^{\eta}\right)^{-\delta}.\label{eq:MT_2}
\end{equation}
Here, $\xi_{ij}$ is an angular-dependent three-body sum
\begin{equation}
\xi_{ij}=\sum_{k\ne i,j}f_{c}(r_{ik})g(\theta_{ijk})\exp\left[\alpha(r_{ij}-r_{ik})^{\beta}\right],\label{eq:MT_3}
\end{equation}
where $\theta_{ijk}$ is the angle between the bonds $ij$ and $ik$.
Physically, $(1+\xi_{ij})$ represent an effective coordination number
of atom $i$. The cutoff function $f_{c}(r_{ij})$ is has the form
\[
f_{c}(r)=\begin{cases}
1, & r\le R_{1}\\
\dfrac{1}{2}+\dfrac{9}{16}\cos\left(\pi\dfrac{r-R_{1}}{R_{2}-R_{1}}\right)-\dfrac{1}{16}\cos\left(3\pi\dfrac{r-R_{1}}{R_{2}-R_{1}}\right), & R_{1}<r<R_{2}\\
0, & r\ge R_{2},
\end{cases}
\]
where $R_{1}$ and $R_{2}$ are the inner and outer cutoff radii.
The angular function $g(\theta_{ijk})$ has the form 
\[
g(\theta)=c_{1}+\dfrac{c_{2}(h-\cos\theta)^{2}}{c_{3}+(h-\cos\theta)^{2}}\left\{ 1+c_{4}\exp\left[-c_{5}(h-\cos\theta)^{2}\right]\right\} .
\]

The original version of this model was proposed by Tersoff \citep{Tersoff88,Tersoff:1988dn,Tersoff:1989wj}
and modified by Kumagai et al.~\citep{Kumagai:2007ly} by generalizing
the angular function $g(\theta)$. It was recently proposed \citep{Purja-Pun:2017aa}
to add the coefficient $c_{0}$ in Eq.(\ref{eq:MT_1}) for better
control of the attractive part of the potential. The potential has
16 free parameters: $A$, $B$, $\alpha$, $h$, $\eta$, $\lambda_{1}$,
$\lambda_{2}$, $R_{1}$, $R_{2}$, $\delta$, $c_{0}$, $c_{1}$,
$c_{2}$, $c_{3}$, $c_{4}$ and $c_{5}$. The power $\beta$ is a
fixed odd integer. In this work, we chose $\beta=1$. 

For strongly covalent elements, such as Si and C, the outer cutoff
$R_{2}$ is chosen between the first and second coordination shells
of the diamond cubic  structure. In this work, we apply the MT model
to a metal (Al) by expanding the outer cutoff to 6.56~$\textrm{\AA}$,
which includes several coordination shells. In this form, MT becomes
a many-body atomic interaction model similar to MEAM and ADP, with
$b_{ij}$ in Eq.(\ref{eq:MT_2}) playing the role of the embedding
function and $\xi_{ij}$ being similar to the electron density function. 

The MT potential was trained on a DFT database similar to that used
to train the MTP potential. The optimized parameters are listed in
Table \ref{tab:Optimized-parameters-MT}. The properties predicted
by the MT potential are summarized in Table \ref{tab:al_properties}
and in Figs. \ref{fig:EOS}, \ref{fig:thermal_expansion}, \ref{fig:strain_pressure},
\ref{fig:phonon}. The potential underestimates the surface energies
relative to the experimental data and DFT calculations, as do many
EAM potentials. The predicted melting temperature of Al is very accurate
without fitting. Overall, the potential demonstrates the accuracy
on par with some of the best EAM and ADP potentials for metals. 

\begin{table}
\centering{}\caption{Optimized parameters of the MT potential for Al developed in this
work.}\label{tab:Optimized-parameters-MT}
\bigskip{}
\begin{tabular}{|l|c|c|c|c|}
\hline 
Parameter & Value &  &  & \tabularnewline
\hline 
$A$ eV & $9.410980\times10^{2}$ &  & $c_{1}$ & $3.412730\times10^{-1}$\tabularnewline
\hline 
$B$ eV & $1.440410\times10^{1}$ &  & $c_{2}$ & $-4.205600\times10^{-1}$\tabularnewline
\hline 
$\lambda_{1}$ (Å$^{-1}$) & $2.860720\times10^{0}$ &  & $c_{3}$ & $4.362630\times10^{0}$\tabularnewline
\hline 
$\lambda_{2}$ (Å$^{-1}$) & $8.590480\times10^{-1}$ &  & $c_{4}$ & $1.032600\times10^{1}$\tabularnewline
\hline 
$\eta$ & $5.199240\times10^{0}$ &  & $c_{5}$ & $1.83167\times10^{0}$\tabularnewline
\hline 
$\eta\times\delta$ & $5.790380\times10^{0}$ &  & $h$ & $2.557150\times10^{-1}$\tabularnewline
\hline 
$\alpha$ & $1.820040\times10^{0}$ &  & $R_{1}$ (Å) & $5.633770\times10^{0}$\tabularnewline
\hline 
$\beta$ & $1$ &  & $R_{2}$ (Å) & $6.588574\times10^{0}$\tabularnewline
\hline 
$c_{0}$ (eV) & $2.146950\times10^{-2}$ &  &  & \tabularnewline
\hline 
\end{tabular}
\end{table}

\section{Effect of stress on the correlation between stacking fault energy
and HCP-FCC energy difference}

\begin{figure}[h]
\includegraphics[width=1\textwidth]{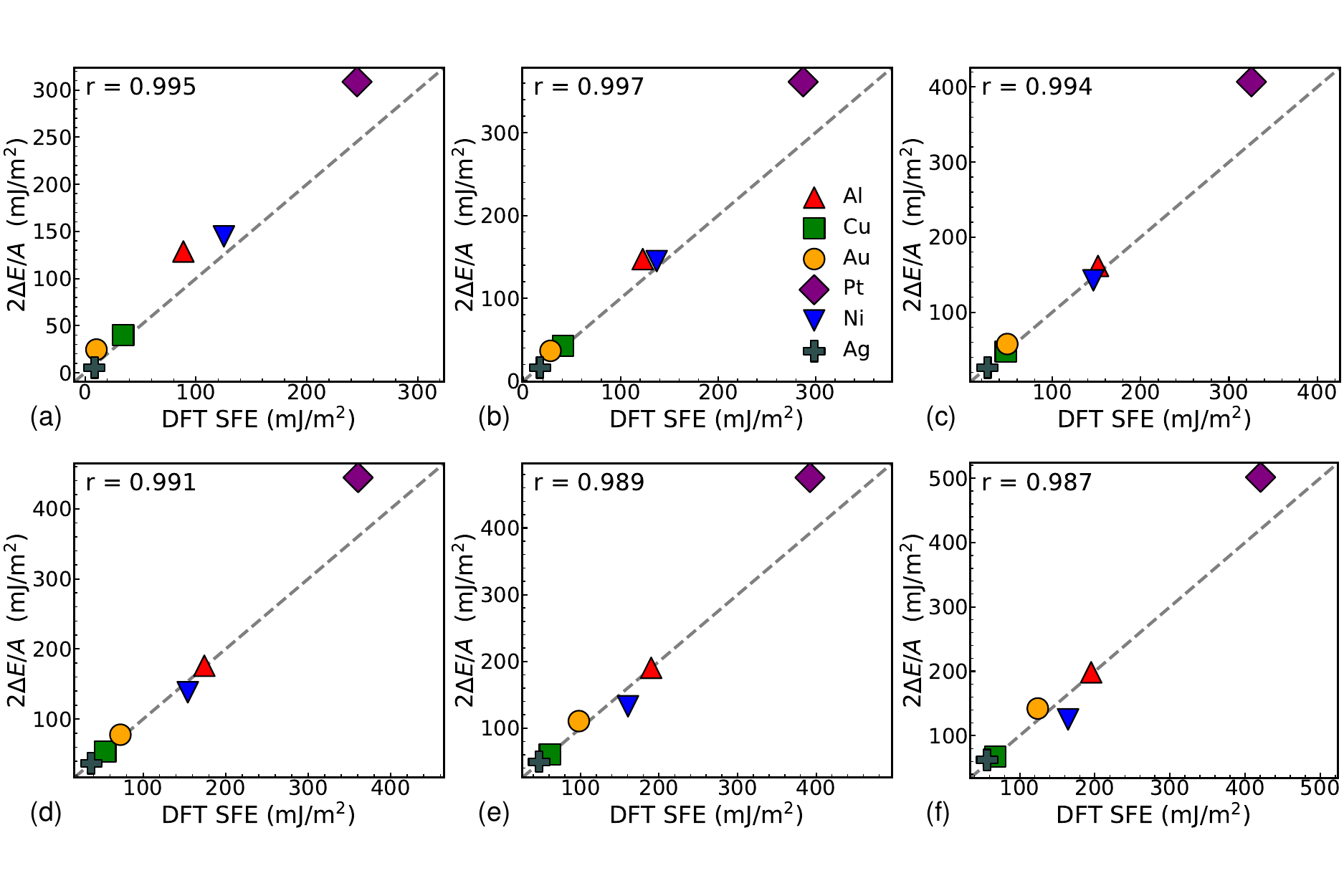}

\caption{SFE for six metals obtained by DFT calculations and predicted by Eq.(\ref{eq:hcp-fcc-1})
using the HCP-FCC energy difference $\Delta E$. The dashed line represents
perfect agreement between the two calculations. The different panels
correspond to the following values of the normal stress $\sigma_{n}$:
(a) $-5$ GPa. (b) 0 GPa. (c) $5$ GPa. (d) $10$ GPa. (e) $15$ GPa.
(f) $20$ GPa. The strength of correlation is measured by Pearson's
correlation factor $r$ indicated in the top left corner of each plot.}\label{fig:SFE-vs-HCP-FCC}
\end{figure}

In the main text, we discussed the correlation between the equilibrium
SFE $\gamma_{\mathrm{SFE}}$ and the HCP-FCC energy difference $\Delta E$
(per atom). The theoretical prediction of this correlation is expressed
by the equation
\begin{equation}
\gamma_{\mathrm{SF}}=2\Delta E/A,\label{eq:hcp-fcc-1}
\end{equation}
where $A$ is the area per atom on the (111) plane of the SF. In the
main text, we compared the DFT values of $\gamma_{\mathrm{SFE}}$
with predictions from Eq.(\ref{eq:hcp-fcc-1}) in the absence of normal
stress. The $\Delta E$ values appearing in this equation were computed
by considering the HCP and FCC phases at the same atomic volume equal
to the stress-free atomic volume of the FCC phase. The results were
shown in Fig.~\ref{fig:HCP-FCC}(b) and reveal a significant correlation
across the six FCC metals studied in this work. Here, we examine this
correlation under applied normal stresses $\sigma_{n}$ within the
stress interval in which both phases remain mechanically stable. $\Delta E$
was computed under the constraint that the HCP structure has the same
in-plane atomic density in the (111) plane as the FCC structure, and
the comparison is made at the same atomic volume. The normal stress
is defined as the stress along the {[}111{]} direction in the FCC
structure. The corresponding stress in the HCP structure is slightly
different but remains very close in magnitude. Figure \ref{fig:SFE-vs-HCP-FCC}
shows the comparison for several representative stresses. The plots
show that the correlation persists and not limited to stress-free
SFs but persists under both tensile and compressive stresses.

\section{Convergence of stacking-fault energy calculations with potentials}

\begin{figure}[h]
\begin{centering}
\includegraphics[width=0.9\textwidth]{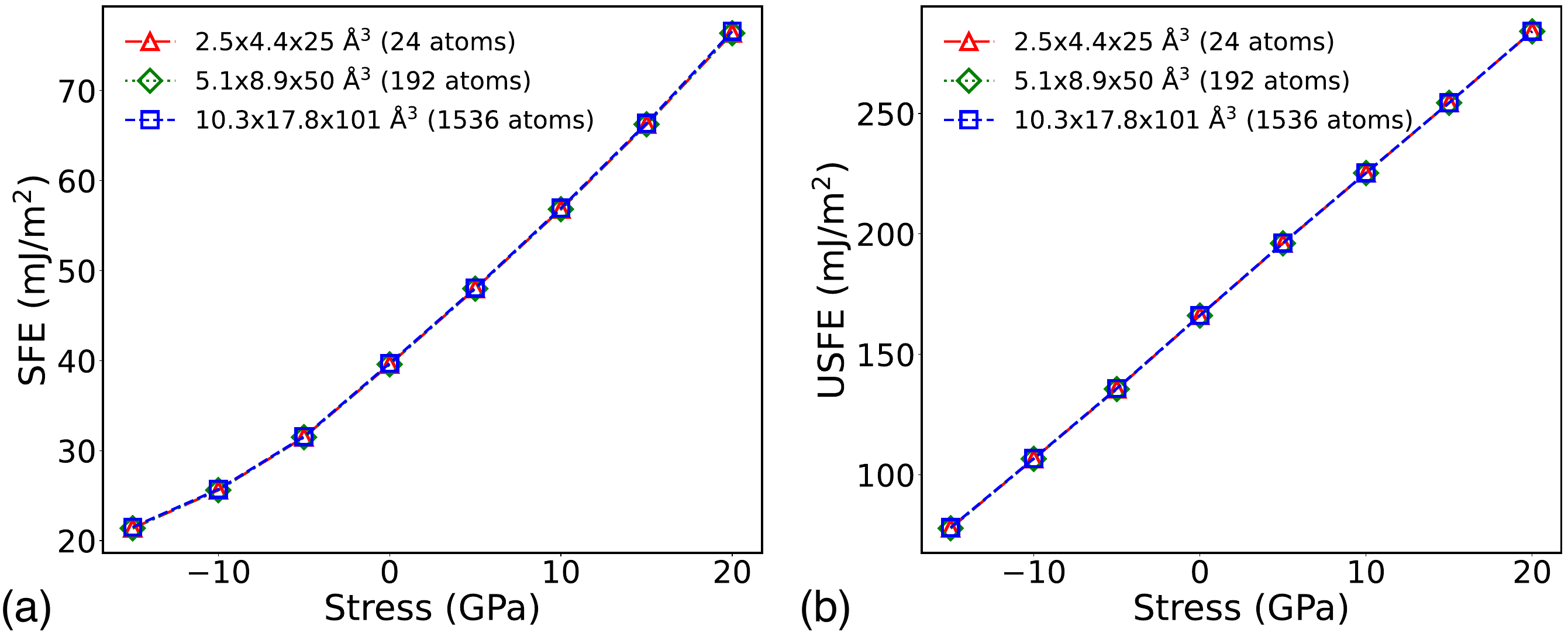}\caption{Stress dependence of (a) SFE and (b) USFE in Cu computed with the
MTP potential for different supercell sizes. The supercell used in
the DFT calculations was $2.5\times4.4\times25\ \textrm{\protect\AA}^{3}$
(24 atoms).}\label{fig:Stress-dependence}
\par\end{centering}
\end{figure}

The SFEs and USFEs predicted by the interatomic potentials were computed
in supercells identical to those used in the DFT calculations, to
ensure consistency in potential testing. However, it was also interesting
to calculate SFE and USFE values with machine-learning potentials
using larger supercells. Assuming that the machine-learning potentials
are sufficiently accurate to serve as a proxy for DFT calculations,
this can be an indirect test of the convergence of DFT values with
respect to system size. As an example, Figure \ref{fig:Stress-dependence}
shows a comparison for MTP Cu, demonstrating excellent size convergence.

\section{Dislocation dissociation width from atomistic simulations}

In Section \ref{subsec:Application-to-dislocation} of the main text,
we presented the results of atomistic simulations of dislocation dissociation
into Shockley partials. The simulations were performed for an edge
dislocation using the MTP \citep{nitol2025evaluating} and EAM \citep{mishin2001structural}
potentials. The table below summarizes the results for different dimensions
of the simulation block to evaluate the size convergence. Although
the dissociation width varies with the block size, this dependence
does not alter the main conclusions of the main text.

\begin{table}[h]
\centering \caption{Simulation results for the partial dislocation separation width $d$
in Cu under the normal stresses of $0$, $20$, and $-10$ GPa using
the MTP \citep{nitol2025evaluating} and EAM \citep{mishin2001structural}
potentials.}
\begin{tabular}{lccc}
\hline 
Model size (nm$^{3}$) & $d$ at 0 GPa (nm) & $d$ at 20 GPa (nm) & $d$ at $-10$ GPa (nm)\tabularnewline
\hline 
\multicolumn{4}{c}{EAM}\tabularnewline
$1.8\times15\times12$ & 3.25 & 5.81 & 3.89\tabularnewline
$0.9\times15\times12$ & 3.25 & 5.94 & 3.89\tabularnewline
$0.9\times30\times25$ & 3.51 & 8.88 & 4.53\tabularnewline
$0.9\times61\times50$ & 3.64 & 10.42 & 4.66\tabularnewline
$0.9\times122\times100$ & 3.64 & 10.66 & 4.66\tabularnewline
\hline 
\multicolumn{4}{c}{MTP}\tabularnewline
$1.8\times15\times12$ & 3.53 & 2.37 & 3.79\tabularnewline
$0.9\times15\times12$ & 3.53 & 2.37 & 3.79\tabularnewline
$0.9\times30\times25$ & 3.91 & 2.63 & 4.56\tabularnewline
$0.9\times61\times50$ & 4.04 & 2.63 & 4.82\tabularnewline
$0.9\times122\times100$ & 3.92 & 2.63 & 4.81\tabularnewline
\hline 
\end{tabular}\vspace{0.3em}
 \label{table:Cu_partial}
\end{table}

\end{document}